\documentclass[12pt,letterpaper]{article}
\usepackage{amsbsy}
\usepackage{amsfonts}
\usepackage{amssymb}
\usepackage{epsfig}
\usepackage[dvips]{color}
\RequirePackage[english]{babel}
\usepackage[verbose]{wrapfig}
\usepackage{bm}
\def\apj{{Astrophys.~J.~}}
\def\nat{{Nature~}}
\def\apjl{{Astrophys.~J. Lett.~}}
\def\apjs{{Astrophys.~J. Suppl.~}}
\def\jgr{J. Geophys. Res. (Space Phys.)~}
\def\ssr{Space Sci. Rev.~}
\def\aara{Ann. Rev. Astron. Astrophys.~}
\def\aap{Astron. Astrophys.~}
\def\solphys{Solar Phys.~}
\def\mnras{MNRAS~}
\def\grl{Geophys. Res. Lett.~}
\def\apss{Astrophys. Space Sci.~}

\begin{document}

\textbf{Heliosheath Processes and the Structure of the Heliopause: Modeling Energetic Particles, Cosmic Rays, and Magnetic Fields}\footnote{Publication resulting from an International ISSI Team of the same name.}


N.~V. Pogorelov,$^{1}$ H. Fichtner,$^2$ A. Czechowski,$^3$ A. Lazarian,$^4$ B. Lembege,$^5$ J.~A. le Roux,$^1$ M.~S. Potgieter,$^6$
K. Scherer,$^2$ E.~C. Stone,$^7$ R.~D. Strauss,$^6$ T. Wiengarten,$^2$ P. Wurz,$^8$ G. P. Zank,$^1$ M. Zhang$^9$

$^1$ Department of Space Science,  University of Alabama in Huntsville, AL 35805, USA

$^2$Institut f\"ur Theoretische Physik IV, Ruhr-Universit\"at Bochum, 44780 Bochum, Germany

$^3$Space Research Centre, Warsaw, Poland

$^4$University of Wisconsin, Madison, USA

$^5$Laboratoire Atmosph\`eres, Milieux, Observations Spatiales (LATMOS), Guyancourt, France

$^6$North-West University, Campus Potchefstroom, South Africa

$^7$California Institute of Technology, Pasadena, USA

$^8$Universit\"at Bern, Switzerland

$^9$Florida Institute of Technology, Melbourne, USA


\begin{abstract}
This paper summarizes the results obtained by the team ``Heliosheath Processes and the Structure of the Heliopause: Modeling Energetic Particles, Cosmic Rays, and Magnetic Fields'' supported by the International Space Science Institute (ISSI)
in Bern, Switzerland. We focus on the physical processes occurring in the outer heliosphere,
especially at its boundary called the heliopause, and in the local interstellar medium.
The importance of magnetic field, charge exchange between neutral atoms and ions,
and solar cycle on the heliopause topology and observed heliocentric distances to different heliospheric discontinuities are discussed. It is shown that time-dependent, data-driven boundary conditions are necessary to describe
the heliospheric asymmetries detected by the \textit{Voyager} spacecraft. We also discuss the structure of the heliopause,
especially due to its instability and magnetic reconnection. It is demonstrated that the Rayleigh--Taylor
instability of the nose of the heliopause creates consecutive layers of the interstellar and heliospheric plasma which are magnetically connected to different sources. This may be a possible explanation of abrupt changes in the
galactic and anomalous cosmic ray fluxes observed by \textit{Voyager~1} when it was crossing the heliopause structure for a period of about one month in the summer of 2012. This paper also discusses the plausibility of fitting simulation results to a number of observational data sets obtained by \emph{in situ} and remote measurements. The distribution of magnetic field
in the vicinity of the heliopause is discussed in the context of \textit{Voyager} measurements. It is argued that a classical heliospheric current sheet formed due to the Sun's rotation is not observed by \emph{in situ} measurements and should not be expected to exist in numerical simulations extending to the boundary of the heliosphere. Furthermore, we discuss the transport of energetic particles in the inner
and outer heliosheath, concentrating on the anisotropic spatial diffusion diffusion tensor and the pitch-angle dependence of
perpendicular diffusion and demonstrate that the latter can explain the observed pitch-angle anisotropies of both the anomalous and
galactic cosmic rays in the outer heliosheath.


\end{abstract}

\section{Introduction}
\label{constraints}
The Sun and the local interstellar medium (LISM) move with respect to each other creating an interaction pattern
likely similar to many other stellar wind collisions with their local interstellar environments.
One would expect differences in details though. E.g., jets and collimated outflows are ubiquitous in astrophysics, appearing in environments
as different as young stellar objects, accreting and isolated neutron stars, stellar mass
black holes, and in supermassive black holes at the centers of Active Galactic Nuclei.
Despite the very different length scales, velocities and composition of these various
types of jets, they share many basic physical principles. They are typically long-duration, supersonically ejected
streams that propagate through and interact with the surrounding medium, exhibiting
dynamical behavior on all scales, from the size of the source to the longest
scales observed. Charged particle flows emitted by stars moving through the interstellar
space form astrotails which can be very different in shape and length, depending on
the astrophysical object under consideration. The Guitar Nebula
is a spectacular example of an H$\alpha$ bow shock nebula observed by the Hubble Space
Telescope (HST) and Chandra \cite{2002ApJ...575..407C}. 
The physics of the interaction is very similar to that of the solar wind (SW)--LISM interaction, but there are substantial differences in
the stellar wind confinement topology. Mira's astrotail observed by the Galaxy Evolution Explorer \cite{2007Natur.448..780M}
extends to 800,000 AU. Carbon Star IRC+10216, on the contrary, exhibits a very wide
astropause and a short astrotail \cite{2010ApJ...711L..53S}. 
The heliotail cannot be observed from outside, but its signatures have been identified in energetic neutral atom (ENA) measurements
with the \emph{Interstellar Boundary Explorer} (IBEX) \cite{2013ApJ...771...77M}.
The heliotail properties have been investigated theoretically \cite{1974ApJ...194..187Y} and numerically
\cite{2015ApJ...800L..28O,2015ApJ...812L...6P,Kleimann-etal-2016}.

The goal of this paper is to demonstrate that a better knowledge of the SW and LISM properties makes it possible to explain, at least qualitatively, a number of observational data. Moreover, it is clearly understood now that proper interpretation of observations is
impossible without taking into account genuine time dependence of the SW--LISM interaction. To reproduce \emph{in situ} and remote observations of the distant SW, we need to take advantage of the full set of observational data in the inner heliosphere.
On the other hand, in situ measurements by \textit{Voyager}~1 (\textit{V1} and remote observations of ENA fluxes from \textit{IBEX}, Ly$\alpha$ backscattered emission from the \emph{Solar Heliospheric Observatories} (\emph{SOHO}) Solar Wind Anisotropy (SWAN) experiment, Ly$\alpha$ absorption profiles in the directions toward nearby stars from the Hubble Space Telescope (HST),
1--10 TeV cosmic ray anisotropy from multiple air shower missions (a number references can be found in \cite{2014ApJ...790....5Z}), and starlight polarization from \cite{2015ApJ...814..112F}
provide us with invaluable information about the LISM properties. The availability of realistic, data driven boundary conditions, makes SW--LISM interaction models a powerful tool to investigate the properties of the heliospheric interface.

It is not the purpose of this paper  to provide an extensive review of the community efforts to investigate the physical processes in the vicinity of the heliospheric interface, for those see, e.g., the reviews \cite{Fahr-etal-1986}, \cite{Suess-1990}, \cite{Zank-1999}, or \cite{Zank-2015}. This is rather a report on the
activity of an international team with a name coinciding with the paper title that was recently supported by
the International Space Science Institute (ISSI) in Bern, Switzerland. For this reason, we mostly address scientific results obtained by the team itself and their relation to other recent studies. We also identify the challenges that emerged in the investigation of the structure of the heliopause (HP) and heliosheath processes, especially related to energetic particles, cosmic rays, and magnetic fields.
In this paper we also address a number of issues discussed in the review article \cite{2016SSRv..200..475O}.

\section{Constraints on the model boundary conditions from observations}
From a purely magetohydrodynamic (MHD) perspective, the global structure of the SW--LISM interaction is clear.
When two plasma streams collide, a tangential discontinuity (here the  HP) should form that separates the SW and LISM plasmas.
This discontinuity can be interpreted as a constituent component of the solution to an MHD Riemann problem \cite{1961JApMM..25..148G}.
Other MHD discontinuities (fast and slow shocks, contact and rotational (Alfv\'en) discontinuities, slow- and fast-mode rarefaction waves) may
or may not form on the LISM and SW sides of the HP, but the presence of a tangential discontinuity is obligatory.
The SW--LISM boundary cannot be a rotational discontinuity, as suggested in \cite{2015ApJ...806L..27G}, because this means the absence of any separation boundary.
Moreover, the SW and LISM velocities at any point on the HP would be in the same direction and have the magnitudes equal to the
Alfv\'en speed on both sides. Early studies of the SW--LISM interaction were mostly theoretical because no boundary conditions were
available either in the SW or the LISM. The seminal paper \cite{1961ApJ...134...20P} proposed a powerful tool to solve the SW--LISM interaction
problem through the application of the MHD equations. The possibility to use continuum equations to model the collisionless SW is supported
by the dramatic decrease in the ion mean free path due to scattering on magnetic field fluctuations caused by numerous
kinetic instabilities typical of the SW flow. Although it is known that only the global, macroscopic structure of the plasma flow can be described using a continuum description, the efficiency of the MHD/hydrodynamic approach cannot be overestimated.

The importance of charge exchange between the LISM hydrogen (H) atoms and SW ions has been known since late 60's
\cite{1969Natur.223..936B,1971NPhS..233...23W,1975Natur.254..202W,1977Holzer,1976Vsyliunas}.
The resonant charge exchange between ions and neutral atoms which have non-zero relative velocity is a death/birth process in which a parent ion-atom pair disappears producing an ion with the parent neutral atom properties and an atom with the properties of the parent ion.
Newly created (secondary) atoms continue to move unaffected by the electromagnetic field, whereas newly born (pickup) ions (PUIs)
are acted upon the motional electric field until their velocity becomes equal to the velocity of the background plasma \cite{1961ApJ...134...20P}. The PUI distribution function is originally a ring-beam, but they quickly scatter onto a shell distribution.
With increasing the heliocentric distance, some particles fill in the shell at lower energies, while other particles are accelerated to
higher energies \cite{Isenberg87}, so PUIs are not in equilibrium with the SW protons. Secondary neutral atoms can propagate
far into the LISM where they may experience charge exchange producing a new population of PUIs, which arguably produce
so-called energetic neutral atoms (ENAs) measured by the \emph{Interstellar Boundary Explorer} (\emph{IBEX}) \cite{2009Sci...326..959M,Jacob10}.
A number of important consequences of such charge exchange had been identified long before the first numerical simulation was made.
These are the SW deceleration and heating, and filtration of interstellar atoms
at the HP, which prevents a substantial fraction of those atoms from entering the heliosphere and results in a so-called
hydrogen wall in front of the HP, and many others. Moreover, charge exchange decreases asymmetries of the three-dimensional (3D) heliosphere
caused by the action of the interstellar magnetic field \cite{2007ApJ...668..611P}. This raises questions about the coupling of the
heliospheric magnetic field (HMF) and ISMF at the HP.

\emph{Voyager 1} (\emph{V1}) crossed the HP in August 2012 \cite{Burlaga147,Burlaga-Ness-2014,2014ApJ...795L..19B,Gurnett-etal-2013}, whereas \emph{Voyager 2} (\emph{V2}) is still in the inner heliosheath  -- the SW region between the HP and the heliospheric termination shock (TS). The spacecraft crossings of the TS and HP, and measurements performed in their vicinity were accompanied
by a number of interesting physical phenomena, which will be addressed in this paper in some detail: (1) the asymmetry of the heliosphere
and the contribution of time-dependent factors; (2) the heliospheric current sheet (HCS) behavior; (3)
anisotropy in the anomalous and galactic cosmic ray (ACR and GCR) fluxes; (4) a prolonged, almost two-year period of sunward flow at V1 before it crossed the HP; (5) puzzling variations in the ACR and GCR flux within a month while V1 was crossing a finely structured HP;
(6) observations of the LISM turbulence spectrum and issues related to the production of an enhanced flux of energetic neutral atoms (ENAs)  originating beyond the HP and propagating toward \emph{IBEX} detectors from directions roughly perpendicular to magnetic field lines draped around the HP \cite{2009Sci...326..959M}; (7) the nature of the ISMF draping and the change of the ISMF
direction as the LISM flow approaches the HP; (8) the HP instability and possible signatures of magnetic reconnection in its vicinity; (9) the ratio between the parallel and perpendicular diffusion coefficients that can be derived from \emph{V1} observations, etc.
We will also discuss our predictions regarding the inner heliosheath width in the \emph{V1} and \emph{V2} directions. Finally, the flow in the heliotail will be discussed together with its possible effect on the anisotropy of the multi-TeV GCR flux observed in a number of air shower experiments.


\section{What is the proper definition of the heliopause and where is it located?}
\label{HP}
When two plasma streams collide, a tangential discontinuity is formed at the collision interface provided that dissipative and finite conductivity effects are absent, i.e., when an ideal MHD approximation is applicable.
In the context of the SW--LISM interaction, this tangential discontinuity is called the HP \cite{Bhatnagar-Fahr-1972}.
The HP separates the LISM and SW flows. The boundary conditions in the HP frame are formulated as follows:
(i) the sum of the thermal and magnetic pressures across the HP is continuous and (ii) the velocity and magnetic field vectors
are tangent to the HP surface. All other quantities may experience arbitrary jumps. The structure crossed by \textit{V1} cannot be a rotational discontinuity because rotational discontinuities are permeable, which means that either the SW or LISM plasma is crossing the surface of this discontinuity at the Alfv\'en velocity. Furthermore, and that there should be a real HP somewhere ahead. In ideal MHD, a tangential discontinuity cannot degenerate into a rotational discontinuity, except for a trivial case with equal densities
on both sides of the HP, because they belong to different classes \cite{Landau}.
The possibility of mixing of the SW and LISM plasmas in the vicinity of the HP, i.e., its dissipative/resitive structure,
has been summarized in \cite{1986SSRv...43..329F}, where it was shown that even anomalous resistivity would likely
result in a structure of about 0.01~AU width. This is 30 times narrower than the structure that was crossed by \emph{V1} within a month.

The applicability of the ideal MHD equations to model the SW--LISM interaction is not obvious. It is mostly based on the assumption
that the ion distribution function is isotropic away from discontinuities. Of major concern is the presence of a nonthermal ion component,
namely PUIs, both in the SW and LISM \cite{2014ApJ...797...87Z}. Most numerical simulations so far have been based on the one-ion-fluid approach where
all ions are treated as a thermal mixture. PUIs are created wherever charge exchange occurs, but they are not distinguished from
the original thermal population of ions. The momentum and energy of thermal ions and PUIs are summed up. This approach is different
from multi-ion approaches, e.g. \cite{Malama}, where several populations of PUIs were introduced depending on the region where they
are created and the population of neutral hydrogen atoms that participates on each charge-exchange process. More precisely,
they used 10 populations of neutral atoms and four types of ions. Regardless of the axisymmetric statement of the problem, the results of \cite{Malama} are of fundamental importance for our understanding of physical processes in the heliosheath and near the HP
because they demonstrate the kinetic behavior of PUIs throughout the heliosphere. Additionally, it was proved
by direct numerical simulations in~\cite{Malama}  that charge exchange between PUIs and hydrogen atoms in the inner heliosheath results in a considerable momentum and energy removal
from plasma to ENAs. As a consequence, the TS moves farther from the Sun, while the heliocentric distance of the HP decreases.
This makes the inner heliosheath thinner, in accordance with \textit{V1} observations (see also 3D, multi-fluid simulations
in \cite{Pogo16}). On the other hand, it is shown in \cite{2015JGRA..120.1516H} that the effect of PUIs may be overestimated if the charge exchange cross-section is assumed constant while calculating the collisional integral -- an approach used in \cite{1995JGR...10021595P,1996JGR...10121639Z} and similar to it. This is especially true if the plasma distribution function is not Maxwellian (e.g., Lorentzian (kappa) distribution). In particular, the charge exchange source term diverges for $\kappa<2$ if the cross-section dependence on energy is ignored. The IHS width can also be decreased by thermal conductivity \cite{2009SSRv..143..139I}. It was shown \cite {Malama} that the presence of
PUIs does not affect the flow near the HP in a topologically dramatic fashion. Theoretical analysis preformed in \cite{2013ApJ...776...79F} and \cite{2015ApJ...806L..27G} suggests that ACRs may be of dynamic importance, possibly creating additional separation surfaces inside the HP. No simulation results have been obtained so far to support or disprove that theory.

While the thickness of the inner heliosheath can be decreased by treating PUIs a separate entity, the heliopause can also exhibit inward excursions due to the HP instability. The latter does not necessarily result in the decrease of the heliosheath width. In fact, there are no
observations that would tell us where the TS is now.
\begin{figure}[t]
\centering
\includegraphics[width=0.49\textwidth]{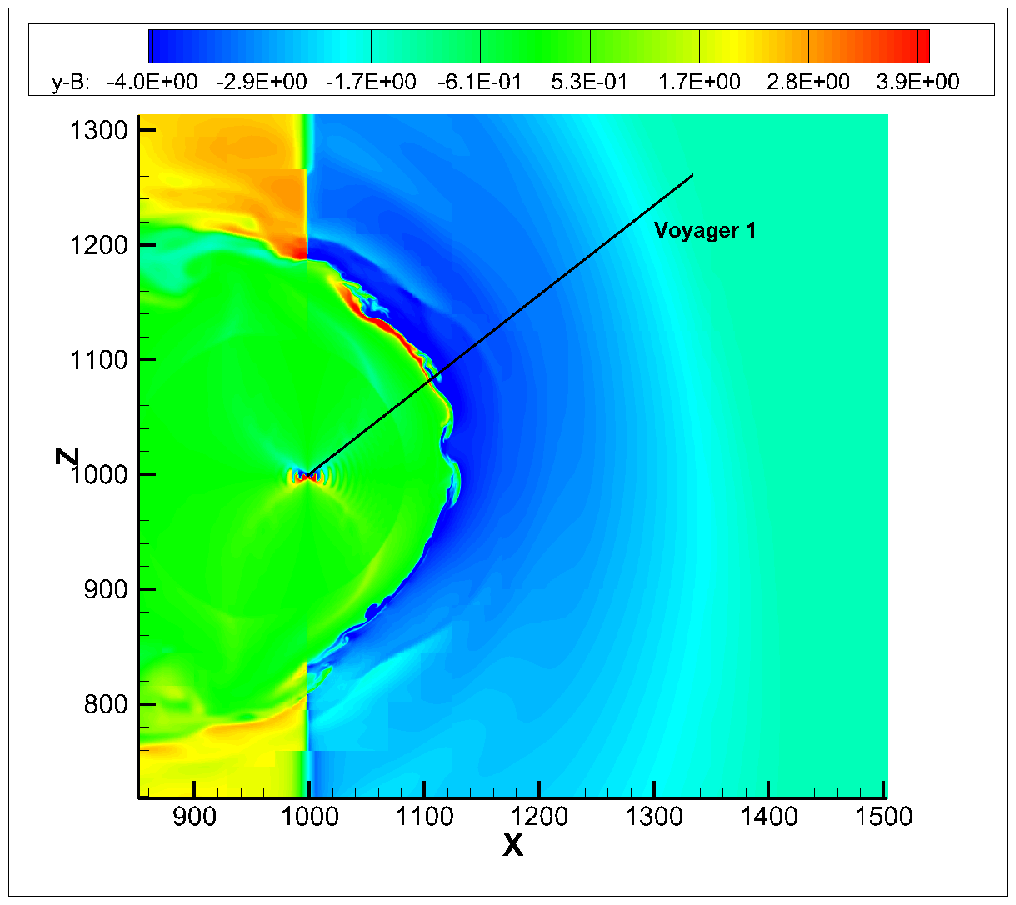}\vspace{1mm}
\includegraphics[width=0.49\textwidth]{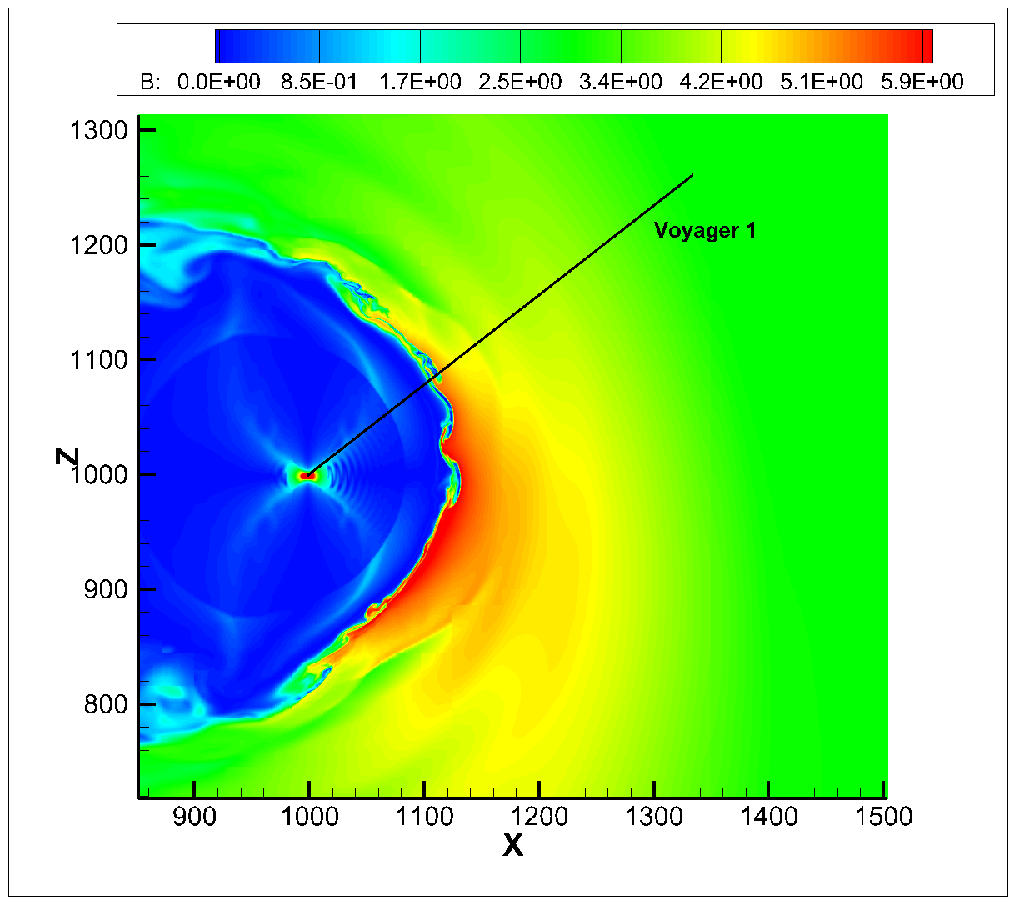}\\
\includegraphics[width=0.49\textwidth]{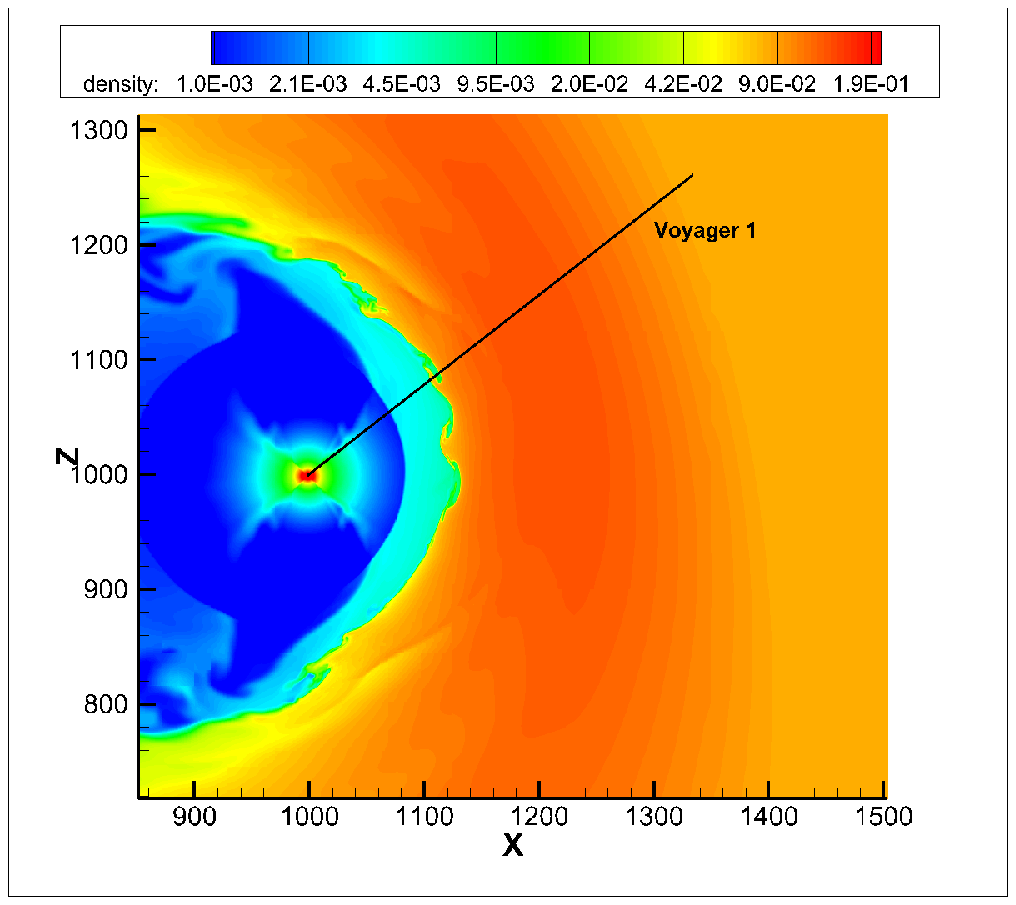}\vspace{1mm}
\includegraphics[width=0.49\textwidth]{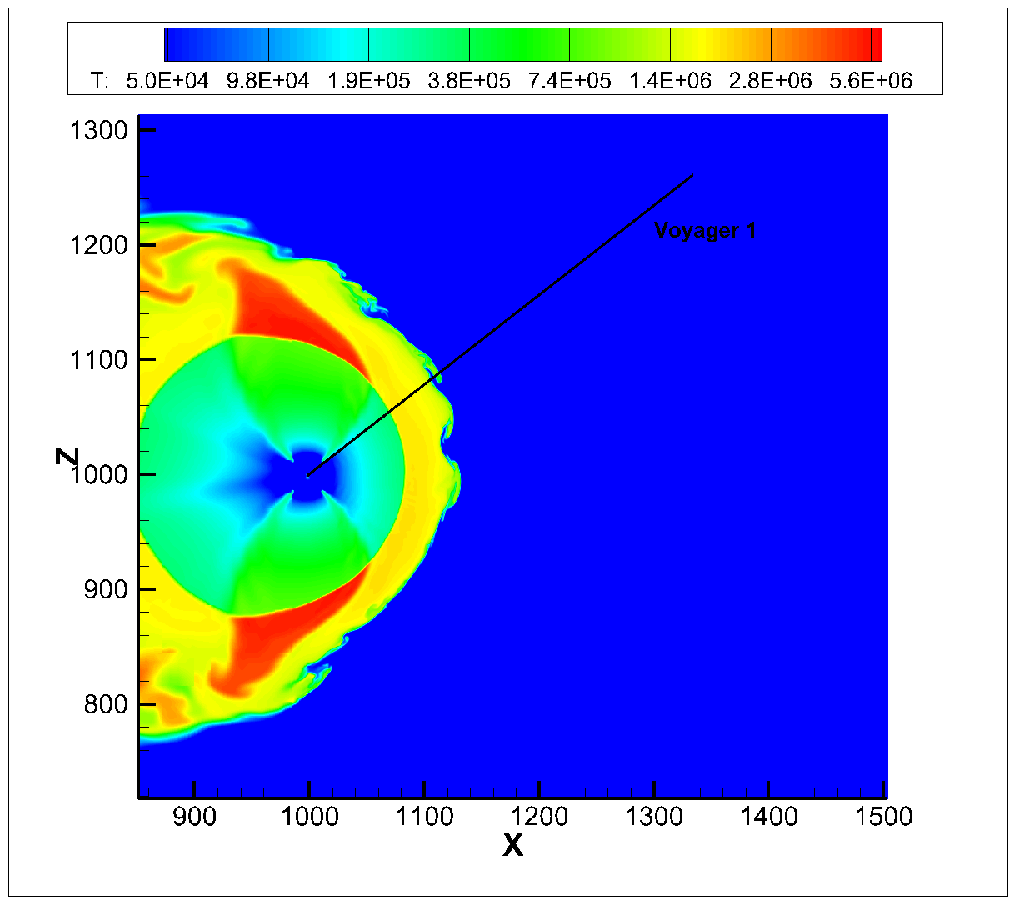}
\caption{Instability of the HP. Clockwise, the distributions of $B_y$, $|\mathbf{B}|$, plasma number density, and pressure
f the magnetic field and its magnitude in the SW--LISM interaction with the solar cycle taken into account.}
\label{Instab}
\end{figure}

Following prior investigations  \cite{Ruderman-Fahr-1993,Ruderman-Fahr-1995,1996JGR...10117119L,1996JGR...10121639Z,1999AIPC..471..783Z,2005JGRA..110.7104F,2008ApJ...682.1404B}, the problem of the HP stability was revisited
by 3D simulations \cite{Borov14} based on a more realistic distribution of the ISMF draping around the HP
(see also an analytical study \cite{Avinash}, where it was shown that there are always perturbations that grow at the linear stage).
In summary, the HP is not a classical MHD discontinuity, but is subject to both Rayleigh--Taylor-like and Kelvin--Helmholtz-like instabilities
caused by charge exchange and shear flows near the HP. According to \cite{1999AIPC..471..783Z}, the mechanism of the Rayleigh--Taylor instability in this case is due to the momentum and energy exchange between protons and neutral atoms.
It has been shown in \cite{Borov14,2015ASPC..498..160P} that such instability may form complicated structures where regions of the SW and LISM plasma follow each other (see Fig.~\ref{Instab}). Similar structures may also be produced by magnetic flux transfer events described in \cite{2013ApJ...778L..33S}. According to \cite{2008ApJ...682.1404B}, the time evolution of the HP instability has no single frequency provided that both primary and secondary neutral atoms are taken into account. In 3D simulations performed in the presence of both HMF and ISMF, deep LISM plasma protrusions related to the HP instability appear at least once per two solar cycles. Their evolution, however, is longer. As mentioned in \cite{Borov14}, the Rayleigh--Taylor instability develops more efficiently when the HMF strength decreases. This situation is typical for the HP when the SW in its vicinity carries a sectored region of alternating HMF polarity, which are subject to magnetic reconnection and turbulence leading to the magnetic field dissipation. Once a protrusion occurred, it further develops
being shaped by the HMF.
As a result of instability, we observe plasma regions that are magnetically connected either to the LISM or the SW. If cosmic ray diffusion perpendicular to magnetic field lines is small compared with the parallel diffusion, consecutive decreases and increases in the GCR flux should be observed.
The fluxes of termination shock ions (ACRs) will have maxima in the interface regions where GCRs have minima.
More detailed simulations of GCR and ACR fluxes are necessary to support this idea.

\section{What is the correlation between the IBEX, SOHO, and Voyager observations?}
Remote sensing observations using ENAs are complementary to  \emph{in situ} observations by the Voyager spacecraft. ENAs are created by charge exchange between neutral atoms and ions in the heliosheath, whereby the momentum exchange is minimal. Thus, the created ENAs keep the velocity and direction of the original ion, but are freed from the electromagnetic forces and therefore follow ballistic trajectories. Thus, ENAs can be used for remote sensing of plasma populations in space \cite{Wurz2000}. These observations can be done from the Earth orbit and
make it possible to investigate the entire sky. However, ENA observations are always line-of-sight observations and therefore have to be interpreted by theory and modeling for a full understanding.

The first ENA observation of the heliosheath were performed with the HSTOF sensor of the CELIAS instrument on the \emph{SOHO} mission for hydrogen ENAs in the energy range from 55--80~keV \cite{1998ApJ...503..916H}. In~\cite{2012A&A...541A..14C},  the final analysis of these data is presented for hydrogen and helium ENAs originating in the heliosheath. At lower energies, in the range from 400~eV to 5~keV,  the first hydrogen observations were done by the  ASPERA-3 and ASPERA-4 ENA instruments on the \emph{Mars Express} and \emph{Venus Express} spacecraft \cite{2006ApJ...644.1317G}. In~\cite{2013ApJ...775...24G}, the final analysis of these data is presented, which is in agreement with the \emph{IBEX} data. ENA energy spectra were compiled already from those first data sets. By considering the charge exchange cross-sections,
the energy spectra of protons in the heliosheath were derived, extending the range covered by the instrumentation of the \emph{Voyager} spacecraft. Before \emph{IBEX} measurements, the covered energy range spanned from 400~eV to 80~keV \cite{2006ApJ...644.1317G,2008ApJ...683..248W,2012A&A...541A..14C,2006phb..conf..203K}, including the HENA data from the IMAGE mission, with the most recent compilation of the ENA energy spectra given in~\cite{2013ApJ...775...24G}. From the fit of the proton spectra, which were derived from these ENA energy spectra, to the \emph{in situ} proton spectra from \emph{Voyager} at higher energies, which is the only fit parameter, the thickness of the heliosheath in the upwind direction can be estimated to be between 35--70~AU.

The \emph{IBEX} mission \cite{2009SSRv..146...11M} is the first space mission dedicated solely to the investigation of the heliospheric interface with the interstellar medium. \emph{IBEX} performs full-sky observation of ENAs with two ENA cameras, \emph{IBEX}-Lo \cite{2009SSRv..146..117F} and \emph{IBEX}-Hi \cite{2009SSRv..146...75F}, combined covering the energy range from 10~eV to 6~keV. The ENA signal recorded by IBEX has its origin in the plasma populations beyond the heliospheric termination shock at distance of more than 100 AU from the Sun, which are explored by the Voyager spacecraft at the same time. \emph{IBEX} full-sky ENA measurements together with \emph{Voyager} \emph{in situ} plasma measurements allowed the space science community to make a major step forward in the scientific investigation of the heliospheric interface.

\begin{figure}
\centering
\includegraphics[width=0.49\textwidth]{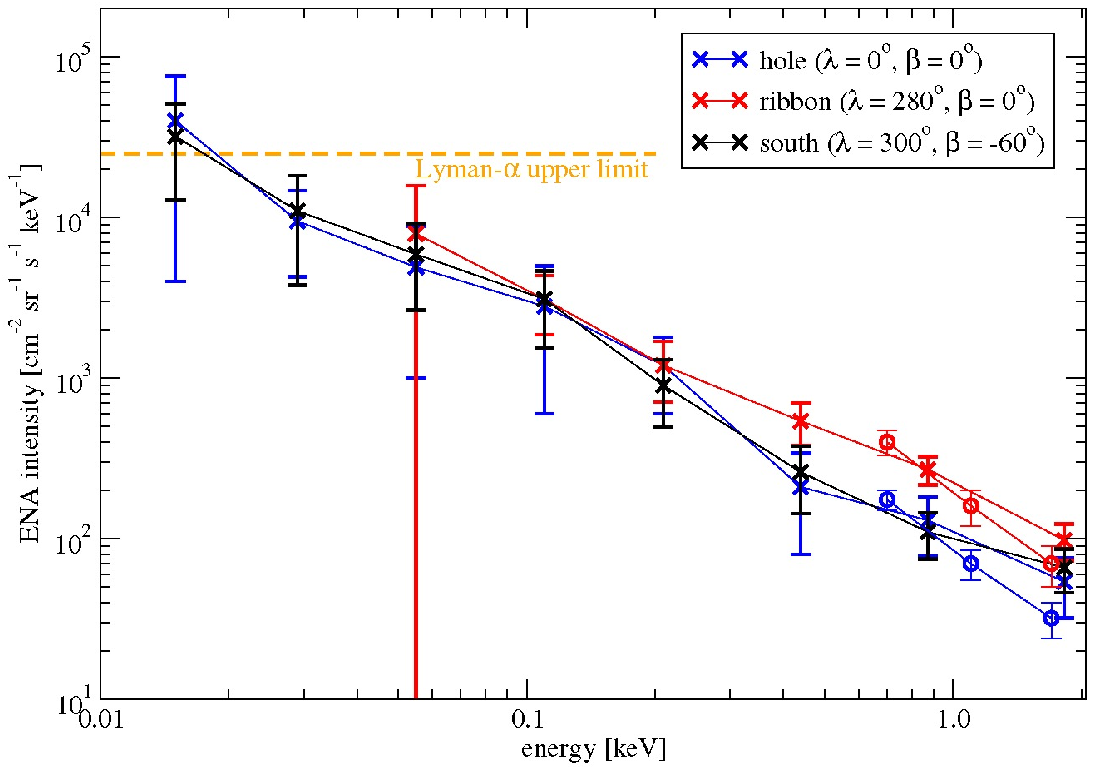}\vspace{1mm}
\includegraphics[width=0.49\textwidth]{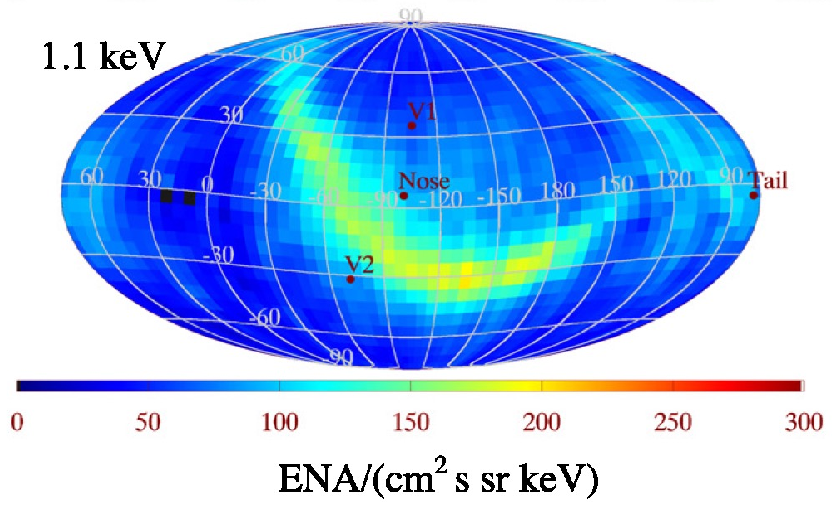}
\caption{\emph{Left panel}: ENA energy spectra for three different regions in sky as indicated by coordinates in the inset \cite{2013ApJ...775...24G}. Blue symbols: boundary of the low ENA intensity region, red symbols: ribbon region in the ecliptic, black symbols: region in the southern hemisphere. The orange dashed line indicates the upper limit on heliospheric ENAs derived from Ly$\alpha$ observations. Crosses indicate IBEX-Lo, circles indicate IBEX-Hi observations. \emph{Right panel}: Full sky map for ENAs at 1.1~keV from IBEX-Hi with the directions of the two Voyager spacecraft and the upwind direction (nose) indicated \cite{2014ApJS..215...13S}.}
\label{Ribbon}
\end{figure}

The first, and completely unexpected, discovery of \emph{IBEX} was the ENA ribbon signal \cite{2009Sci...326..959M,2009Sci...326..962F}, which is a narrow band in the ENA sky maps, about 20$^\circ$--40$^\circ$ wide, of enhanced ENA fluxes, initially observed over the energy range from 0.7 to 2.7~keV (see Fig.~\ref{Ribbon}). The ribbon is best seen in the energy range between 0.5~keV and 4.7~keV \cite{2014ApJS..215...13S}.  The ENA ribbon is a stable signal that has been observed in every IBEX map recorded since 2009 in the \emph{IBEX}-Hi images \cite{2014ApJS..215...13S}. The  fluxes in the ENA ribbon are up to about 2--3 times larger than the surrounding ENA fluxes, the globally distributed ENA fluxes, with the ribbon fluxes peaking around 0.7~keV, which corresponds to a flow velocity of 350 km/s (left panel of Fig.~\ref{Ribbon}). At higher energies above 2~keV the ribbon starts to become more fragmented and the ribbon structure at energies of 4.7~keV and above is difficult to identify. By significantly improving the identification and removal of the background in the \emph{IBEX}-Lo ENA images this ribbon could be identified down to energies of 100~eV, also finding at the lowest energies increasing spatial fragmentation \cite{2014ApJ...796....9G}.

The origin of the ribbon is still debated. From comparisons between  the outer heliosheath and  ribbon models, it was  surmised already at the time of the ribbon discovery that ISMF in the outer heliosheath is roughly perpendicular to the directions toward the \emph{IBEX} ribbon, that is where $\vec{B} \cdot \vec{r} = 0$, where $\vec{r}$ is the radial line-of-sight (LOS) direction and $\vec{B}$ is the interstellar magnetic field \cite{2009Sci...326..959M,2014ApJS..215...13S}. Unfortunately, both \emph{Voyager} trajectories do not overlap with the ENA ribbon (see Fig.~\ref{Ribbon}), so no \emph{in situ} data are available for these parts of the sky to assist the interpretation of the ENA observations. The proposed location of origin of the ribbon ENAs ranges from the heliospheric termination shock, the inner and outer heliosheath, all the way to the nearby edge of the local interstellar cloud  \cite{2009Sci...326..959M,Jacob10,2010ApJ...715L..84G,2010ApJ...725.2251G,2010ApJ...719.1097F,2010ApJ...716L..99C,2013ApJ...776..109K,2013A&A...551A..58S}. All models involve charge exchange between ions and neutral atoms. Alternatively, a density fluctuation in the neutral interstellar gas passing over the heliosphere has been proposed, which would cause a localized increase of the charge exchange, thus locally increasing the production of the globally distributed ENA flux \cite{2014A&A...561A..74F}.

The energy distribution of  ions is one of the important quantities for every plasma population, because it affects the definition of boundaries between different plasma populations. Thus, deriving the energy spectra in the heliosheath from the ENA observations is an important science objective. First analyses of \emph{IBEX} ENA data covered the energy range down to 100~eV \cite{2014ApJ...784...89F,2014ApJ...796....9G}, where the observed spectral shape are power laws with indices of $\gamma = -1.4 \pm 0.1$ for all sky directions. Because the observed energy spectra are power laws with negative exponents (see Fig.~\ref{Ribbon} for example), the lowest energies contribute the most to the pressure, and when the energies down to 100~eV are considered the pressure is already dominated by the lowest energy measured. In~\cite{2014ApJS..215...13S},
full sky maps of the LOS-integrated-pressure from the measurements of the globally distributed ENA flux were derived, peaking in the nose direction at about 40~pdyn~AU~cm$^{-2}$ for the energy range from 0.2 to 4.7~keV (where 1 pdyn~AU~cm$^{-2}$  = 0.015~N~m$^{-1}$).

\begin{figure}
\centering
\includegraphics[width=0.49\textwidth]{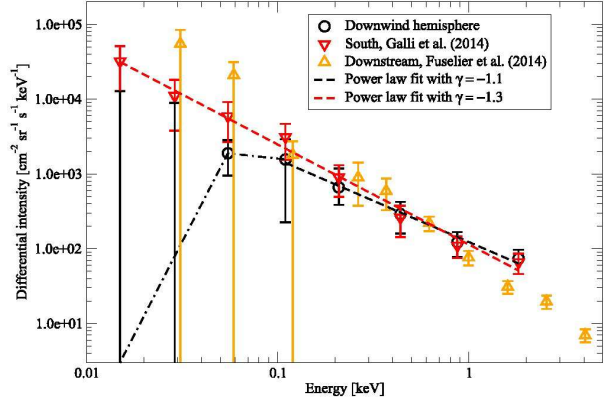}\vspace{1mm}
\caption{Energy spectra of heliospheric ENAs in the downwind hemisphere. Black symbols are data from \cite{2016ApJ...821..107G},
red triangles down from \cite{2014ApJ...796....9G}, and orange triangles up are from \cite{2014ApJ...784...89F} for similar regions in the sky. The black line shows the power law with slope $\gamma = -1.1$, which describes  the energy spectra at energies above 0.1~keV well.
For lower energies, the earlier energy spectrum  \cite{2016ApJ...821..107G} was consistent with a uniform power law continuing to the lowest energies (red dashed line);
 the newest study shows that the energy spectrum rolls over and the signal vanishes at low energies (black dashed-dotted line).}
\label{ENA_spectrum}
\end{figure}

In the latest \emph{IBEX} analysis of the IBEX-Lo data, the identification and removal of background sources was significantly improved and the ENA energy range could be extended down to almost 10~eV for selected locations in the sky \cite{2016ApJ...821..107G}, as shown in Fig.~\ref{ENA_spectrum}. It was found that the power law shape of the ENA energy spectrum continues to the lowest energies accessible to \emph{IBEX}-Lo, for some directions in the sky, with a with slope of $\gamma = -1.2 \pm 0.1$ for most of the sky directions. However, there is a roll-over of the ENA energy spectrum at the downwind hemisphere. This has important consequences for the pressure balance in the heliosheath: for the downwind hemisphere the LOS-integrated-pressure is 304~pdyn~AU~cm$^{-2}$   and  for the \emph{V1} region it is
66~pdyn~AU cm$^{-2}$. Moreover, from this measurement the ``cooling thickness'' of the heliosheath at the downwind side of 220$\pm$110~AU could be derived assuming pressure balance across the termination shock, while the heliosheath thickness in the \emph{V1} direction is 50~AU.
The term of ``cooling length'' was introduced in \cite{2014ApJS..215...13S} to emphasize that ENAs of any particular energy should have a maximum line-of-sight integration length. ENAs born beyond this length cannot return to \emph{IBEX}.

\section{Plasma and magnetic field modeling in the context of observational data}
\subsection{Interplay between charge exchange, ISMF draping, and time-dependence}
\begin{figure}
\centering
\includegraphics[width=\textwidth]{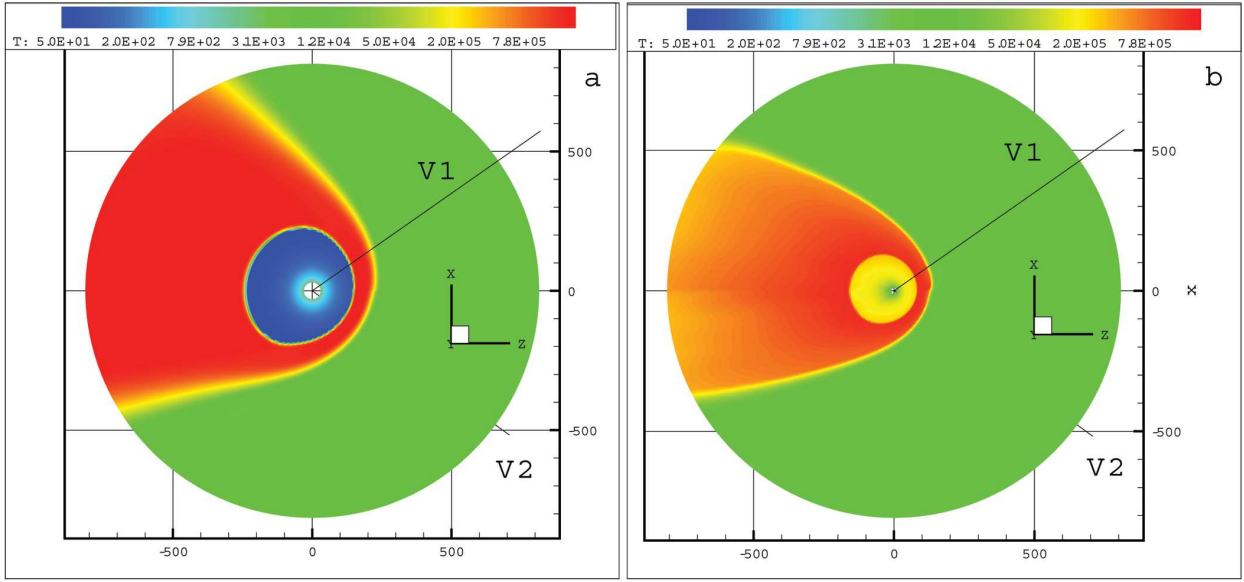}
\caption{Plasma temperature distributions in the meridional plane for the ISMF vector, $\vec{B}_\infty$, belonging to this plane
with a tilt of 45$^\circ$ to the LISM velocity vector, $\vec{B}_\infty$, and $B_\infty=2.5\ \mu\mathrm{G}$: (a) the ideal MHD calculation without an IMF; (b) the plasma-neutral (two-fluid) model with $n_{\mathrm{H}\infty}=0.15\ \mathrm{cm}^{-3}$. The straight lines in the northern and southern hemispheres correspond to the V1 and V2 trajectories, respectively. The TS asymmetry is considerably smaller in case (b) due to
the symmetrizing effect of charge exchange. [From \cite{2007ApJ...668..611P} with permission of the AAS.]}
\label{cx}
\end{figure}

Charge exchange of the LISM neutral atoms with both LISM ions decelerated by the HP and SW ions
makes the heliosphere more symmetric (see Fig.~\ref{cx}). The reason
of this is simple. Since the unperturbed ISMF vector, $\mathbf{B}_\infty$, is directed to the southern hemisphere
at an angle of $45^\circ$ to the LISM velocity vector, $\mathbf{V}_\infty$, which is directed from the right to the left in the figure,
the magnetic pressure rotates the HP clockwise. This rotation exposes the northern side of the HP to the LISM plasma
shifting the LISM stagnation point and the corresponding maximum of the plasma number density northward.
As a result, more charge exchange occurs in that region creating more ions with the velocity of the parent neutral atoms,
which should be decelerated by the HP and exert additional thermal pressure on the HP rotating it counterclockwise.
In summary, while the ISMF tends to make the heliosphere asymmetric, charge exchange, on the contrary, symmetrizes it.
For this reason, a squashed shape TS in the left panel of Fig.~\ref{cx} disappears when charge exchange is taken into account,
as seen from the right panel of Fig.~\ref{cx}. Thus, the difference of 10~AU in heliocentric distances at which V1 and V2
crossed the TS can easily be explained by the action of the ISMF draped around the HP \cite{Poma98,Roma98,Opher-etal-2006,Isenberg-etal-2015,Roeken-etal-2015,Kleimann-etal-2016} if charge exchange is ignored, but it becomes very small once charge exchange is taken into account \cite{2007ApJ...668..611P}.
\begin{figure}[t]
\centering
\includegraphics[width=0.7\textwidth]{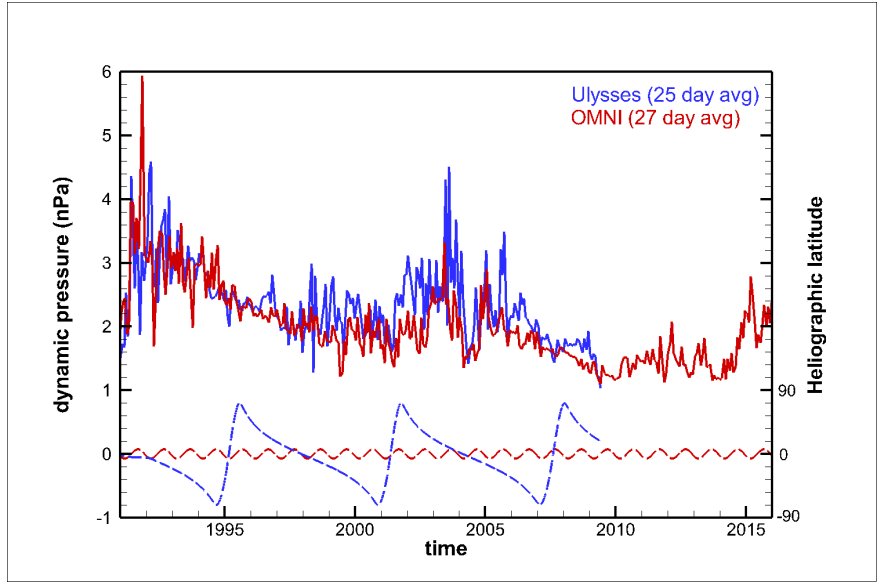}
\caption{Time evolution of the SW ram pressure at \textit{Ulysses} and \textit{OMNI}
is shown together with the spacecraft latitudes as functions of time. [Data courtesy of the SPDF COHOWeb database.]}
\label{UOMNI}
\end{figure}

\begin{figure}[t]
\centering
\includegraphics[width=0.5\textwidth]{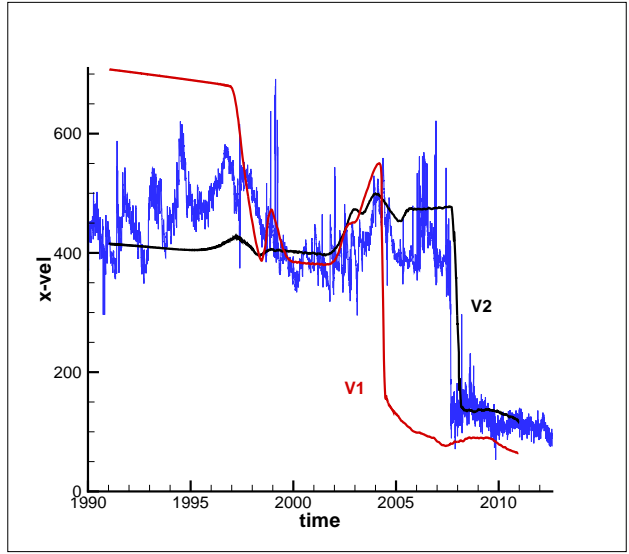}
\caption{The distribution of the radial component of the plasma velocity vector along the
\emph{V2} (black line) and \emph{V1} (red line) trajectories.
\emph{Voyager~2} observations are shown with the blue lines. [From \cite{Pogorelov-etal-2013} with permission of the AAS.]}
\label{UR_compar}
\end{figure}

If charge exchange symmetrizes the heliosphere in general and the TS, in particular, the question arises about the reason of the
observed difference. The TS is responding to changes in the ratio between the SW and LISM ram pressures ($\rho V_R^2/ \rho_\infty V^2_\infty$).
\textit{Ulysses} measurements \cite{2000JGR...10510419M} identified the presence of slow wind at low latitudes and fast wind at high latitudes. The boundary between slow and fast winds is a function of solar cycle: the latitudinal extent of the slow wind is the smallest at solar minima and it can be as large as $90^\circ$ (no direct measurements have ever been done at latitudes well above $80^\circ$) at solar maxima.
In \cite{Sokol15}, the SW ram pressure is assumed to be same in the slow and fast SW. Indeed, a comparison of \textit{Ulysses} and \textit{OMNI} data made in \cite{2008GeoRL..3518103M} resulted in the conclusion that those are in quantitative agreement. We reproduce observational data from \textit{Ulysses} and \textit{OMNI} in Fig.~\ref{UOMNI} on linear scale as functions of time. In addition to the ram pressure, we also show the \emph{Ulysses} and \textit{Earth} latitudes. Clearly there are deviations between observational data at non-coinciding latitudes, some of them should likely be attributed to such transient phenomena as coronal mass ejections and corotating interaction regions. However, such deviations are important once we are interested in realistic boundary conditions for SW--LISM simulations. Another, possibly better, ``latitudinal invariant'' was considered in \cite{LeChat}. This is the SW energy flux $W$. However, although the average $W$ is very close at \textit{Ulysses} and \textit{OMNI}, there are substantial deviations due to the presence of transients.
It is interesting that, according to the Ulysses data analysis in \cite{Ebert09,Pogorelov-etal-2013}, the ram pressure in the genuine slow wind (not only the velocity magnitude but also the SW composition was taken into account to discriminate between the fast and slow winds) was $\sim 0.8$ of that in the genuine fast wind during solar cycle 22 (SC22), but became $\sim 1.1$ during solar cycle 23 (SC23). Notice that the slow wind ram pressure became larger than that in the fast wind during SC23. The ram pressure of the slow wind decreased by $\sim 12\%$ between SC22 and SC23, while the decrease in the fast wind was $\sim 37\%$. As seen from Fig.~\ref{UR_compar}, the simulation that takes into account this effect reproduces both the time and distance at which \textit{Voyagers} crossed the TS \cite{Pogorelov-etal-2013}. This shows that time-dependence effects are important for the explanation of the observed asymmetry of the heliosphere. On the other hand, the HP in that simulation, which was performed only for the period of time when the boundary conditions from \textit{Ulysses} measurements were available, decreased its heliocentric distance in the \textit{V1} direction only by $\sim 2$~AU and ultimately reached distance of $\sim 140$~AU
in 2010. The heliosphere was clearly decreasing in size at the end of simulation and it is possible that it continued decreasing in response to the decrease in the SW ram pressure to the value of 122~AU, when the HP was crossed by \textit{V1}.
On the other hand, the simulations in \cite{2011MNRAS.416.1475W}, where \emph{V2} observations were extended in a spherically symmetric manner
over a moving spherical boundary with the radius equal to the \emph{V2} heliocentric distance, show considerably larger
excursions of the HP.  A possible reason for this may be that the plasma quantities oscillate in unison over the inner boundary with the amplitude of spacecraft observations. The \textit{Ulysses}-based solar cycle simulations in \cite{Pogorelov-etal-2013} show that
the HP motion is mostly determined by the differences between solar cycles rather than by the changes on the latitudinal extent
of the slow wind.
\begin{figure}
\centering
\includegraphics[width=0.48\textwidth]{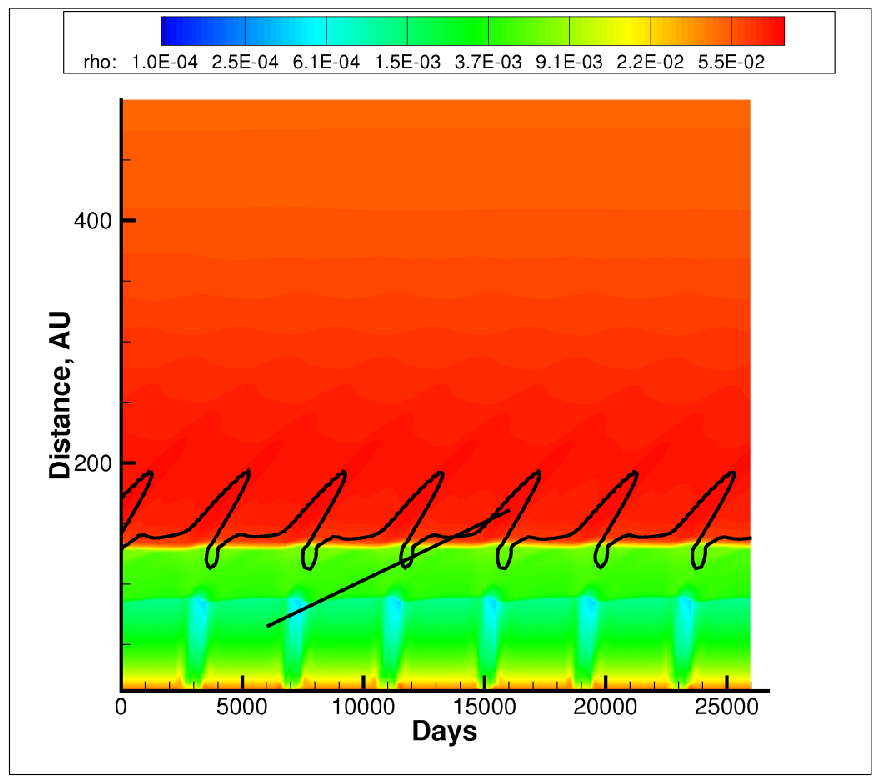}\hspace{2mm}
\includegraphics[width=0.48\textwidth]{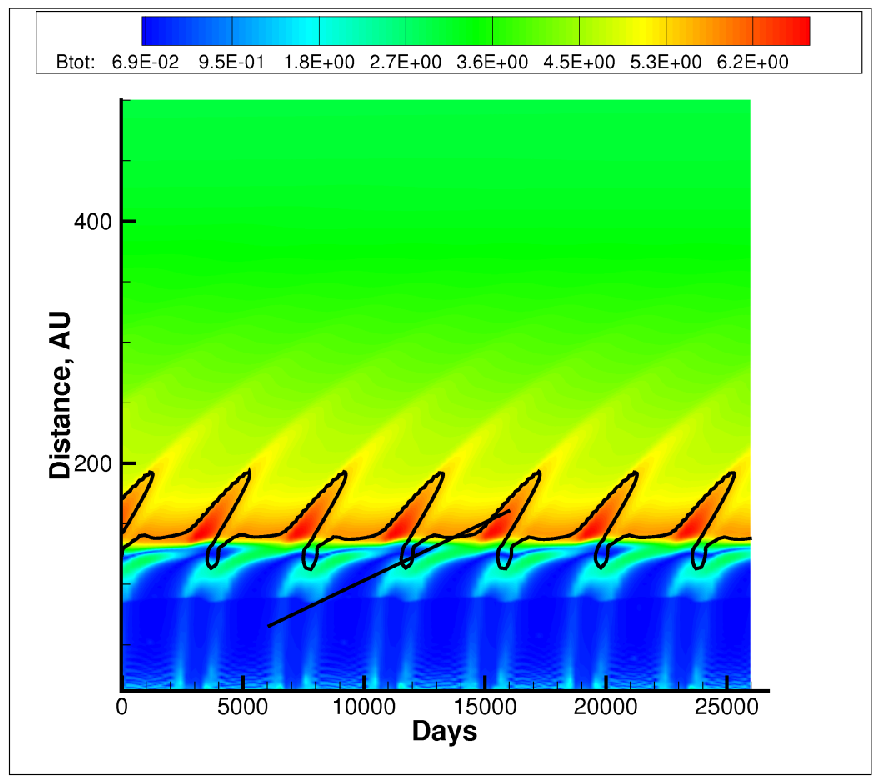}
\caption{Space-time plots of (\textit{left}) plasma number density and (\textit{right}) magnetic field magnitude in a direction
imitating the \textit{Voyager 1} trajectory.
The black curve shows the line where $v_R=0$. The black straight line is a possible trajectory of a spacecraft
moving at the V1 velocity. [From \cite{Pogo12} with permission of the AAS.]}
\label{negative}
\end{figure}

\begin{figure}
\centering
\includegraphics[width=0.48\textwidth]{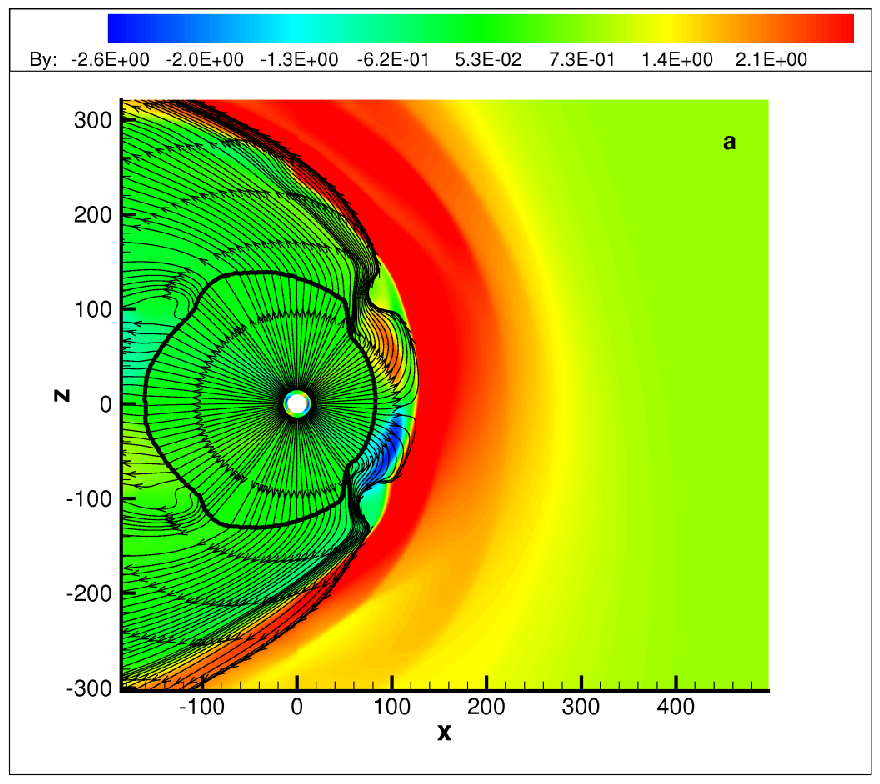}\hspace{2mm}
\includegraphics[width=0.48\textwidth]{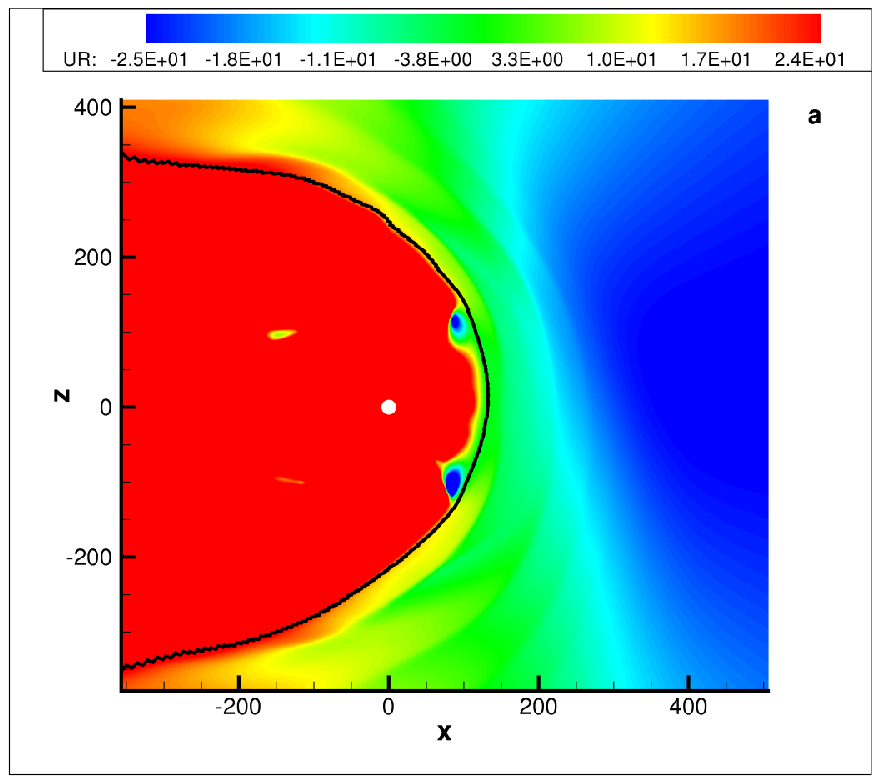}
\caption{Magnetic barriers (left panel) and related negative values of the SW radial component (right panel).
The streamlines start on a heliocentric circle of 15 AU radius and are shown neglecting the out-of-plane velocity component.
The TS is shown with a thick black line. Distances are given in AU. The $y$-axis is directed into the figure plane. [From \cite{Pogo12} with permission of the AAS.]}
\label{barrier}
\end{figure}

The HP motion closer toward the Sun also results in negative values of the SW radial velocity component, $V_R$, near the HP.
A question is about how large those components can be. The radial velocities that were derived from \textit{V1} LECP data in the inner heliosheath \cite{2011Natur.474..359K,2012Natur.489..124D} are smaller than the value of approximately $- 40$~km/s which may have resulted from a HP shift from 140~AU to 122~AU in 2 years. Another possibility has been proposed in \cite{Pogo12}. As seen in Fig.~\ref{negative},
which shows the space-time plots of the plasma number density and magnetic field magnitude in a direction imitating the \textit{V1}
trajectory, such behavior of the SW velocity is typical if the solar cycle is taken into account (see also, e.g., \cite{Scherer-Fahr-2003}). It is also possible that \textit{V1} may cross a LISM region with positive $v_R$. The latter regions extend into the LISM as far as 50~AU. In the inner heliosheath, the regions of negative $V_R$ are smaller ($\sim 7$~AU). The existence of both regions had been predicted in \cite{Pogo09}, two years before they were measured by \textit{V1}. Magnetic barriers are created due to the interaction of slow and fast streams in the SW
(see, e.g., \cite{Tanaka}). However,  only in \cite{Pogo09,Pogo12} was it noticed that SW streamlines that start near the equatorial plane become occasionally  concentrated between a magnetic barrier and the HP. Since such a barrier has finite
latitudinal extent, those streamlines diverge towards the Sun when the barrier disappears. This is seen in
Fig.~\ref{barrier}.

An additional conclusion of \cite{Pogo12} is that \textit{V2}, because of solar cycle parameters,  is unlikely to see backward SW flow if it was observed by \textit{V1}. The reasons are as follows: (1) its velocity is less than \emph{V1} and (2) it crossed the TS later, within a solar cycle, than \emph{V1}. As a result, the V2 trajectory should miss the region of substantial negative velocity. Another interesting consequence is that \textit{V1} may ultimately observe positive radial velocity components in the LISM approximately in 2020--2021.

\subsection{Magnetic field in the inner heliosheath and beyond}
\begin{figure}[t]
\centering
\includegraphics[width=0.48\textwidth]{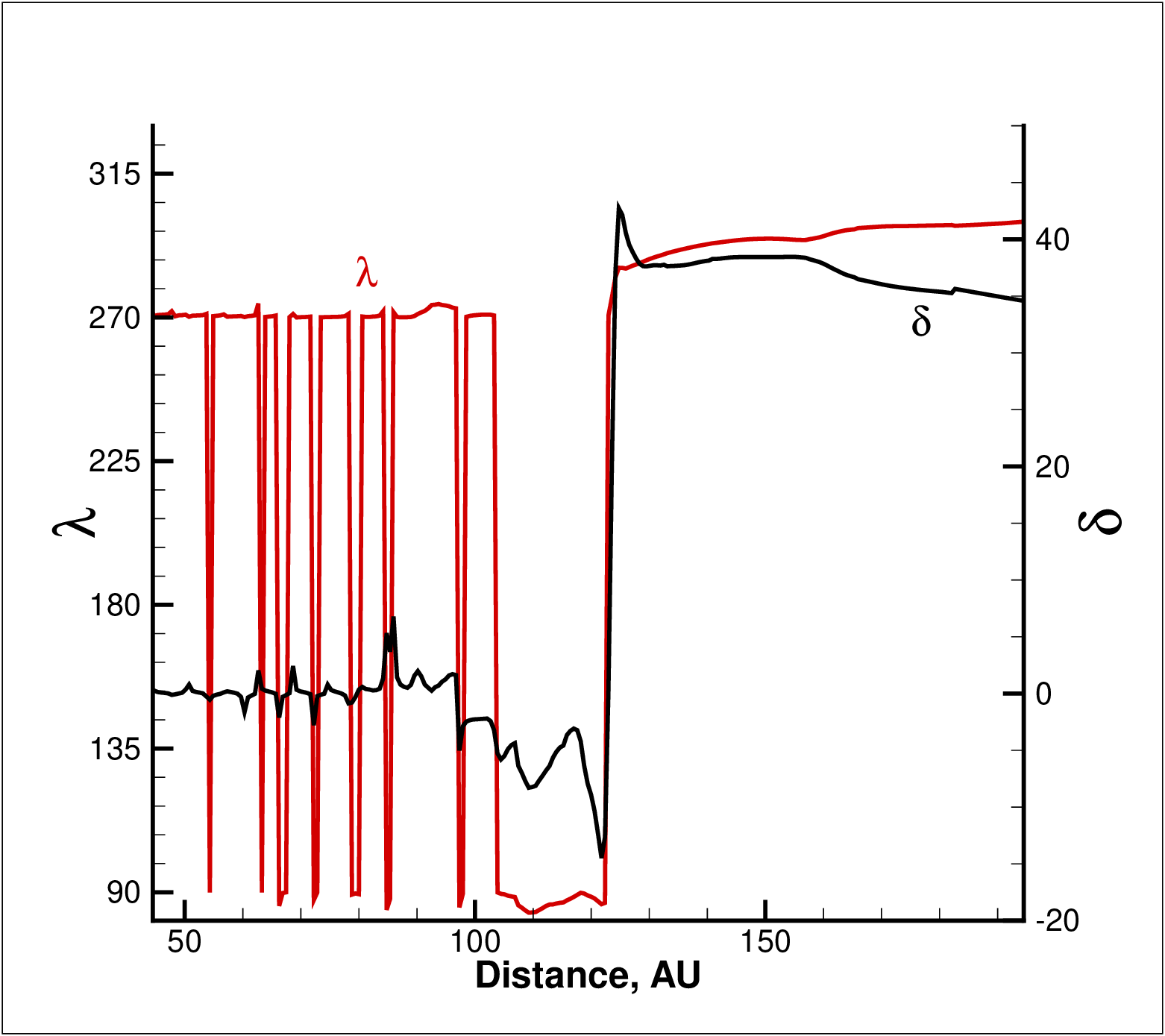}\hspace{2mm}
\includegraphics[width=0.48\textwidth]{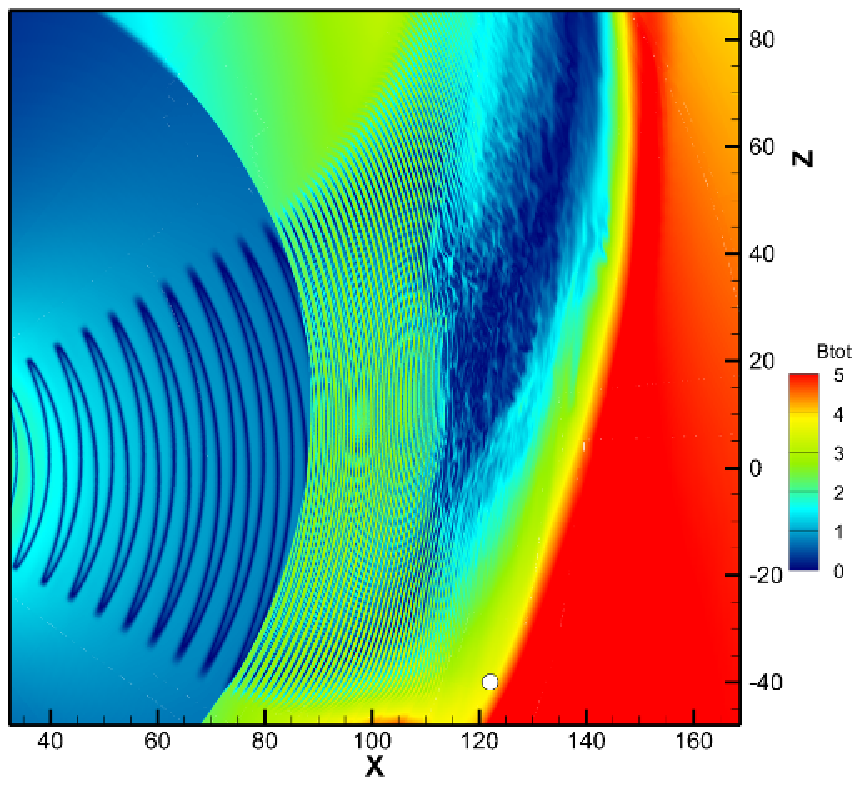}
\caption{\emph{Left panel}) Instantaneous distributions of the $\mathbf{B}$ elevation and azimuthal angles ($\delta$ and $\lambda$). (\emph{Right panel}) Transition to chaotic behavior in the inner heliosheath. Magnetic field strength
distribution (in $\mu$G) is shown in the meridional plane. The angle between the Sun's
rotation and magnetic axes is $30^\circ$. [From \cite{Borov14} and \cite{Pogorelov-etal-2013} with permission of the AAS.]}
\label{HCS}
\end{figure}

The \textit{Voyager} magnetic field instrument (MAG) provided us with invaluable distributions of the HMF at \emph{V1} and \emph{V2}. The HMF exhibits turbulent fluctuations on both kinetic and small scales. It is seen from \cite{2012ApJ...744...51B,Burlaga-Ness-2014} that the variability and especially the number of  HMF vector reversals at sector boundaries was much greater before each of the spacecraft crossed the TS.
This is puzzling if we assume that the sectors are due to the global heliospheric current sheet (HCS). In this case, the number of sector crossings should gradually increase to very large values while the velocity component normal to the HP tends to zero. We should recall here that the radial velocity component was zero to negative for about 8~AU before the HP crossing, which makes it doubtful that the existence of the HCS structure is determined entirely  by the tilt between the Sun's rotation and magnetic axes. This pattern can be seen qualitatively in Fig.~\ref{HCS} (the right panel), where
the disruption of the HCS structure is due to the tearing mode instability caused by numerical resistivity.
It is worth noticing that the figure shown in this panel is drastically different from similar figures in \cite{2011ApJ...734...71O,2012ApJ...751...80O}, although the boundary conditions were chosen to be identical.
In particular, in \cite{2012ApJ...751...80O} (Figs.~2 and~3), one can see something resembling a radially-oriented discontinuity crossing the IHS. This discontinuity is not related to the boundary between the slow and fast SW, and its presence therefore has no explanation. In contrast to \cite{Pogo09,2012ApJ...751...80O}, where the heliospheric magnetic field dissipates in the IHS completely, \cite{Pogorelov-etal-2013} rather observe a chaotic disruption of the HCS, which is a likely fate for it regardless of the actual mechanism, turbulence or magnetic reconnection, responsible for this phenomenon.
On the other hand, sector crossings were observed by \textit{V1} and are being observed by \textit{V2} in the inner heliosheath, although the sector widths are not as small as one would expect. Additionally, numerous sector crossings seem to have been observed when the HMF strength was close to or below the MAG accuracy. Clearly, current sheets can be created not only due to the above-mentioned tilt. This can be due to stream interactions, which are observed throughout the heliosphere.
Additionally, observations of the magnetic equator of the Sun from the Wilcox Solar Observatory show small-scale non-monotonicity. Any change in the sign of the tilt derivative
at the latitudes of \textit{Voyager} spacecraft creates a current sheet with a sector size considerably greater that those due to the Sun's rotation.
These issues are of importance because they tell us what to expect from the magnetic field distribution as the SW approaches the HP. Is the sector structure of the HMF destroyed by SW turbulence, as shown in \cite{Borov14}, while other current sheets still exist and are detected by spacecraft? Answering this question is of importance not only
to understand the heliosheath flow, but also the flow in the heliotail \cite{2015ApJ...800L..28O,2015ApJ...812L...6P}.

The \emph{V1} crossing of the heliospheric boundary was accompanied by a
change in the magnetic field \cite{Burlaga147}. Before the crossing,
the magnetic field direction was consistent with the Parker spiral. After the crossing the
direction of the field changed, but only by a small amount ($\sim 20^\circ$). Since there is no particular reason for the ISMF
direction to remain close to that of the HMF,
this observation was for some time regarded as an indication that \emph{V1}
might not yet be in the LISM. However, a similar set of the magnetic field elevation and azimuthal angles
in the LISM was reported before the crossing in \cite{Pogo09} (see also \cite{Borov14}). On the other hand,
numerical simulations in \cite{Borov14} demonstrated (see Fig.~\ref{HCS}, \emph{left panel}) that the elevation angle was greater than the observed value
when the LISM properties, especially the direction of the LISM velocity, were taken from \cite{2012ApJS..198...12B,2012ApJS..198...11M}.
The updated properties of the LISM proposed on the basis of \textit{IBEX} observations in \cite{2015ApJ...801...28M}
are in better agreement with \textit{V1} observations and, as in \cite{Pogo09}, make it possible to
reproduce the ISMF draping around the heliopause \cite{Eric16}.

A simple explanation of the \emph{V1} measurements of the draping angles was proposed in \cite{2014ApJ...789L..43G}.
It relies on the fact that the \emph{V1} trajectory direction and the direction of the unperturbed ISMF,
assuming that the ISMF is directed into the center of the IBEX ribbon \cite{Funsten} have
almost the same heliographic latitude ($\sim 34.5^\circ$). The deviation of the ISMF direction from the ribbon center
increases with decreasing the ISMF strength \cite{Eric16}.
The draped magnetic field line must ultimately become parallel to its unperturbed direction at large distances from the HP.
If a magnetic field line passing through \emph{V1} has a shape close to a great
circle in the projection of the celestial sphere, it may become nearly parallel to the Parker HMF.

Before reaching the heliopause \emph{V1} encountered two ``precursors,''
where the flux of heliospheric energetic particles dropped sharply,
although by a smaller amount that at the heliopause, while the
magnetic field strength sharply increased. Clearly, this is related to the HP structure discussed earlier in
Section~\ref{HP}.
To explain these observations, a model is presented in \cite{2014ApJ...782L...7S}, which is based on 2.5D MHD, in-the-box simulations
(the computational box was chosen to be 20 AU wide and 4 AU deep). The initial
distribution  includes two discontinuities (current sheets)
corresponding to the polarity changes observed by \emph{V1}.
One of these singularities represented the heliopause, with the magnetic field
strength and plasma density higher on the LISM side.
Magnetic reconnection was initiated at the HP by introducing
random noise. As a result, magnetic islands started forming, growing, and merging.
These simulations showed that magnetic field compressions created in such reconnection model
may be interpreted as the observed ``precursors'' accompanied by
the penetration of the LISM plasma into the heliosheath.

As the HP is a tangential discontinuity separating the SW from the LISM,
both the HMF and the ISMF must be parallel to the HP on its surface. The process of topological changes in the ISMF that
result in its rotation from the direction of $\mathbf{B}_\infty$ to some direction parallel to the surface of the HP is called
draping. A simple model of such draping may be developed by assuming that the HP  is stationary and impenetrable both to the LISM
and ISMF. Analytical solutions for such simple cases as a spherical or cylindrical
obstacles were used to estimate the ``draping factor,'' i.e., the ratio of the maximum draped
field strength to the strength of the unperturbed field (see \cite{2008JGRA..113.4102M}).

\begin{figure}[t]
\centering
\includegraphics[width=0.48\textwidth]{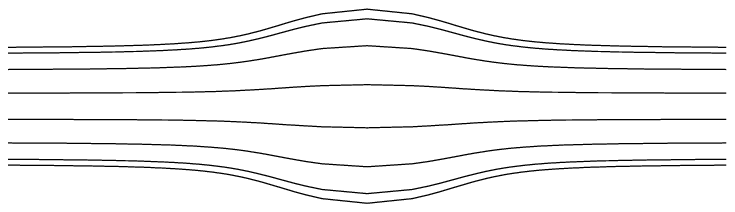}\hspace{2mm}
\includegraphics[width=0.48\textwidth]{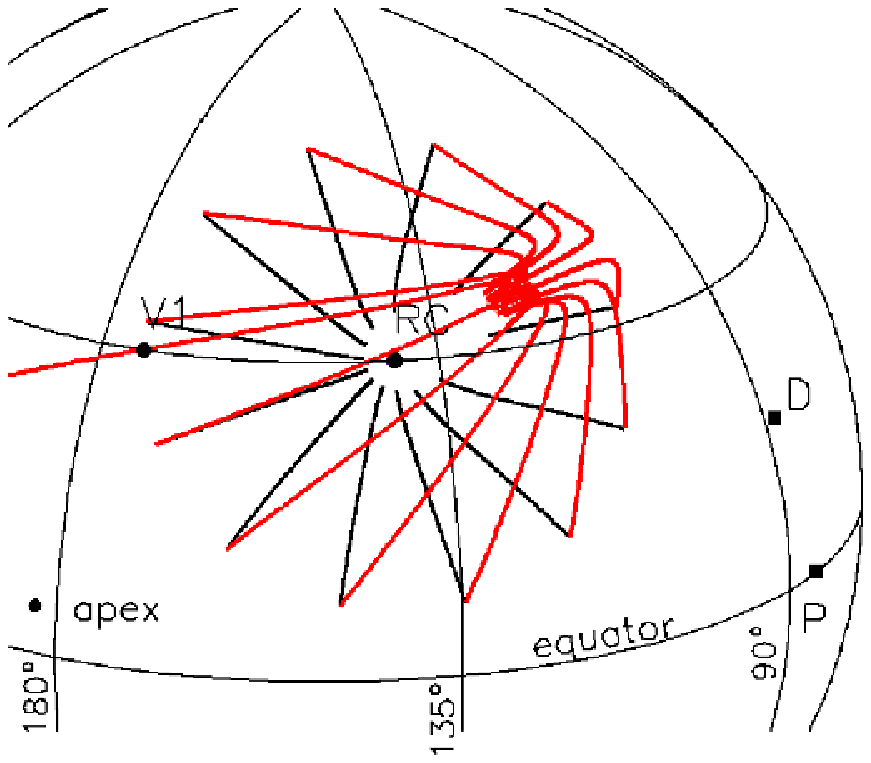}
\caption{\emph{Left panel}) Magnetic field lines in the Parker model of the astrosphere
confined by the magnetic field. When projected on the celestial sphere, the
field lines become great circles connecting the unperturbed field
and anti-field directions. (\emph{Right panel}) Projections of the magnetic field lines in heliographic
coordinates for two models of the heliosphere, corresponding to
the ISMF strength of 20 $\mu$G (thick black lines) and
4 $\mu$G (red lines) \cite{2014ApJ...789L..43G}.
Also shown are the directions of the V1 trajectory, the
IBEX ribbon center (RC), the magnetic field measured by V1 before
(P) and after (D) it crossed the heliopause, and the interstellar
helium inflow (apex).}
\label{ac}
\end{figure}

One simplified solution to the SW--LISM interaction was proposed in \cite{1961ApJ...134...20P}
who considered the propagation of the spherically symmetric SW into a strongly magnetized, high plasma $\beta$ surrounding medium at rest.
An astrosphere is formed in this case with the shape of the astropause determined by the equality
of total pressures on its surface. The external magnetic field confines the stellar wind creating
a central cavity with two oppositely directed channels parallel and
antiparallel to the magnetic field (see Fig.~\ref{ac}, \emph{left panel}).

In \cite{Roeken-etal-2015}, an analytical solution was proposed
for a magnetic field frozen into the plasma flow, corresponding to another model of Parker: the
incompressible axially-symmetric flow with the scalar velocity potential in the
form $\Phi(r)=u_0 (z+q/r)$, where $r$ and $z$ are two cylindrical coordinates, and $u_0$ is constant and equal to the
LISM velocity at $r\to\infty$. To remain in the framework of the analytical solution, the effect of the magnetic field on the plasma flow was neglected. For a slightly more general form of the flow potential \cite{Suess90}, the solution for
the magnetic field frozen into the flow was reduced to a single ordinary differential equation \cite{Suess91,Suess93}.
However, these solutions are not fully consistent: at a distance $d$ from the boundary of the model
astrosphere the field strength diverges as $1/d^{1/2}$ leading to infinite energy. This issue is caused by the presence of a stagnation
point in the flow \cite{1977AnG....33..423P,1977AnG....33..429P}.

Clearly, more realistic models for the description of the plasma flow and magnetic field in the vicinity of the heliospheric boundary
are based on numerical solutions of MHD equations with proper source terms describing charge exchange between ions and neutral atoms.
A number of references are given in this paper (see also \cite{Pogo11} and references therein).
It should be understood, however, that certain care is required to interpret numerical simulations
of the magnetic field draping if the HP is smeared by numerical viscosity and resistivity.
This is especially true because of the necessity to correctly identify the neutral atom populations inside the HP structure.
This is the case, of course, only for multi-fluid (non-kinetic) models that describe the neutral atom transport throughout the heliosphere
(in \cite{2011ApJ...728L..21B}, this is done by tracking the HP with a level-set method).
The idea that the ISMF always becomes nearly equatorial at the heliopause in the \textit{V1} trajectory direction \cite{2013ApJ...778L..26O}
is not supported by other numerical simulations \cite{2011ApJ...728L..21B,2014ASPC..484..174P,Borov14}.
From this viewpoint, exact solutions, however simplified, provide a useful supplement
to numerical simulations.
Parametric simulations are of importance to understand the evolution of numerical solutions.
This approach was used recently in \cite{2014ApJ...789L..43G} to explain the
puzzling observation of a very small change in the magnetic field elevation angle by \emph{V1}  while
crossing the heliospheric boundary \cite{Burlaga147}.
The approach was chosen to track individual magnetic field lines
and analyze them in projection on the celestial sphere.
Consider a magnetic field line passing through a chosen point just outside
the HP. As long as this line remains close to the HP it represents the draped magnetic field.
Ultimately, the line departs from the vicinity of the heliopause and starts to approach the direction of the unperturbed field.
As a consequence, the projection of such line onto the celestial sphere
approaches the points representing the inward and outward directions of the unperturbed
field.
For the strong-field Parker's model of the astrosphere,  the projections of magnetic field lines are great circles on the
celestial sphere. If this model were applicable to the heliosphere,
it would provide an immediate explanation to the small change in the magnetic field direction
across the HP. As the \emph{V1} trajectory and the unperturbed magnetic field direction
are very close in latitude and not widely separated in
longitude, it is argued in \cite{2014ApJ...789L..43G} that the angle between the HMF and ISMF at the HP should be small.

The Sun is moving relative to the LISM. However, a hypothetical heliosphere
obtained under the assumption of a very strong ISMF (20 $\mu$G) will have draped magnetic field lines
deviating only slightly from great circles (see black lines in the right panel of Fig.~\ref{ac}.
The angle between the projection of the draped field line and the heliographic parallel at
\emph{V1} are still small. For an ISMF strength of 3--4 $\mu$G,
consistent with \emph{V1} observations, the draped magnetic field lines obtained from the simulation deviate from the
simple Parker model-like structure (the right panel in Fig.~\ref{ac}, \emph{red
lines}). However, this deviation remains small in the nose of the HP,  as well as in the \emph{V1}
trajectory direction. The projection of the draped magnetic field
line passing through \emph{V1} is at a small angle with respect to the
heliographic parallel at this point, and this angle is close to the one observed by \emph{V1}.
It is argued in \cite{2014ApJ...789L..43G} that this is because the shape of the heliopause at its nose is roughly similar to
a spherical shell resembling the stellar wind cavity in the Parker model.
This is clearly not true in the heliotail.

Another region where the draped ISMF lines should be expected to have
similar structure regardless  of the ISMF magnitude, $|\mathbf{B}_\infty$, is the vicinity of the so-called $BV$-plane
\cite{2008ApJ...675L..41P,Pogorelov-etal-2009}, which is determined by the  velocity and magnetic field vectors
in the unperturbed LISM. The direction of $\mathbf{V}_\infty$ is determined from the neutral He observations
\cite{Witte_2004,2015ApJS..220...28B,2015ApJ...801...28M}.
If $\mathbf{B}_\infty$ is directed into the IBEX ribbon center (according to \cite{Eric16}, the accuracy of this statement increases with $B_\infty$), the $BV$-plane is approximately coincides with  the interstellar hydrogen deflection
plane (HDP, see \cite{Lallement05,Lallement10}), which is formed by the H-atom flow directions in the unperturbed LISM and in the inner heliosphere. In the projection onto the celestial sphere, the $BV$-plane is a great circle linking
the unperturbed magnetic field and anti-field directions and passing
through the helium inflow direction. If the ISMF-HMF coupling across the HP is ignored, the symmetry would require that magnetic field lines that start close to the $BV$-plane create a symmetric pattern only weakly dependent on the ISMF strength.

\subsection{The possibility of a data-driven model of the outer heliosphere}
The possibility of developing a data-driven model of the outer heliosphere was not even considered 10--15 years ago. Now, because of the observations performed by the \textit{Voyagers}, \textit{SOHO}, and \textit{IBEX}, this has become a possible, albeit very challenging, task for theorists. Paper~\cite{Eric16} is an example of a systematic approach to fit multiple data sets. Earlier efforts have focused mostly on one or two challenging questions raised by observational data, e.g., negative radial velocity component at \textit{V1} in the inner heliosheath before the HP crossing \cite{Pogo09,Pogo12}, fitting the \textit{IBEX} ribbon \cite{2010ApJ...716L..99C,Jacob10,Jacob11,Eric15,Sylla-Fichtner-2015}, using the HDP  to constrain the orientation of the $BV$-plane and the distribution of radio emission sources observed by the plasma wave instrument (PLS) onboard \textit{Voyagers} \cite{Kurth04,Gurnett06,Gurnett2015},
using the ISMF draping results from \textit{V1} measurements to adjust the angle between $\mathbf{B}_\infty$ and $\mathbf{V}_\infty$ as well as $|\mathbf{B}_\infty|$ in simulations \cite{Izmod05,Opher07,2008ApJ...675L..41P,2009ApJ...695L..31P,2015MNRAS.446.2929K},
or trying to adjust the SW and LISM properties in order to fit time-dependent observations along the spacecraft trajectories.
In \cite{Eric16}, the boundary conditions in the SW and LISM were chosen to (1) get the best fit to the \textit{IBEX} ribbon; (2) reproduce the magnetic field angles observed by \textit{V1} in the HP draping region; (3) obtain the HP at the heliocentric distance consistent with \textit{V1} observations; (4) reproduce the density of the neutral hydrogen atoms at the heliospheric termination shock, which can be derived from \textit{Ulysses} observations of PUIs \cite{2009SSRv..143..177B}; (5) ensure that the $BV$-plane is in agreement with \textit{SOHO} observations (uncertainties in the HDP determination are discussed in \cite{2007ApJ...668..611P}).
The model used in \cite{Eric16} is based on the kinetic treatment of hydrogen atom transport throughout the heliosphere, which is very important to have a more realistic filtration ratio of the LISM hydrogen atoms near the HP.
In \cite{2009SSRv..143...31P}, a detailed comparison of the 5-fluid and kinetic models of the SW--LISM interaction was made. It showed
that the results are qualitatively agreeable, with only a slight shift in the quantity distributions along different lines of sight.
On the other hand,
kinetic modeling of a realistic solar cycle is more time-consuming. To improve statistics and reduce numerical noise typical of the Monte Carlo simulations, one needs either assume the presence of a longer cycle (in multiples of the usual solar cycle) and perform averaging based on the repeated simulation of such cycle \cite{2005A&A...429.1069I} or perform averaging over multiple implementation of the same period inside the solar cycle period \cite{Eric15}. We note in this connection that
a solar cycle model \cite{Pogorelov-etal-2013} based on \textit{Ulysses} observations was successful in reproducing both the heliocentric distance and the time at which \textit{V1} and \textit{V2} crossed the TS. This means that taking into account solar cycle effects is of major importance. Additionally, the model of \cite{Eric16} used the solution of the SW--LISM interaction problem based on a single plasma fluid model
where PUIs born in the process of charge exchange with neutral atoms were added to the mixture of ions preserving the conservation of mass, momentum, and energy. The separation of PUIs and thermal SW ions was made at a post-processing stage which involved a sophisticated
procedure to fit \textit{IBEX} observations in different energy bands covered by the spacecraft. This procedure is very important for understanding the energy separation between ions (see, e.g., \cite{2014ApJ...797...87Z,Mihir1,Desai_etal_2014}), but ignores the dynamical effect of PUIs on the heliospheric interface. While the necessary improvements to the fitting procedure are well understood, their implementation will be rather laborious. It is known that treating PUIs as a separate ion population results in a narrow heliosheath: the TS heliocentric distance increases, while the HP moves closer to the Sum \cite{Malama,Pogo16}. In \cite{Eric16}, the HP stand-off distance in the \textit{V1} trajectory direction was adjusted by choosing the SW/LISM stagnation pressure ratio and the HMF and ISMF strengths and direction. In the future, \textit{V1} and \textit{V2} measurements should be used to improve the quality of the MHD-kinetic fitting of data from multiple sources.

\subsection{The heliotail}
An additional constraint on the LISM properties is provided by multiple air shower observations of the 1--30 TeV GCR anisotropy
\cite{Abbasi,Abdo,Amenomori,2012JPhCS.375e2008D,Guillian}. According to \cite{2014ApJ...790....5Z,2014Sci...343..988S}, this anisotropy is affected by the presence of the heliosphere, especially due to the ISMF modifications
in the heliotail and bow-wave regions. It is clear that the heliotail should be very long to produce an observable anisotropy of 10 TeV cosmic rays whose gyro radii, assuming protons, may be as large as 500~AU.
\label{plasma}
\begin{figure}
\centering
\includegraphics[width=0.49\textwidth]{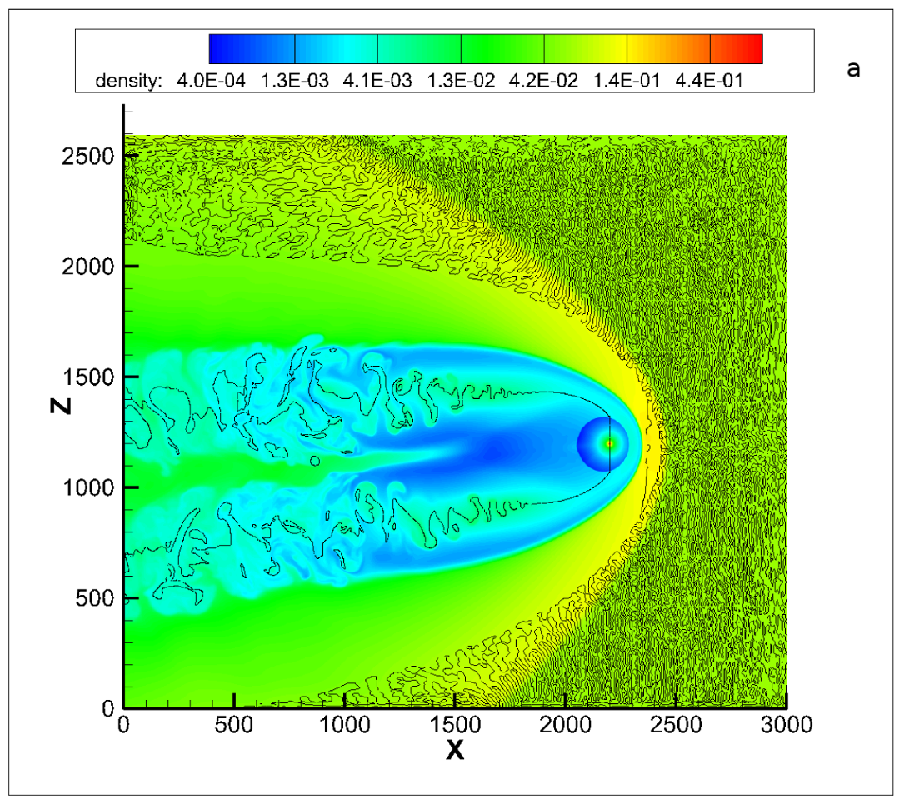}\vspace{1mm}
\includegraphics[width=0.49\textwidth]{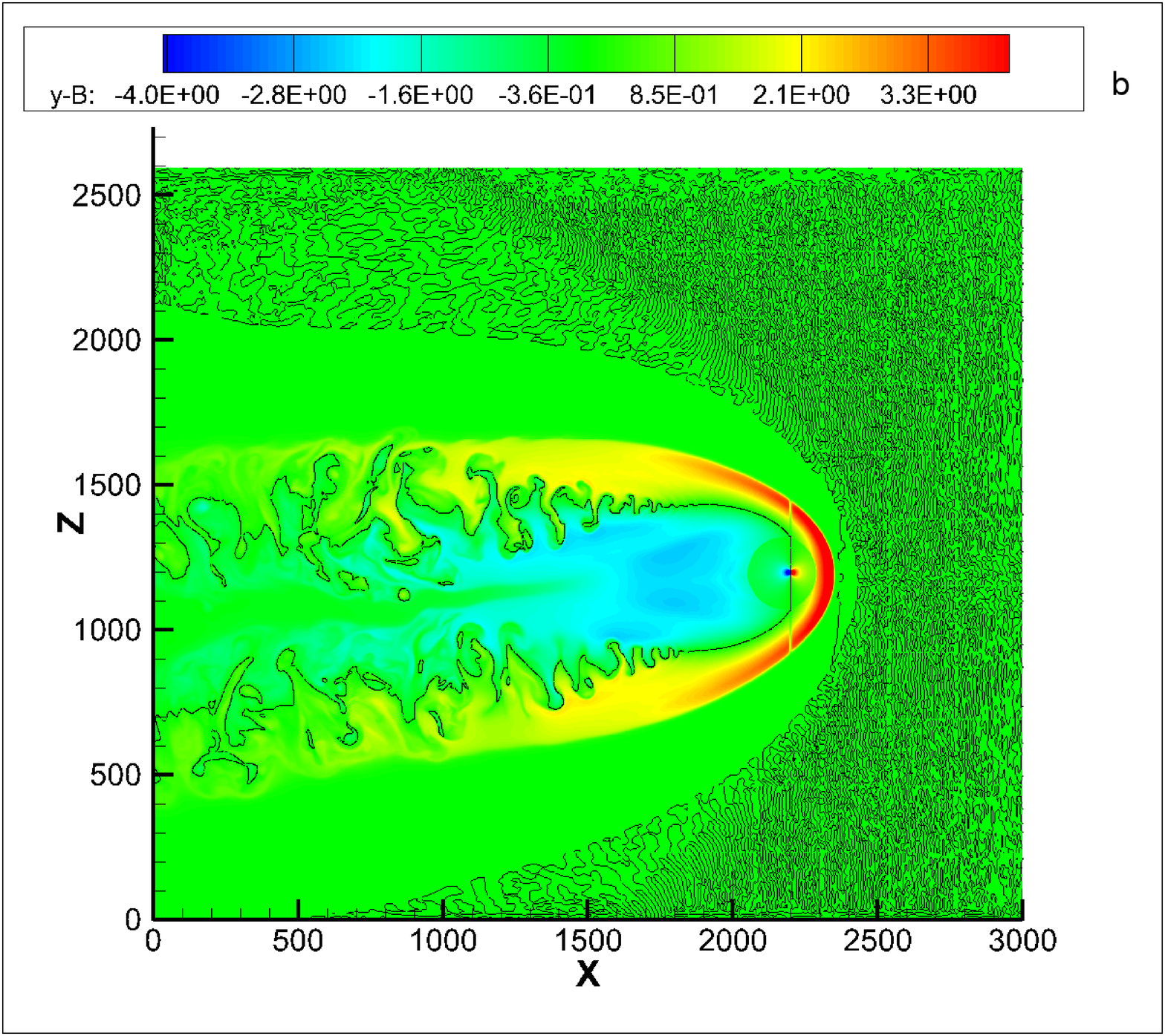}\\
\includegraphics[width=0.49\textwidth]{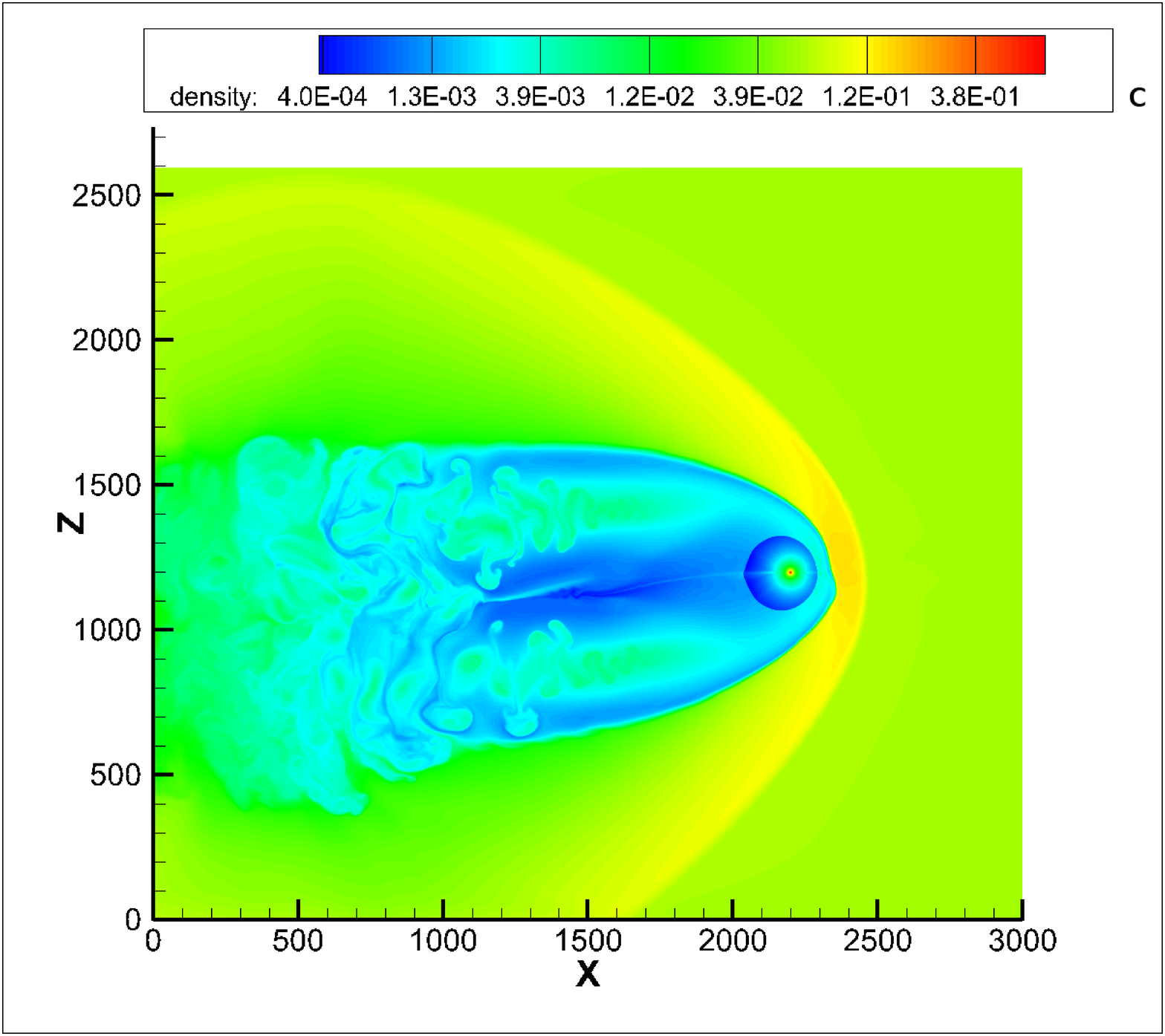}\vspace{1mm}
\includegraphics[width=0.49\textwidth]{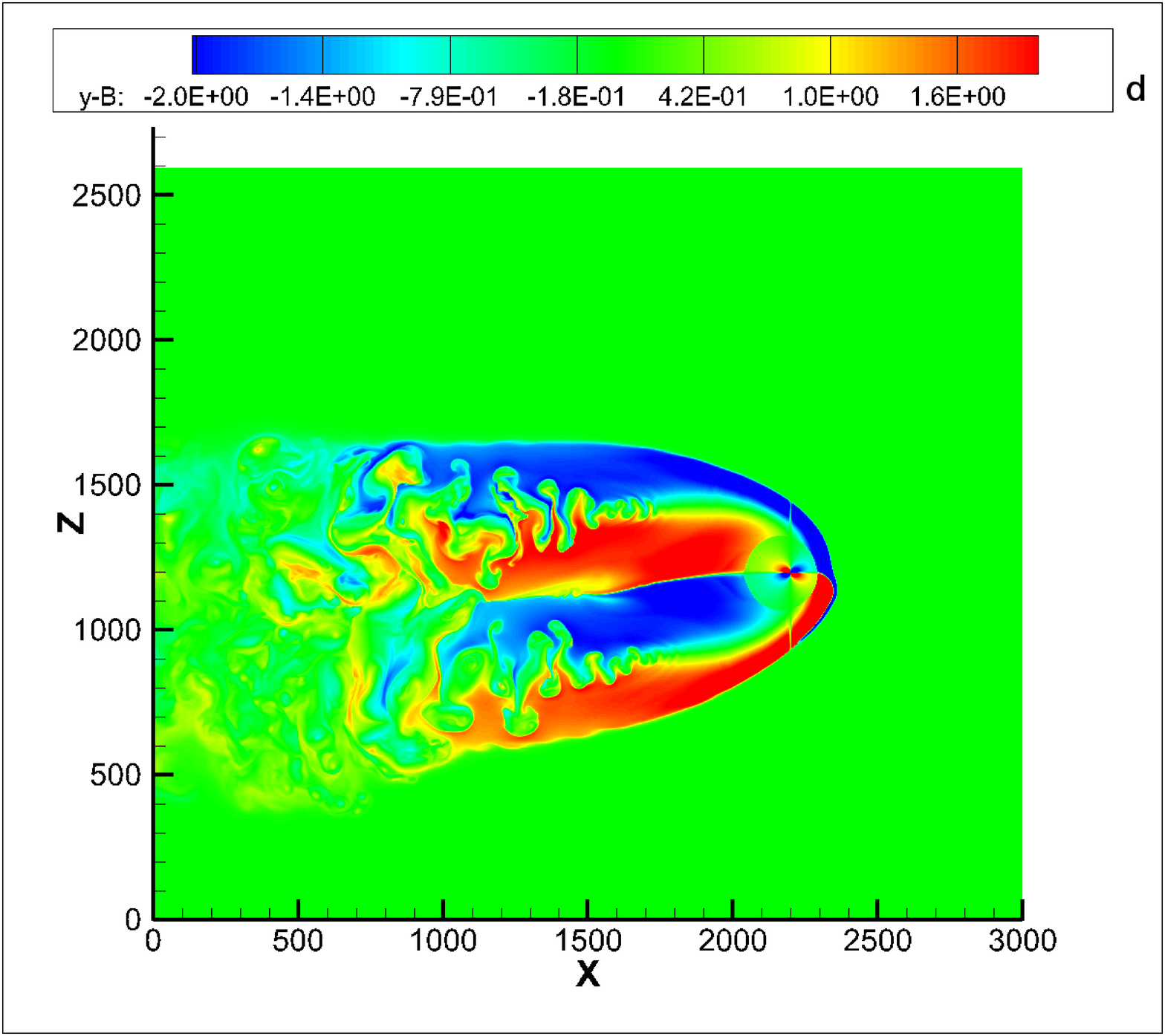}
\caption{(\emph{Top row.}) The distributions of the (\emph{a}) plasma number density and (\emph{b}) out-of-plane component, $B_y$, of the magnetic field vector in the meridional plane in the \emph{multi-fluid} simulations without interstellar magnetic field, unipolar heliospheric magnetic field, and all other parameters from \cite{2015ApJ...800L..28O}. (Bottom row.) The same as in the top row, but assuming the helisphereic current sheet is flat, i.e., there is no angle between the Sun's rotation and magnetic axes. Densities are in particles per cm$^3$ }
\label{Opher}
\end{figure}

\begin{figure}[t]
\centering
\includegraphics[width=0.48\textwidth]{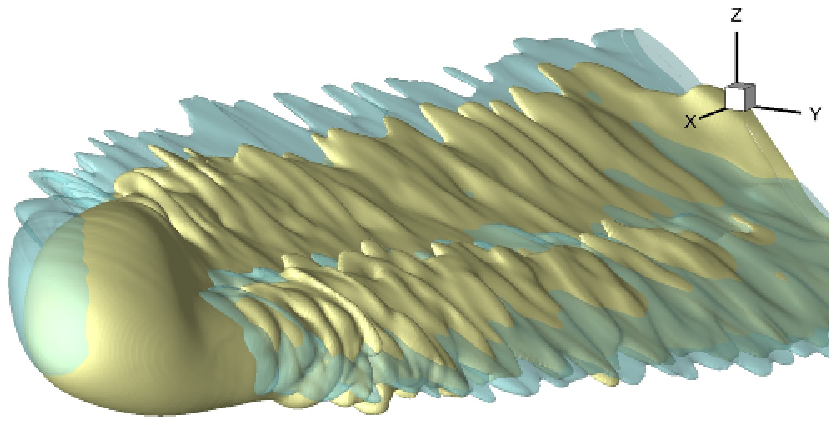}
\includegraphics[width=0.48\textwidth]{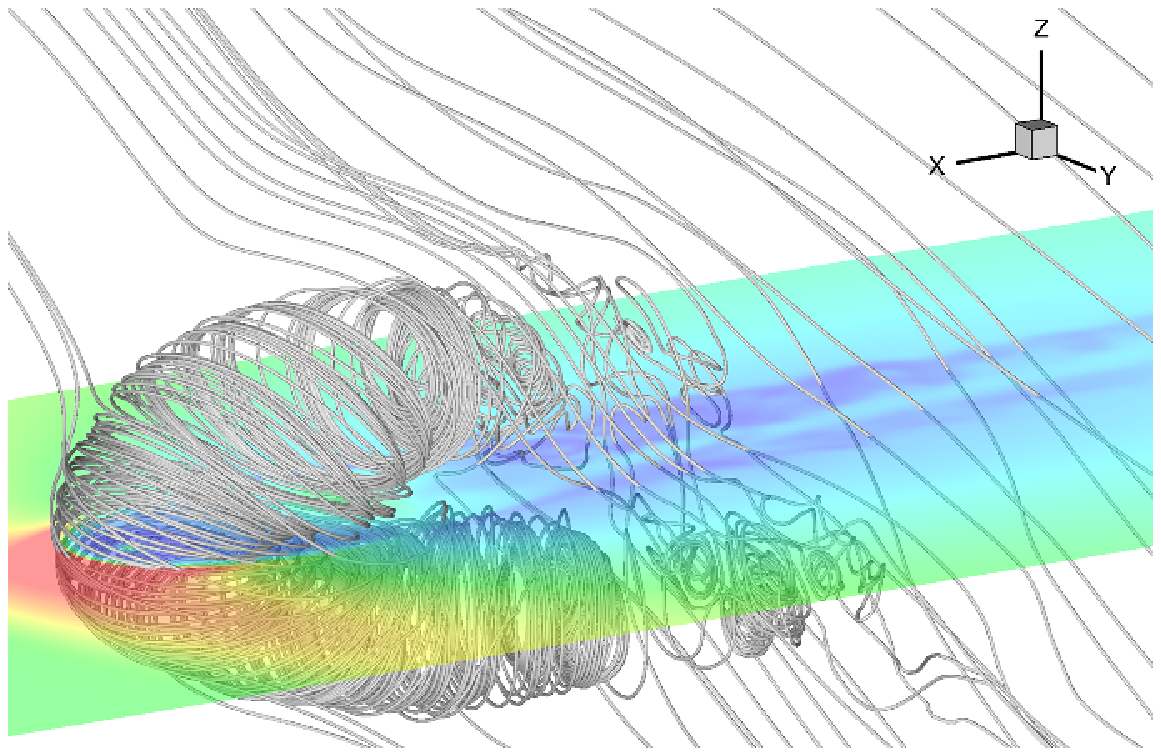}
\caption{\scriptsize MHD-plasma/kinetic-neutrals simulation of the SW--LISM interaction from \cite{2015ApJ...812L...6P}.
(Top panel) The shape of the heliopause for two different ISMF strengths
is shown (yellow and blue for $B_\infty= 3\ \mu$G and 4 $\mu$G, respectively). (Bottom panel) HMF line behavior initially exhibits a Parker spiral,
but further tailward becomes unstable. Also shown are ISMF lines draping around the heliopause.
The distribution of the plasma density is shown in the semi-transparent equatorial plane. [From \cite{2015ApJ...812L...6P} with permission of AAS].
}
\label{Kinetic}
\end{figure}

Paper~\cite{2015ApJ...812L...6P} considered the flow in the heliotail and compared simulation results with theoretical predictions \cite{1974ApJ...194..187Y,Jaeger-Fahr-1998} and numerical modeling \cite{2015ApJ...800L..28O,Izmod15}. The main conclusion is that the heliotail is very long, likely about $2\times 10^4$~AU. If the LISM is superfast magnetosonic (the flow velocity is greater than tghe fast magnetosonic speed), which happens if $B_\infty$ is not too strong (less that $\sim 3\ \mu$G), the SW flow becomes superfast at distances of about $4\times 10^3$~AU  along the tail. It was found that a kinetic treatment on neutral hydrogen atoms becomes critical. This is not surprising since multi-fluid approaches (see, e.g., \cite{1996JGR...10121639Z,1997JGR...10219779P,2009SSRv..143...31P}) are more likely to produce artifacts at larger distances. In multi-fluid models, the flow of neutral hydrogen atoms is described by multiple sets of the Euler gas dynamics equations, each for every population of neutrals born in thermodynamically different regions. In particular, it was found in \cite{2015ApJ...812L...6P} that there is a region in the heliotail where the SW flow remains subfast magnetosonic in contrast to the kinetic-neutrals solution. The reason can be understood if we look at the distribution of plasma number density in multi-fluid simulations from \cite{2015ApJ...812L...6P} shown in Figs.~\ref{Opher}a,b. These figures show the density and the out-of-plane component, $B_y$, of the magnetic field vector.
In both panels, there appear to exist two lobes of enhanced SW plasma number density, which are separated at $x\approx 1,500$~AU by a region with substantially different parameters attributed to the LISM in \cite{2015ApJ...800L..28O}. It was shown long ago in \cite{1974ApJ...194..187Y} that these lobes are due to the concentration of the SW plasma inside the Parker spiral field line diverted to the tail when the SW interacts with the HP.
The central spiral originates where the $z$-axis crosses the inner boundary. Both $B_x$ and $B_y$ are zero along this line, shown in Figs.~\ref{Opher}a,b, until it exits the supersonic SW outside the TS. This critical magnetic field line deflects tailward with other spiral field lines.
According to \cite{1956ApJ...124..430R,1974ApJ...194..187Y}, the plasma inside the spiral field is subject to a kink instability. As a result, the line $B_y=0$ exhibits rather chaotic behavior. As shown in \cite{2015ApJ...812L...6P}, the above line carries an electric current, which  increases considerably when the plasma distribution becomes unstable. Once the Parker field is destroyed by the kink instability, the necessity of plasma concentration inside the lobes disappears. However, as seen in Figs.~\ref{Opher}a,b, they still exist at $x=0$, although their width increases.
This behavior is in a drastic contrast with the solution where the transport of neutral hydrogen is treated kinetically, by solving the kinetic Boltzmann equation with a Monte Carlo method \cite{2015ApJ...812L...6P,Izmod15} (see Fig.\ref{Kinetic}). When neutral atoms are treated using a multi-fluid approach,
there is little charge exchange in the region separating the lobes. This is because the LISM neutral atoms,
whose flow is governed by the pressure gradient, do not cross this region. On the other hand, kinetic neutrals always cross the separation region because of their thermal velocity. Notice that although the simulations in ~\cite{2015ApJ...812L...6P} demonstrate some separation between the lobes, it is much smaller than in \cite{2015ApJ...800L..28O}, and the heliotail is considerably longer.

It is interesting to notice in this connection that short heliotails, such as observed in solutions \cite{2015ApJ...800L..28O}, are not favorable for
creating flux anisotropies in 1, and especially 10, TeV GCRs. A heliotail of less than 1,000~AU long would have little effect on those GCRs because of their large gyroradius. The assumption of the unipolar heliospheric magnetic field made in \cite{2015ApJ...800L..28O} requires special discussion. While it is clear
that the region of the SW swept by the HCS is impossible to resolve when the sector width becomes small, which is inevitable when the SW is decelerated by the HP to very small velocities, it is not quite clear why the solution with the removed HCS is better. {In \emph{Ulysses}-based, time-dependent simulations of \cite{2015ApJ...812L...6P}, the HMF along the \emph{Voyager} trajectories is reproduced on the average, even though the HCS dissipates, which would not be possible if the magnetic field was assumed unipolar.}
It is worth noticing that \textit{V1} was in a region of very small, even sunward, radial velocity component for two years before it crossed the HP.
{As previously mentioned, when the numerical resolution is sufficiently high, the HCS does not simply dissipate due to numerical effects. The plasma and magnetic field behavior in the region swept by the HCS becomes chaotic likely due to the tearing mode instability, which is inevitably numerical in MHD simulations.} As a consequence, the magnetic field strength becomes rather weak and the sector structure disappear. This is in agreement with \textit{V1} observations
which otherwise would show sector crossings much more frequently. {In our opinion, it is possible that} the sectors observed by \textit{V1} are more likely due to stream interaction and solar cycle effects. Such sectors are much less frequent than those related to the
Sun's rotation. It is possible that the spacecraft are crossing such sectors even in regions where the classical HCS does not exist. When the heliospheric field is assumed to be unipolar, its strength may be greater than in \textit{V1} observations. Further, assuming a unipolar field necessarily assigns an incorrect sign to the HMF below or above the magnetic equator. Additionally, solar cycle effects disappear, despite being an important ingredient of the SW flow.
\begin{figure}
\centering
\includegraphics[width=0.49\textwidth]{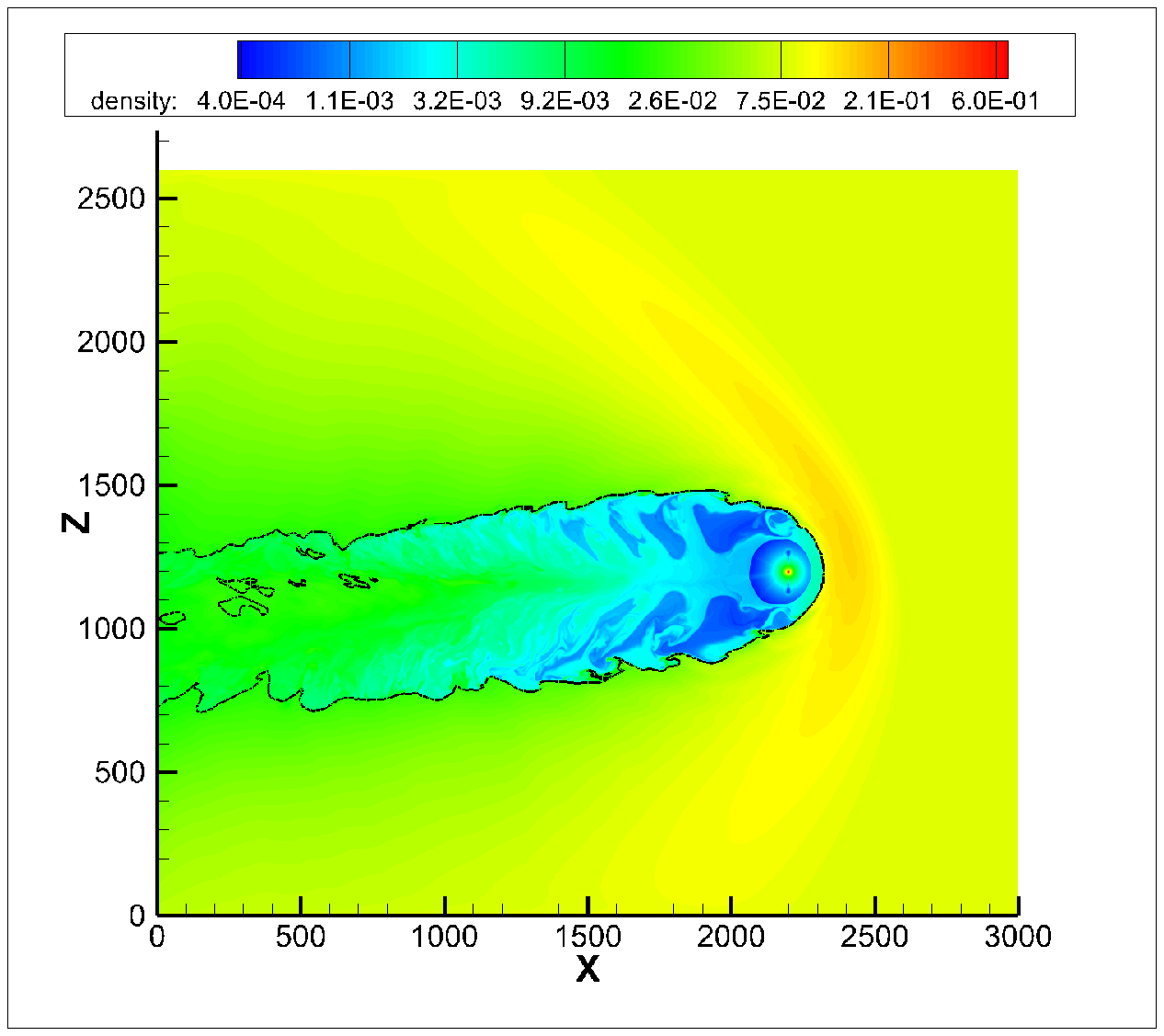}\vspace{1mm}
\includegraphics[width=0.49\textwidth]{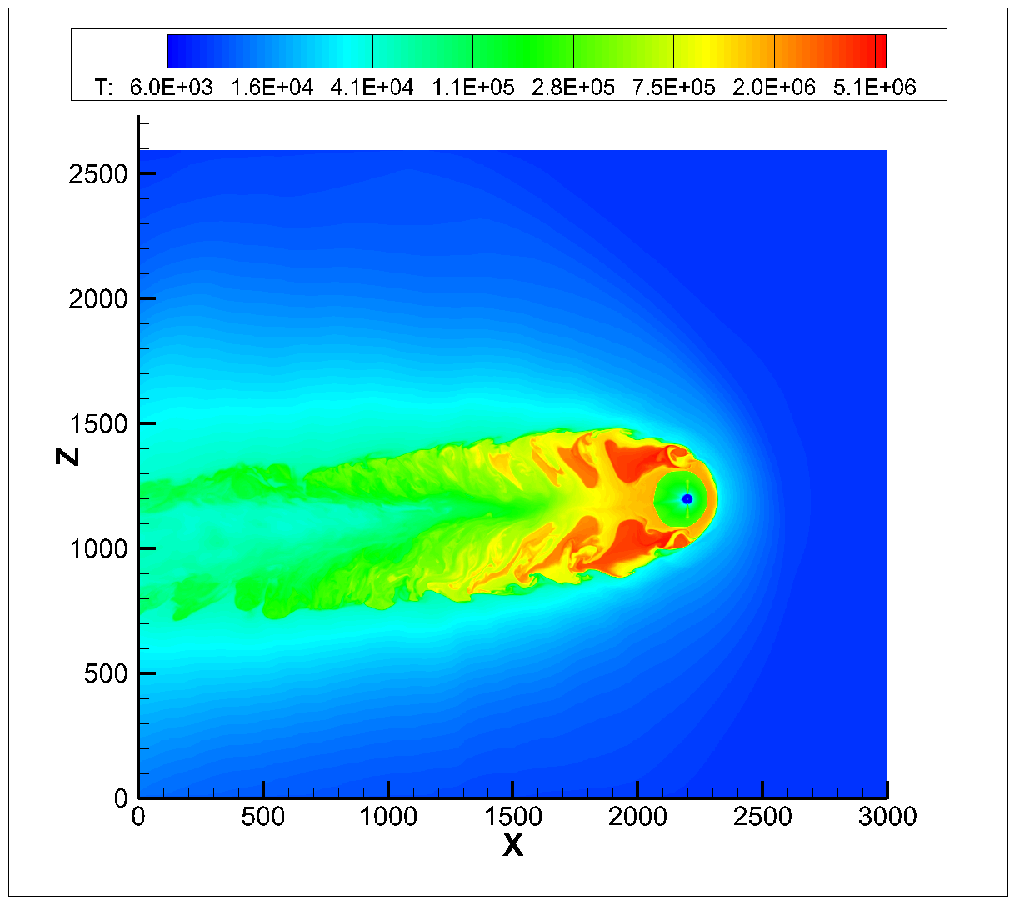}\\
\includegraphics[width=0.49\textwidth]{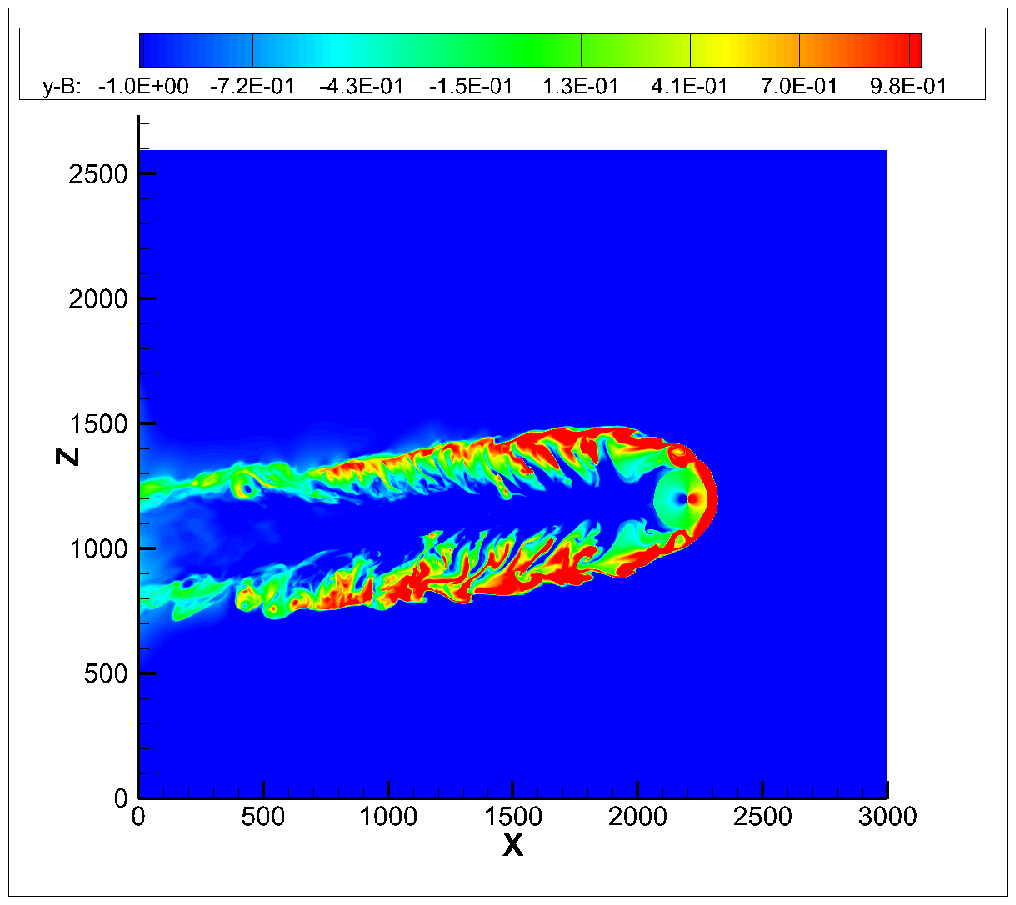}\vspace{1mm}
\includegraphics[width=0.49\textwidth]{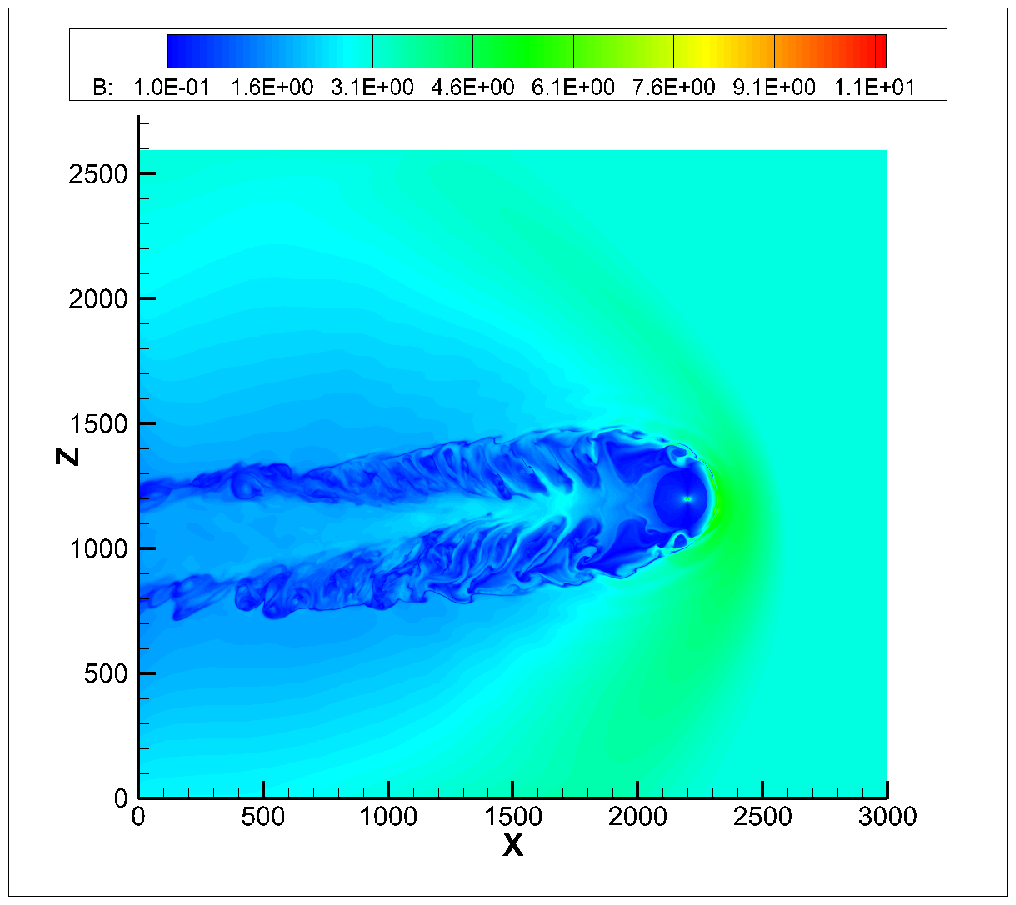}
\caption{Clockwise, the distributions of plasma number density (in cm$^{-3}$) and temperature (in K),
the $y$-component of the magnetic field and its magnitude in the SW--LISM simulation
in our solar-cycle simulation assuming unipolar heliospheric magnetic field. The top left panel also outlines the HP.}
\label{cycletail}
\end{figure}

The HCS is nearly flat close to solar minima. As seen from \cite{Pogo04}, it bends into one of the hemispheres depending of the direction and strength of the ISMF. The flat-HCS case easily can be treated numerically and is therefore a good test for unipolar simulations. Figures~\ref{Opher}c,d show the
solution similar to that shown in the top row of this figure, except that the HCS is flat in the supersonic SW. It is seen that although the lobes do reveal themselves at small distances from the Sun, there is no separation between them farther along the tail. This happens because the HCS in the tail is affected by the unstable SW flow.

Another test for the unipolar HMF assumption would be to  allow the SW variations related to the solar cycle. The solution obtained under these assumptions is shown in Fig.~\ref{cycletail}. We see here a drastic change in the entire structure of the heliotail flow. The lobes disappear completely. On the contrary, the SW plasma is more dense near the equatorial plane. This is not surprising because the slow SW is denser than the fast wind near the poles. This solution makes questionable the idea of a short, ``croissant''-like heliotail shown in \cite{2015ApJ...800L..28O}. In other words, the heliotail structure becomes completely different from that described in the analytical studies of~\cite{1974ApJ...194..187Y,Drake15}. The latter also did not take into account charge exchange, while it is known that even the original Parker solution \cite{1961ApJ...134...20P}, which described the SW propagation into the magnetized vacuum, is only partially valid in the presence of interstellar neutrals
(see \cite{Pogo11}). This is because charge exchange does not allow
the SW to propagate upstream indefinitely. One can see from Fig.~\ref{cycletail} that the solar cycle smears out more subtle effects related to the SW plasma collimation within the Parker magnetic field swept by the flow into the tail. As shown in \cite{Pogo04,2015ApJ...812L...6P}, the HP usually
rotates to become nearly aligned with the $BV$-plane. {Black lines in the tail show that the instability of the HP flanks may produce local
protrusion that cross the meridional plane. }

\begin{figure}
\centering
\includegraphics[width=0.49\textwidth]{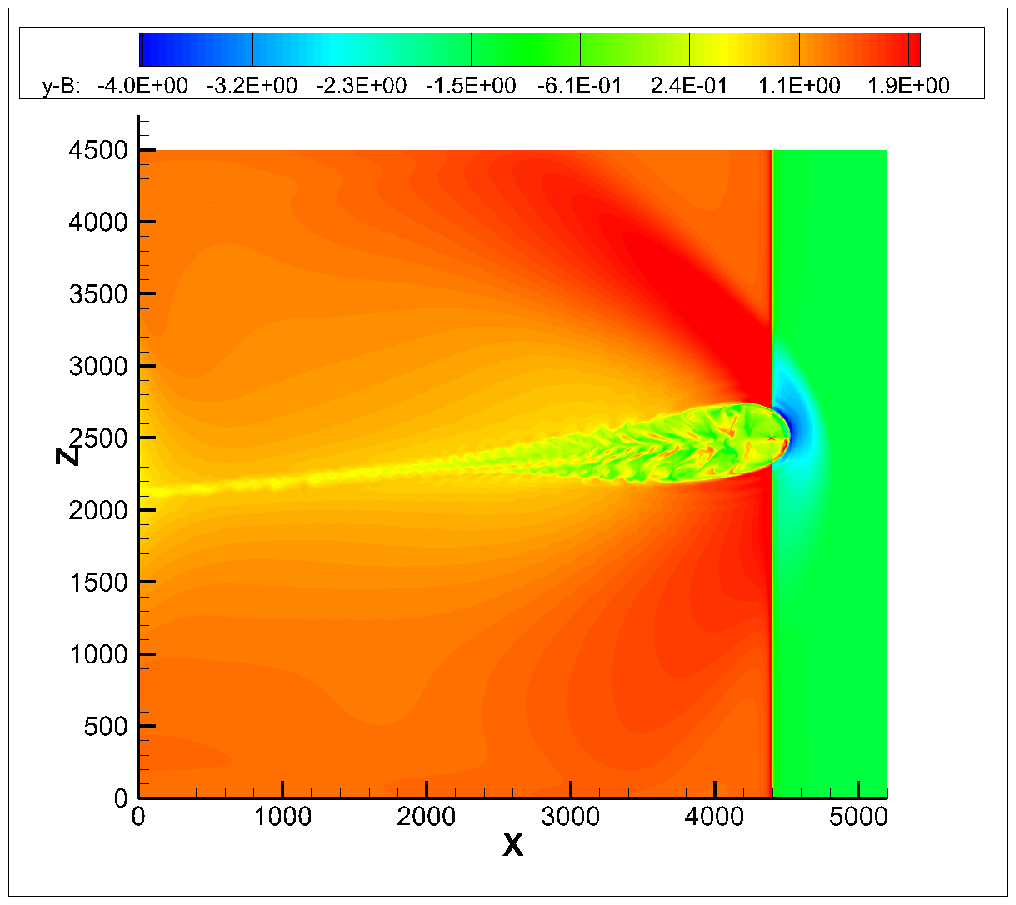}\vspace{1mm}
\includegraphics[width=0.49\textwidth]{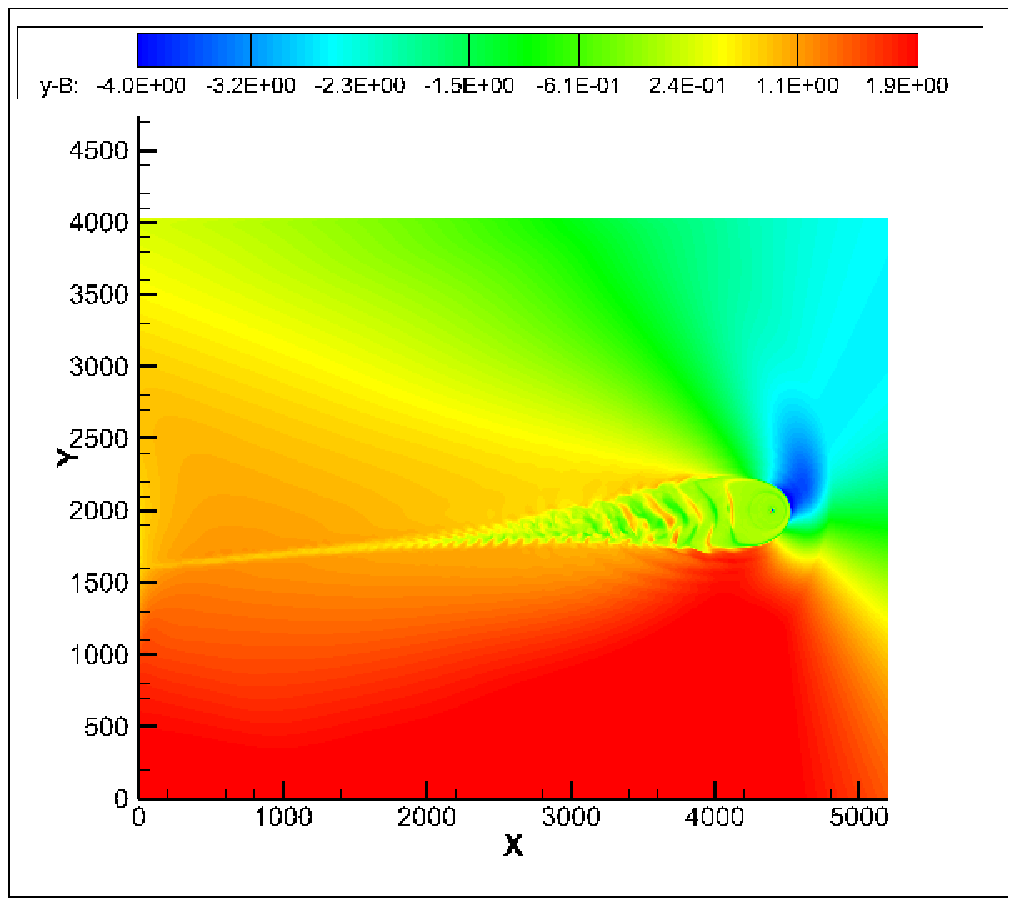}
\caption{The heliotail in the multi-fluid simulation which takes into account solar cycle effects.
The distributions of $B_y$ are shown in the meridional (left panel) and ecliptic (right panel)
planes. The HP looks rather thin beyond 2,000~AU. In reality it is rather wide latitudinally in the $BV$-plane, but very thin
in the direction perpendicular to that plane. The LISM boundary conditions for this problem are taken from \cite{Eric16}.}
\label{tilt}
\end{figure}

As shown in \cite{2015ApJ...812L...6P}, the effects of the solar cycle are not only due to the changes in the latitudinal extent of the slow wind.
Of importance are also changes in the angle between the Sun's rotation and magnetic axes, as well as the change of the magnetic polarity of the Sun every solar cycle at maxima. In Figure~\ref{tilt}, we show the distribution of the $y$-component of the magnetic field vector
in the meridional and ecliptic planes for a simulation using parameters from \cite{Eric16} for $B_\infty = 3\ \mu$G. Note the similarity of the shape of the heliotail to that estimated earlier by \cite{Jaeger-Fahr-1998}.

{Solar cycle simulations of the heliotail presented in Figs.~\ref{Opher} and \ref{cycletail}--\ref{tilt} are obtained with a multi-fluid model. No characteristic wave reflections have been observed from the exit boundary. Time-dependence creates conditions where no fluid dynamics artifacts in the neutral H flow are observed.}

The numerical analysis of \cite{Eric_tail} demonstrates that solar cycle effects, especially the presence of slow and fast wind regions,
are seen in the ENA fluxes observed by \textit{IBEX} from the tail direction. This requires no collimation of the SW plasma that is
observed in simplified models of the heliosphere. Additionally, as mentioned above, the short heliotail obtained in numerical simulations
\cite{2015ApJ...800L..28O} is incompatible with the idea that the multi-TeV cosmic ray anisotropy is affected by a large perturbation of the ISMF due to the presence of the heliotail.

By fitting the anisotropy of multi-TeV cosmic rays observed in air shower observations by the Tibet,
Milagro, Super-Kamiokande, IceCube/EAS-Top, and ARGO-YGB teams (see references in \cite{2014ApJ...790....5Z}),
we can derive restrictions on the LISM properties as found in \cite{2013ApJ...762...44D,2014Sci...343..988S,2014ApJ...790....5Z}.
Additionally,  it is suggested in \cite{2010ApJ...722..188L} that ion acceleration  due to reconnection in the heliotail may affect observed anisotropies.

{The main result of our heliotail study is three-fold:}
\begin{itemize}
\item
{Even our multi-fluid model, when run with the unipolar heliospheric magnetic filed assumption, shows results different from \cite{2015ApJ...800L..28O}. This is shown in Figure 10. One can only guess about the reasons for that. A possibility is the implementation of the subsonic exit boundary conditions.}

\item
{In \cite{2015ApJ...812L...6P}, we have found that in agreement with \cite{2003AstL...29...58I} and \cite{Izmod15}, the SW flow becomes superfast magnetosonic again at distances of about 4,000 AU. In such cases, no boundary conditions are necessary at the exit boundary. In the absence of solar cycle effect, this happens only if neutral atoms are treated kinetically, but never if they are treated with a multi-fluid approach. This is our explanation of the qualitative difference between MHD-kinetic and multi-fluid results.}

\item
{All of the above conclusions become irrelevant when solar cycle effects are taken into account. As shown in Fig.~\ref{cycletail}, the collimation of the SW within two polar lobes disappears even if the heliospheric magnetic field is assumed unipolar, which is the necessary condition for obtaining a ``croissant''-shaped heliosphere with the LISM between the lobes. We obtain one single heliosphere. Instead of concentrating inside the lobes, the SW has higher density near the equatorial plane, where the slow SW is.
From this standpoint, the above two conclusions have only theoretical importance because they do not take into account one of the basic features of the SW flow: the solar cycle.}

\item
{As the SW propagates tailward, both thermal and nonthermal ions continue to experience charge exchange which substitutes them with the cool LISM ions until the plasma temperature in the tail becomes uniform and the heliopause disappears. As seen from \cite{2015ApJ...812L...6P}, the heliopause should become very narrow, while being aligned with the $BV$-plane. Newly created neutral atoms, because of their large mean free path will be leaking through the HP surface into the LISM and ultimately reach thermodynamic equilibrium with the pristine LISM.}

{The assumption of a unipolar field in the tail is damaging for determination of GCR fluxes coming from the heliotail. There is no imperative to running the code with the variable tilt between the Sun's magnetic and rotation axis. This inevitably results in the HMF dissipation in initially sectored regions of the SW. Clearly, only models that involve SW turbulence can correctly address this issue. Local kinetic simulations may be useful  to establish the dissipation rate and in this way supplement global models. On the other hand, as shown in
\cite{Pogorelov-etal-2013}, the HMF at \textit{Voyagers} can be reproduced on the average even if some sector structure is lost.}
\end{itemize}

\section{Significance of the multi-species structure of the LISM}
\subsection{The effect of helium charge exchange}
It is well-known that the LISM not only consists of protons and hydrogen but
that it contains a non-negligible amount of singly-charged and neutral helium.
So far the significance of helium for the large-scale structure of the
heliosphere has been discussed in \cite{Izmodenov-etal-2003,Malama}, as well as in \cite{Scherer-Fichtner-2014,Scherer-etal-2014}, while it is not yet standardly incorporated in self-consistent multi-fluid modelling.  It has been demonstrated in \cite{Scherer-Fichtner-2014}
 that including the charged helium component of the LISM is crucial
for the comparison of the LISM flow speed with the wave speeds and, thus, for
the answer to the question whether or not the interstellar flow is
super-Alfv\'enic and/or superfast magnetosonic. Moreover, the
presence of helium ions influences the characteristic wave speeds, which
are crucial in numerical models.

{In most self-consistent models of heliospheric dynamics, so far, only
the influence of neutral hydrogen is considered by taking into account its
charge exchange with solar wind protons and its ionization by the solar
radiation (e.g., \cite{Fahr-Rucinski-2001,Pogo09,Alouani-Bibi-etal-2011} and
references therein).
The dynamical relevance of both the electron impact ionization of hydrogen,
although recognized by \cite{Fahr-etal-2000}, \cite{Malama} as well as
\cite{Gruntman_2015}, and the photo-ionization of helium, although
recognized as being filtered in the inner heliosheath
\cite{Rucinski-Fahr-1989,Cummings-etal-2002}, have not yet been
explored in detail. There is only one attempt to include helium
self-consistently in the heliospheric modeling \cite{Malama}, in which the
emphasis is on the additional ram pressure due to the charged helium ions.}
\begin{figure}[t!]
\begin{center}
  \includegraphics[width=0.71\columnwidth]{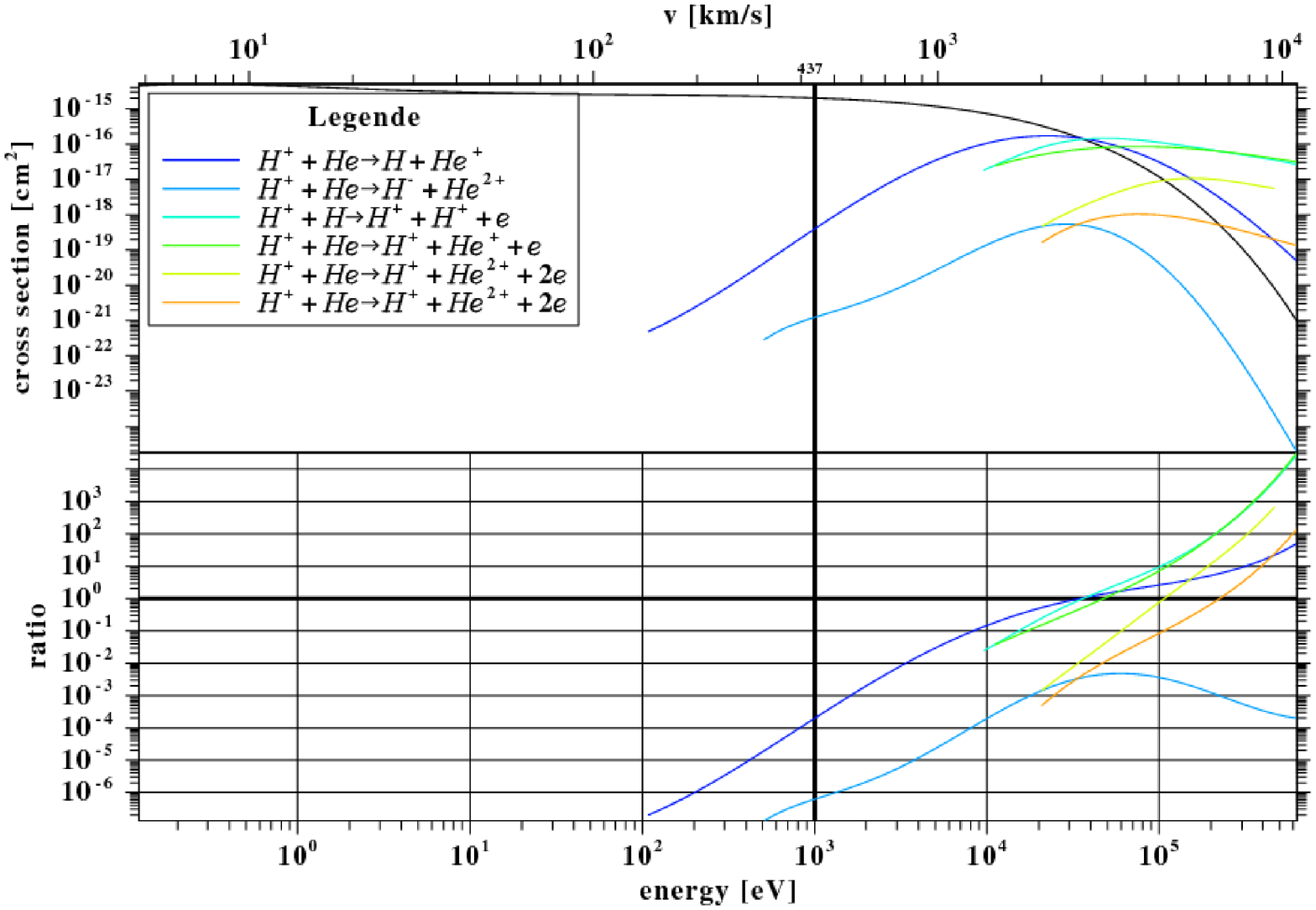}
  ~\\
  ~\\
  \includegraphics[width=0.71\columnwidth]{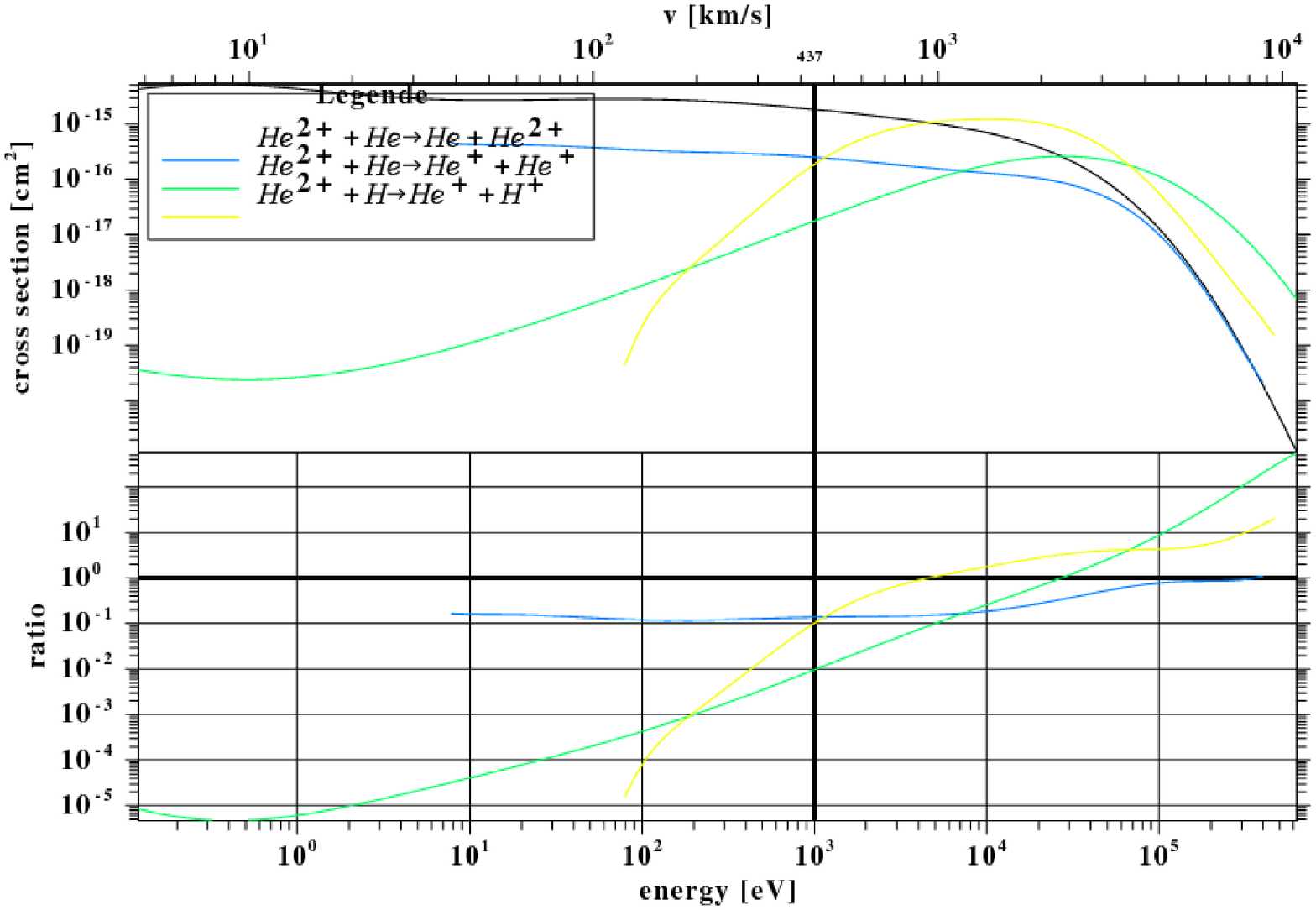}
\end{center}
  \caption{The charge-exchange cross section as function of energy per
    nucleon for protons (upper panel) as well as $He^{+}$-ions and
    $\alpha$-particles (lower panel) of the solar wind with
    interstellar helium and hydrogen. In the upper part of both panels the
    cross sections are shown, while the lower parts show the ratio to
    $\sigma_{cx}(H^{+}+H\rightarrow H+H^{+})$. The black curve in both
    panels is the reaction $H+H^{+}\rightarrow H^{+}+H$. As can be
    seen in the lower panel the reactions He$^{+}$+He,
    He$^{2+}$+He, and He$^{2+}$+He$^{+}$ have similar cross sections
    than that of H+p, and thus are important in modeling the dynamics
    of the large-scale astrospheric structures. Note the different y-axis
  scales between different panels, taken from \cite{Scherer-etal-2014}.}
  \label{Fig2}%
\end{figure}

One interesting feature of the heliosphere and some nearby
astrospheres is their hydrogen walls, which are built beyond the
helio-/astropauses by charge exchange between interstellar hydrogen
and protons. The feature can be observed in Lyman-$\alpha$ absorption
\cite{Linsky96,Gayley96,Wood-etal-2007,2009SSRv..143...21W,2014ApJ...780..108W,2014ApJ...789...80K}, which in turn allows the determination of the stellar
wind and interstellar parameters at some nearby stars \cite{2014ASTRP...1...43L}. Because the
hydrogen wall forms in the shocked interstellar medium, where
the temperature is low ($<10^{5}$\,K) and in case of the heliosphere
increases only by approximately a factor of two towards the heliopause, the
charge-exchange process involved is that between protons and hydrogen. In
addition some helium reactions, like He$^{+}$+He, He$^{2+}$+He and
He$^{+}$+He$^{+}$ have large cross sections even at low energies.  A
helium wall as a result of helium-proton charge exchange was found not to
exist \cite{Mueller-Zank-2004b}. However, in \cite{Scherer-etal-2014}
 a helium wall was predicted, based on helium-helium reactions with
sufficiently large cross sections.

{Note, that the sum of the number densities of the proton and helium charges
derived in \cite{Slavin-Frisch-2008} corresponds nicely to the recently
observed electron number density $n_{e}=0.08$\,cm$^{-3}$ observed with the
plasma wave instrument onboard Voyager \cite{Gurnett-etal-2013}.}

In Fig.~\ref{Fig2} it can be seen that the charge-exchange cross section
$\sigma_{cx}(H^{+}+H\rightarrow H+H^{+})$ is roughly in the range of
$10^{-15}$\,cm$^{2}$ below 1\,keV, i.e.\ the range of interest for
heliospheric models.  All other cross sections $\sigma_{cx}$ between
protons and neutral hydrogen or helium are orders of magnitude smaller for slow
solar/stellar wind conditions.  In high-speed streams and
especially in coronal mass ejections, cross sections like
$\sigma_{cx}(H^{+}+He\rightarrow H+He^{+})$,
$\sigma_{cx}(H^{+}+He\rightarrow H^{+}+He^{+}+e)$ and
$\sigma_{cx}(H^{+}+He\rightarrow H^{+}+He^{+}+e)$ can be of the
same magnitude as $\sigma_{cx}(H^{+}+H\rightarrow H+H^{+})$.  For
astrospheres with stellar wind speeds of the order of a few thousand
km/s, the energy range is shifted toward 10\,keV up to 100\,keV and
other interactions, like non-resonant charge-exchange processes, need
to be taken into account.

While the cross section $\sigma_{cx}$ between $\alpha$-particles and
neutral hydrogen or helium compared to the
$\sigma_{cx}(H^{+}+H\rightarrow H+H^{+})$ reaction seem not to be
negligible in and above the keV-range (Fig.~\ref{Fig2}), the solar abundance of
$\alpha$-particles is only 4\% of that
of the protons, so that the effect seems to be small. Nevertheless,
the mass of helium or its ions is roughly four times that of (charged) hydrogen,
and thus may play a role in mass-, momentum-, and energy loading.

\subsection{Modeling a pickup ion mediated plasma}
As mentioned earlier, the outer heliosphere beyond the ionization cavity (i.e., $\ge \sim 8$ AU) is dominated thermally by PUIs (see, e.g.,
the observational results in \cite{Burlaga_etal_1994,Richardson_etal_1995a}).
As reported in~\cite{Decker_etal_2008,Decker_etal_2015}, the inner heliosheath pressure contributed
by energetic PUIs and anomalous cosmic rays far exceeds that of the thermal background plasma and magnetic field.

Coulomb collisions are necessary to equilibrate a background thermal plasma and PUI protons. For a background Maxwellian plasma comprised of thermal protons, the relative ordering of the thermal speed of ``hot'' PUIs can be exploited \cite{2014ApJ...797...87Z} to determine equilibration time scales in the supersonic and subsonic solar wind and LISM. The equilibration time scale can then be compared to the convection time scale and the size of the region under consideration to determine with PUIs and thermal background plasma will equilibrate. For the supersonic solar wind, \cite{Isenberg_1986} showed that a multi-fluid model is necessary to describe a coupled SW--PUI plasma since neither proton nor electron collisions can equilibrate the PUI-mediated supersonic solar wind plasma \cite{2014ApJ...797...87Z}.

The inner heliosheath is  complicated by the microphysics of the TS. The supersonic solar wind is decelerated on crossing the quasi-perpendicular TS. The flow velocity is directed away from the radial direction and is $\sim 100$ km/s. The HMF remains approximately perpendicular to the plasma flow. \emph{Voyager~2} measured the downstream solar wind temperature to be in the range of $\sim 120,000$--180,000~K \cite{Richardson_2008GeoRL,Richardson_etal_2008Nature}, which was much less than predicted by MHD-neutral models.
This is because \textit{Voyager} isntruments are not designed to measure the PUI contribution. In reality, the thermal energy in the inner heliosheath is dominated by PUIs. There are two primary sources of PUIs in the inner heliosheath. One is interstellar neutrals that move freely across the HP and charge exchange with hot SW plasma. Newly created ions are picked up in the inner heliosheath plasma in the same way that ions are picked up in the supersonic SW. The characteristic energy for PUIs created in this way is $\sim 50$ eV or $\sim 6 \times 10^5$~K, which is about five times hotter than the inner heliosheath SW protons. The second primary source is PUIs created in the supersonic SW that are convected across the TS into the inner heliosheath. PUIs convected to the TS are either transmitted immediately across the TS or reflected before transmission \cite{1996JGR...10121639Z}. PUI reflection was predicted in~\cite{1996JGR...10121639Z} to be the primary dissipation mechanism at the quasi-perpendicular HTS, with the thermal solar wind protons experiencing comparatively little heating across the TS. The transmitted PUIs downstream of the HTS have temperatures $\sim 9.75 \times 10^6$~K ($\sim 0.84$ keV) and the reflected protons have a temperature of $\sim 7.7 \times 10^7$~K ($\sim 6.6$ keV) \cite{Zank_etal_2010ApJ}. PUIs, those transmitted, reflected, and injected, dominate the thermal energy of the inner heliosheath, despite being only some 20\% of the thermal subsonic solar wind number density. {The ionization rate in the outer heliosphere (both in the supersonic and subsonic solar wind - inner heliosheath) is very slow due to the extremely low proton number densities in this region. Most of the accumulation occurs closer to the Sun, and since the ionization time scale is $\sim 10^{-6}$~s$^{-1}$, the net change in the PUI density is small. More rapid changes in the SW density (shocks, MIRs, etc.) are very minor, and even factors of 2--4 will make little difference. Possible change in the neutral number density is even slower, due in part to the very slow response of neutral H to changes in the boundary regions as it drifts slowly ($\sim 20$~km s$^{-1}$) through the inner heliosheath and the outer regions of the supersonic SW.}

To simplify the kinetic approach of \cite{Malama} based on the kinetic treatment of multiple populations of PUIs, the inner heliosheath proton distribution function can be approximated by a 3-component \cite{Zank_etal_2010ApJ,Burrows_etal_2010} or 4-component distribution function \cite{Zirnstein_etal_2014}, with a relatively cool thermal solar wind Maxwellian distribution and 2 or 3 superimposed PUI distributions. In \cite{Mihir1,Zirnstein_etal_2014} and \cite{Desai_etal_2014}, this decomposition of the inner heliosheath proton distribution function was exploited  in modeling ENA spectra observed by the \emph{IBEX} spacecraft at 1 AU. They identified multiple proton distribution functions in the inner heliosheath and the LISM, these being the various PUI populations described above and the thermal SW proton population \cite{Zank_etal_2010ApJ}.
In \cite{2014ApJ...797...87Z}, it is shown that neither proton nor electron collisions can equilibrate a PUI-thermal SW plasma in the subsonic SW or inner heliosheath on scales smaller than at least 10,000 AU. This requires treating PUIs as a separate component of the plasma flow. This issue will be addressed in the nest section.

The interstellar plasma upwind of the heliopause is also mediated by energetic PUIs. As seen from \cite{Bama93,1996JGR...10121639Z} that energetic neutral H atoms created via charge exchange in the inner heliosheath and fast solar wind could ``splash'' back into the LISM where they would experience a secondary charge exchange. The secondary charge exchange of hot and/or fast neutral H with cold {($\sim 6300$ K, as in \cite{McComas_etal_2012Sci}, or $\sim 8000$~K as in \cite{2015ApJS..220...28B,2015ApJS..220...22M})}  LISM protons leads to the creation of a hot or suprathermal PUI population locally in the LISM.
The presence of the charge-exchange source terms in the system of MHD equations cannot change the Hugoniot jump conditions. This is possible only for source terms involving delta-functions. However, charge exchange can modify upstream and downstream quantities at a hypothetical
bow shock. Since this modification reveals itself only after the problem is solved in its entirety, it impossible to say whether
any shocked transition should be expected for a chosen set of LISM boundary conditions \cite{2006ApJ...644.1299P,Jacob11}.
The LISM is known to be supersonic, but it can be subfast magnetosonic ($V_\infty$ less than the fast magnetosonic speed, $c_{\mathrm{f}\infty}$, in the unperturbed LISM). If $V_\infty<c_{\mathrm{f}\infty}$, no fast-mode bow shock is possible. If the angle between $\mathbf{V}_\infty$
and $\mathbf{B}_\infty$ is small, slow-mode bow shocks remain possible \cite{2004ApJ...604..700F,Pogo11,2010ApJ...716L..99C,2013GeoRL..40.2923Z}.
It was noted in~\cite{Zank_etal_2013} that heating of the LISM induced by charge exchange may result in an increase of the fast magnetosonic
speed in the outer heliosheath with a concomitant weakening or even elimination of the subshock in a structure which is now called a bow wave. PUIs form a tenuous ($n_p \simeq 5 \times 10^{-5}$ cm$^{-3}$) \cite{Zirnstein_etal_2014} suprathermal component in the LISM.
It is shown in~\cite{2014ApJ...797...87Z} that neither proton nor electron collisions can equilibrate a PUI-thermal plasma in the LISM on scales smaller than at least 75~AU.
The observational results by \cite{Mihir1,Desai_etal_2014} confirm that indeed the inner heliosheath and LISM are multi-component non-equilibrated plasmas. Simplified single-fluid MHD plasma descriptions, while preserving the total mass, momentum, and energy balances, do not capture the complexity of the plasma. On the other hand, fully kinetic model \cite{Malama} is rather complicated for realistic time-dependent simulations.
PUIs were treated as a separate fluid in \cite{2011JGRA..116.3105D,2012ApJ...754...40U,2014ApJ...788...43U} in the supersonic SW.
MHD equations for plasma were coupled to a kinetic treatment of PUIs, also for the supersonic SW only, in \cite{Gamayunov}. Some of the above models take into account the transport  of turbulence. In \cite{Usmanov16}, the model applicable to the supersonic SW was used to all
regions of the SW--LISM interaction. However, such application causes serious questions because (1) the turbulence transport equations derived for the super-Alfv\'enc plasma are invalid in the inner heliosheath, (2) the charge exchange source term formulae used in \cite{Usmanov16} are applicable only in cold plasma and are very inaccurate in the inner heliosheath, and (3) the application of non-conservative equations across discontinuities creates uncontrollable mistakes in their speeds and strengths. Moreover, the boundary conditions for PUIs at shocks are too complicated to be
modeled by approximating derivatives in the governing equations straightforwardly \cite{Zank_etal_2010ApJ,2012Ap&SS.341..265F,2013A&A...558A..41F}.
In contrast, \cite{2014ApJ...797...87Z} is the first rigorous attempt to extend basic continuum-mechanics (non-kinetic)  models to incorporate the physics of non-thermal PUI distributions.

\subsection{Multi-component model}
In deriving a multi-component plasma model that includes PUIs, we shall assume that the distribution functions for the background protons and electrons are each Maxwellian, which ensures the absence of heat flux or stress tensor terms for the background plasma. The exact continuity, momentum, and energy equations governing the thermal electrons ($e$) and protons ($s$) are therefore given by
\begin{eqnarray}
\frac{\partial n_{e,s} }{\partial t} + \nabla \cdot \left( n_{e,s} {\bf u}_{e,s} \right) = 0, \label{eq:4}  \\
m_{e,p} n_{e,p} \left( \frac{\partial {\bf u}_{e,s} }{\partial t} + {\bf u}_{e,s} \cdot \nabla {\bf u}_{e,s} \right) = -\nabla P_{e,s} + q_{e,s} n_{e,s} \left( {\bf E} + {\bf u}_{e,s} \times {\bf B} \right), \label{eq:5}  \\
\frac{\partial P_{e,s}}{\partial t} + {\bf u}_{e,s} \cdot \nabla P_{e,s} + \gamma_{e,s} P_{e,s} \nabla \cdot {\bf u}_{e,s} = 0. \label{eq:6}
\end{eqnarray}
Here $n_{e,s}$, ${\bf u}_{e,s}$, and $P_{e,s}$ are the macroscopic fluid variables  for the electron/proton number density, velocity, and pressure respectively, $\gamma_{e,s}$ the electron/proton adiabatic index, ${\bf E}$ the electric field, ${\bf B}$ the magnetic field, and $q_{e,s}$ the charge of particle.

The streaming instability for the unstable PUI ring-beam distribution excites Alfv{\'e}nic fluctuations. The self-generated fluctuations and {\it in situ} turbulence serve to scatter PUIs in pitch-angle. The Alfv{\'e}n waves and magnetic field fluctuations both propagate and convect with the bulk velocity of the system. The PUIs are governed by the Boltzmann transport equation with a collisional term $\delta f/\delta t |_c$, due to wave-particle scattering,
\begin{equation}
\frac{\partial f}{\partial t} + {\bf v} \cdot \nabla f + \frac{e}{m_p} \left( {\bf E} + {\bf v} \times {\bf B} \right) \cdot \nabla_v f = \left. \frac{\delta f}{\delta t} \right|_c,, \label{eq:10}
\end{equation}
for average electric and magnetic fields ${\bf E}$ and ${\bf B}$. On transforming the transport equation (\ref{eq:10}) into a frame that ensures there is no change in PUI momentum and energy due to scattering, assuming that the cross-helicity is zero, and introducing the random velocity ${\bf c} = {\bf v} - {\bf U}$, we obtain
\begin{eqnarray}
&&\frac{\partial f}{\partial t} + \left( U_i +  c_i \right) \frac{\partial f}{\partial x_i} + \left[ \frac{e}{m_p} \left( {\bf c} \times {\bf B} \right)_i - \frac{\partial U_i}{\partial t} - \left( U_j + c_j \right) \frac{\partial U_i}{\partial x_j}  \right] \frac{\partial f}{\partial c_i} \nonumber\\
&&= \frac{\partial }{\partial \mu} \left( \nu_s ( 1 - \mu^2 ) \frac{\partial f}{\partial \mu} \right), \label{eq:15}
\end{eqnarray}
where we have introduced the guiding center frame to eliminate the motional electric field and $\mu = \cos \theta$ is the cosine of the particle pitch-angle $\theta$, and $\nu_s = \tau_s^{-1}$ is the scattering frequency. The scattering operator is the simplest possible choice, and corresponds to isotropic pitch-angle diffusion.

By taking moments of (\ref{eq:15}), we can derive the evolution equations for the macroscopic PUI variables, such as the number density $n_p = \int f d^3c$, momentum density $n_p {u_p}_i = \int c_i f d^3c$, and energy density. Moments of the scattering term are zero. The zeroth moment of (\ref{eq:15}) yields the continuity equation for PUIs,
the next moment the momentum equation for PUIs,
\begin{eqnarray}
\frac{\partial }{\partial t} \left( n_p \left( U_j + {u_p}_j \right) \right) + \nabla \cdot \left[ n_p {\bf U} \left( U_j + {u_p}_j \right) + n_p {\bf u}_p U_j \right] \nonumber \\
\mbox{} + \frac{\partial }{\partial x_i} \int c_i c_j f d^3c = \frac{e}{m_p} n_p \varepsilon_{jkl} {u_p}_k B_l , \label{eq:17}
\end{eqnarray}
where $\varepsilon_{ijk}$ is the Levi--Civita tensor. Note the presence of the term $\int c_i c_j f d^3c$, which is the momentum flux or pressure tensor.

To close equation (\ref{eq:17}), we need to evaluate the momentum flux, which requires that we solve (\ref{eq:15}) for the PUI distribution function $f$. In solving (\ref{eq:15}), we assume 1) that the PUI distribution is gyrotropic, and 2) that scattering of PUIs is sufficiently rapid to ensure that the PUI distribution is nearly isotropic. We can therefore average (\ref{eq:15}) over gyrophase, obtaining the ``focused transport equation'' for non-relativistic PUIs. The second-order correct solution to the gyrophase-averaged form of equation (\ref{eq:15}) is
\begin{eqnarray}
f &\simeq& f_0 + \mu f_1 + \frac{1}{2} (3 \mu^2 - 1) f_2  ; \label{eq:19} \\
f_0 &=& f_0 ({\bf x}, c, t) ; \label{eq:20} \\
f_1 &=& -\frac{c \tau_s}{3} b_i \frac{\partial f_0}{\partial x_i} + \frac{DU_i}{Dt} \frac{\tau_s}{3} b_i \frac{\partial f_0}{\partial c} ; \label{eq:21}  \\
f_2 &\simeq& \frac{c \tau_s}{15} \left( b_i b_j \frac{\partial U_j}{\partial x_i} - \frac{1}{3} \frac{\partial U_i}{\partial x_i} \right) \frac{\partial f_0}{\partial c} ,  \label{eq:21a}
\end{eqnarray}
where $c = |{\bf c}|$ is the particle random speed, ${\bf b} \equiv {\bf B}/B$ is a directional unit vector defined by the magnetic field, and $D/Dt \equiv \partial /\partial t + U_i \partial /\partial x_i$ is the convective derivative. The expansion terms $f_0$, $f_1$, and $f_2$ are functions of position, time, and particle random speed $c$ i.e., independent of $\mu$ (and of course gyrophase $\phi$). Of particular importance is the retention of the large-scale acceleration, and shear terms. These terms are often neglected in the derivation of the transport equation describing $f_0$ (for relativistic particles, the transport equation is the familiar cosmic ray transport equation). In deriving a multi-fluid model, retaining the various flow velocity terms is essential to derive the correct multi-fluid formulation for PUIs.

Following \cite{Zank_2016}, the pressure tensor is found to be the sum of an isotropic scalar pressure $P_p$ and the stress tensor, i.e.,
\begin{eqnarray}
 \left( P_{ij} \right) &=& P_p \left( \delta_{ij} \right) + \left( \begin{array}{ccc} 1 & 0 & 0 \\
 													     0 & 1 & 0 \\	
													     0 & 0 & -2
													     \end{array} \right)
\frac{\eta_{k \ell} }{2} \left( \frac{\partial U_{pk} }{\partial x_{\ell} } + \frac{\partial U_{p\ell} }{\partial x_k } - \frac{2}{3} \delta_{k \ell} \frac{\partial U_{pm} }{\partial x_m } \right) \nonumber \\
&\equiv& P_p {\bf I} + \Pi_p.  \label{eq:2.2.11}
\end{eqnarray}

The stress tensor is a generalization of the ``classical'' form in that several coefficients of viscosity are present, and of course the derivation here is for a collisionless charged gas of PUIs experiencing only pitch-angle scattering by turbulent magnetic fluctuations. Use of the pressure tensor (\ref{eq:2.2.11}) yields a ``Navier--Stokes-like'' modification of the PUI momentum equation,
\begin{eqnarray}
\frac{\partial }{\partial t} \left( \rho_p {\bf U}_p \right) + \nabla \cdot \left[ \rho_p {\bf U}_p {\bf U}_p + {\bf I} P_p \right] = e n_p \left( {\bf E} + {\bf U}_p \times {\bf B} \right) \nonumber \\
\mbox{} - \nabla \cdot \left( \begin{array}{ccc} 1 & 0 & 0 \\
 													     0 & 1 & 0 \\	
													     0 & 0 & -2
													     \end{array} \right)
\frac{\eta_{k \ell} }{2} \left( \frac{\partial {U_p}_k }{\partial x_{\ell} } + \frac{\partial {U_p}_{\ell} }{\partial x_k } - \frac{2}{3} \delta_{k \ell} \frac{\partial {U_p}_m }{\partial x_m } \right) ,  \nonumber \\
=  e n_p \left( {\bf E} + {\bf U}_p \times {\bf B} \right) - \nabla \cdot  \Pi_p    \label{eq:24}
\end{eqnarray}
where we used the transformation ${\bf U}_p = {\bf u}_p + {\bf U}$ for the remaining velocity terms in (\ref{eq:17}) and $\rho_p = m_p n_p$.

To close the PUI energy equation requires the evaluation of the corresponding moments using the expressions (\ref{eq:19})--(\ref{eq:21a}). In so doing, we obtain the total energy equation for the PUIs from (\ref{eq:15}),
 \begin{eqnarray}
\frac{\partial }{\partial t} \left( \frac{1}{2} \rho_p U_p^2  + \frac{3}{2} P_p  \right) + \frac{\partial }{\partial x_i} \left[ \frac{1}{2} \rho_p U_p^2   {U_p}_i  + \frac{5}{2} P_p {U_p}_i + \Pi_{ij} {U_p}_j + q_i \right] \nonumber \\
\mbox{} = e n_p {U_p}_i \left( E_i  + \left( {\bf U}_p \times {\bf B} \right)_i \right) , \label{eq:29}
\end{eqnarray}
after transforming to ${\bf U}_p$. To evaluate the heat flux, we use
\begin{equation}
\frac{m_p}{2} \int {c^{\prime} }^2 c_i^{\prime} \mu f_1 d^3c^{\prime} = -\frac{2 \pi}{3} m_p \int {c^{\prime} }^2 \kappa_{ij} \frac{\partial f_0}{\partial x_j} {c^{\prime} }^2 dc^{\prime}
= -\frac{1}{2} \bar{\kappa}_{ij} \frac{\partial P_p }{\partial x_j} = q_i ({\bf x} ,t) , \label{eq:26}
\end{equation}
after introducing the spatial diffusion coefficient
\begin{equation}
\kappa_{ij} \equiv b_i \frac{c^2 \tau_s}{3} b_j , \label{eq:27}
\end{equation}
together with PUI speed-averaged form $\bar{\kappa}_{ij} \equiv K_{ij}$. The collisionless heat flux for PUIs is therefore described in terms of the PUi pressure gradient and consequently the averaged spatial diffusion introduces a PUI diffusion time and length scale into the multi-fluid system.

For continuous flows, the transport equation for the PUI pressure $P_p$ can be derived from (\ref{eq:29}), yielding
\begin{equation}
\frac{\partial P_p}{\partial t} + {U_p}_i \frac{\partial P_p}{\partial x_i}  + \frac{5}{3} P_p \frac{\partial {U_p}_i }{\partial x_i} = \frac{1}{3} \frac{\partial}{\partial x_i} \left( K _{ij} \frac{\partial  P_p}{\partial x_j}  \right) - \frac{2}{3} \Pi_{ij} \frac{\partial {U_p}_j}{\partial x_i} , \label{eq:28}
\end{equation}
illustrating that the PUI heat flux yields a spatial diffusion term in the PUI equation of state together with a viscous dissipation term. The PUI system of equations is properly closed and correct to the second-order. Note the typo in~\cite{2014ApJ...797...87Z} since they mistakenly omitted the viscous term of equation (\ref{eq:28}) in the corresponding pressure equation.

The full system of PUI equations can be written in the form
\begin{eqnarray}
\frac{\partial \rho_p}{\partial t} + \nabla \cdot \left( \rho_p {\bf U}_p \right) = 0 ;  \label{eq:30} \\
\frac{\partial }{\partial t} \left( \rho_p {\bf U}_p  \right) + \nabla \cdot \left[ \rho_p {\bf U}_p  {\bf U}_p + {\bf  I} P_p +  \Pi \right] = e n_p \left( {\bf E} + {\bf U}_p \times {\bf B} \right) ;  \label{eq:31}  \\
\frac{\partial }{\partial t} \left( \frac{1}{2} \rho_p U_p^2  + \frac{3}{2} P_p   \right) + \nabla \cdot \left[ \frac{1}{2} \rho_p U_p^2 {\bf U}_p + \frac{5}{2} P_p{\bf U}_p + \Pi  \cdot  {\bf U}_p - \frac{1}{2} {\bf K}\cdot \nabla  P_p  \right]   \nonumber  \\
= e n_p {\bf U}_p \cdot {\bf E}  .
 \label{eq:32}
\end{eqnarray}
The full thermal electron-thermal proton-PUI multi-fluid system is therefore given by equations (\ref{eq:4})--(\ref{eq:6}) and (\ref{eq:30})--(\ref{eq:32}) or (\ref{eq:28}), together with Maxwell's equations
\begin{eqnarray}
\frac{\partial {\bf B} }{\partial t} = - \nabla \times {\bf E} ;  \label{eq:34} \\
\nabla \times {\bf B} = \mu_0 {\bf J} ; \label{eq:35} \\
\nabla \cdot {\bf B} = 0 ; \label{eq:36} \\
{\bf J} = e \left( n_s {\bf u}_s + n_p {\bf U}_p - n_e {\bf u}_e \right), \label{eq:37}
\end{eqnarray}
where ${\bf J}$ is the current and $\mu_0$ the permeability of free space.

\subsubsection{Single-fluid model}

For many problems, the complete multi-component model derived above is far too complicated to solve. The multi-fluid system (\ref{eq:4})--(\ref{eq:6}) and (\ref{eq:30})--(\ref{eq:32}) or (\ref{eq:28}), together with Maxwell's equations can be considerably reduced in complexity by making the key assumption that ${\bf U}_p \simeq {\bf u}_s$. The assumption that ${\bf U}_p \simeq {\bf u}_s$ is quite reasonable since i) the bulk flow velocity of the plasma is dominated by the background protons since the PUI component scatters off fluctuations moving with the background plasma speed and ii) the large-scale motional electric field forces newly created PUIs to essentially co-move with the background plasma flow perpendicular to the mean magnetic field. Accordingly, we let ${\bf U}_p \simeq {\bf u}_s = {\bf U}_i$ be the bulk proton (i.e., thermal background protons and PUIs) velocity.

We can combine the proton (thermal plus PUI) equations  with the electron equations (\ref{eq:4})--(\ref{eq:6}) to obtain an MHD-like system of equations. On defining the macroscopic variables,
\begin{eqnarray}
\rho \equiv m_e n_e + m_p n_i ; \quad
q \equiv -e (n_e - n_i ) ; \quad
\rho {\bf U} \equiv m_e n_e {\bf u}_e + m_p n_i {\bf U}_i  ; \nonumber \\
{\bf J} \equiv -e \left( n_e {\bf u}_e - n_i {\bf U}_i \right) ,  \label{eq:2.3.5}
\end{eqnarray}
we can express
\begin{eqnarray}
n_e &=& \frac{\rho - (m_p /e) q }{m_p ( 1 + \xi ) } \simeq \rho /m_p ; \quad
n_i = \frac{\rho + \xi (m_p /e) q }{m_p ( 1 + \xi ) } \simeq \rho /m_p ; \nonumber \\
{\bf u}_e &=& \frac{\rho {\bf U} - (m_p/e) {\bf J} }{\rho - (m_p/e) q } \simeq {\bf U} - \frac{m_p}{e} \frac{\bf J}{\rho} ; \quad
{\bf u}_i = \frac{\rho {\bf U} + \xi (m_p/e) {\bf J} }{\rho + \xi (m_p/e) q }  \simeq {\bf U}  ,  \label{eq:2.3.6}
\end{eqnarray}
where the smallness of the mass ratio $\xi \equiv m_e / m_p \ll 1$ has been exploited. We can also assume that the current density is much less than the momentum flux, i.e., $|{\bf J}| \ll |\rho {\bf U}|$, and combine the thermal proton and electron equations in a single thermal plasma pressure equation with $P \equiv P_e + P_s$. After deriving a suitable Ohm's law \cite{2014ApJ...797...87Z,Zank_2016}, we obtain a reduced single-fluid model equations that may be summarized as
\begin{eqnarray}
\frac{\partial \rho}{\partial t} + \nabla \cdot \left( \rho {\bf U} \right) = 0 ;   \label{eq:2.3.13} \\
\rho \left( \frac{\partial {\bf U} }{\partial t} + {\bf U} \cdot \nabla {\bf U} \right) = -\nabla (P + P_p ) + {\bf J} \times {\bf B} - \nabla \cdot \Pi ; \label{eq:2.3.14} \\
\frac{\partial }{\partial t} \left( \frac{1}{2} \rho U^2  + \frac{3}{2} (P + P_p) + \frac{1}{2\mu_0} B^2  \right) + \nabla \cdot \left[ \frac{1}{2} \rho U^2 {\bf U} + \frac{5}{2} (P + P_p) {\bf U}  \right. \nonumber \\
\left. \mbox{} + \frac{1}{\mu_0} B^2 {\bf U} - \frac{1}{\mu_0} {\bf U} \cdot {\bf B} {\bf B} + \Pi \cdot {\bf U}_p - \frac{1}{2} {\bf K}\cdot \nabla  P_p  \right]  = 0 ;  \label{eq:2.3.16} \\
\frac{\partial P }{\partial t} + {\bf U} \cdot \nabla P + \gamma P \nabla \cdot {\bf U} = 0 ;  \label{eq:2.3.15} \\
{\bf E} = -{\bf U} \times {\bf B}  ; \quad
\frac{\partial {\bf B} }{\partial t} = - \nabla \times {\bf E}  ;  \quad
\mu_0 {\bf J} = \nabla \times {\bf B} ;  \quad
\nabla \cdot {\bf B} = 0 .  \label{eq:2.3.20}
\end{eqnarray}
The single-fluid description (\ref{eq:2.3.13})--(\ref{eq:2.3.20}) differs from the standard MHD model in that a separate description for the PUI pressure is required. Instead of the conservation of energy equation (\ref{eq:2.3.16}), one could use  the PUI pressure equation (\ref{eq:28}) for continuous flows. PUIs introduce both a collisionless heat conduction and viscosity into the system.

The model equations (\ref{eq:2.3.13})--(\ref{eq:2.3.20}), despite being appropriate to non-relativistic PUIs, are identical to the so-called two-fluid MHD system of equations used to describe cosmic ray mediated plasmas \cite{webb_1983_structureobliqueshocks}. However, the derivation of the two models is substantially different in that the cosmic ray number density is explicitly neglected in the two-fluid cosmic ray model and a Chapman--Enskog derivation is not used in deriving the cosmic ray hydrodynamic equations. Nonetheless, the sets of equations that emerge are the same indicating that the cosmic ray two-fluid equations do in fact include the cosmic ray number density explicitly.

The single-fluid-like model may be extended to include e.g., ACRs, as well as PUIs. In this case, ACRs are relativistic particles. The same analysis carries over, and one has an obvious extension of the model equations (\ref{eq:2.3.13})--(\ref{eq:2.3.20}) with the inclusion of the ACR pressure. Thus, the extension of
(\ref{eq:2.3.13})--(\ref{eq:2.3.20}) is \cite{Zank-2015,Zank_2016}
\begin{eqnarray}
\frac{\partial \rho}{\partial t} + \nabla \cdot \left( \rho {\bf U} \right) = 0 ;   \label{eq:2.3.21} \\
\rho \left( \frac{\partial {\bf U} }{\partial t} + {\bf U} \cdot \nabla {\bf U} \right) = -\nabla (P + P_p + P_A) + {\bf J} \times {\bf B} - \nabla \cdot \Pi_p - \nabla \cdot \Pi_A ; \label{eq:2.3.22} \\
\frac{\partial P }{\partial t} + {\bf U} \cdot \nabla P + \gamma P \nabla \cdot {\bf U} = 0 ;  \label{eq:2.3.23} \\
\frac{\partial P_p }{\partial t} + {\bf U} \cdot \nabla P_p + \gamma_p P_p \nabla \cdot {\bf U} = \frac{1}{3} \nabla \cdot \left( {\bf K_p} \cdot \nabla P_p \right) - (\gamma_p - 1) \Pi_p :
(\nabla {\bf U}) ;  \label{eq:2.3.23a} \\
\frac{\partial P_A }{\partial t} + {\bf U} \cdot \nabla P_A + \gamma_A P_A \nabla \cdot {\bf U} =  \frac{1}{3} \nabla \cdot \left( {\bf K_A} \cdot \nabla P_A \right) - (\gamma_A - 1) \Pi_A :
(\nabla {\bf U})  ; \label{eq:2.3.23b} \\
{\bf E} = -{\bf U} \times {\bf B}  ; \quad
\frac{\partial {\bf B} }{\partial t} = - \nabla \times {\bf E}  ;  \quad
\mu_0 {\bf J} = \nabla \times {\bf B} ;  \quad
\nabla \cdot {\bf B} = 0 ,  \label{eq:2.3.24}
\end{eqnarray}
where we have introduced the ACR pressure $P_A$, the corresponding stress tensor $\Pi_A$, the ACR diffusion tensor ${\bf K}_A$ and adiabatic index $\gamma_A$ ($4/3 \leq \gamma_A \leq 5/3$). The coupled system (\ref{eq:2.3.21})--(\ref{eq:2.3.24}) is the simplest continuum model to describe a non-equilibrated plasma comprising a thermal proton-electron plasma with suprathermal particles (e.g., PUIs or even solar energetic particles) and relativistic energy (anomalous) cosmic rays. The system includes both the collisionless heat flux and viscosity associated with the suprathermal and relativistic particle distributions.

On reverting to equations (\ref{eq:2.3.13})--(\ref{eq:2.3.20}), we can recover the standard form of the MHD equations if we set the heat conduction spatial diffusion tensor ${\bf K} = 0$ and the coefficient of viscosity $(\eta_{kl}) = 0$, which corresponds to assuming $\tau_s \rightarrow 0$. If the total thermodynamic pressure $P_{total} = P + P_p$ is introduced,  then we recover the standard MHD equations (dropping the subscript ``total'') i.e.,
\begin{eqnarray}
\frac{\partial \rho}{\partial t} + \nabla \cdot \left( \rho {\bf U} \right) = 0 ;   \label{eq:2.3.25} \\
\rho \frac{\partial {\bf U} }{\partial t} + \rho {\bf U} \cdot \nabla {\bf U} + (\gamma - 1) \nabla e + (\nabla \times {\bf B}) \times {\bf B} = 0 ; \label{eq:2.3.26}  \\
\frac{\partial }{\partial t} \left( \frac{1}{2} \rho U^2 + e + \frac{B^2}{2\mu_0} \right) + \nabla \cdot \left[ \left( \frac{1}{2} \rho U^2 + \gamma e \right) {\bf U} + \frac{1}{\mu_0} {\bf B} \times ({\bf U} \times {\bf B} ) \right] = 0 ;  \label{eq:2.3.27}  \\
\frac{\partial {\bf B} }{\partial t} = \nabla \times ({\bf U} \times {\bf B} ); \quad \nabla \cdot {\bf B} = 0 ,  \label{eq:2.3.28}
\end{eqnarray}
with an equation of state $e = \alpha n k_B T /(\gamma - 1)$. The choice of $\alpha = 2$ (or greater if incorporating the contribution of cosmic rays etc.) corresponds to a plasma population comprising protons and electrons.

In setting ${\bf K} = 0$ and $(\eta_{kl}) = 0$, we have implicitly assumed that PUIs are completely coupled to the thermal plasma. With ${\bf K} \neq 0$, heat conduction reduces the effective coupling of energetic particles to the thermal plasma, and their contribution to the total pressure is not as large. This will have important consequences for numerical models of e.g., the large-scale heliosphere since they incorporate PUIs into the MHD equations, without distinguishing PUIs from thermal plasma and therefore neglect heat conduction.  Consequently the total pressure is over-estimated.

\section{Energetic Particles}
\label{particles}
In the following, various aspects of the transport of energetic particles in the inner and outer heliosheath are discussed, with an emphasis on ACRs and GCRs. The outer heliosheath is the region of the LISM perturbed by the presence of the heliosphere.
The corresponding subsection headings are formulated as the currently crucial questions that need to be
answered to make further significant progress in the field.
\subsection{What is the propagation tensor in the heliosheath?}
First, it should be emphasized that determining this tensor throughout the heliosphere,
not just in the heliosheath, is still a work in progress. This is despite the progress
that has been made since the millennium change in 2000, see, e.g., the comprehensive
overview by \cite{Shalchi-2009}. As in most research fields in physics, there are two
ways of how progress is made: An empirical, phenomenological approach driven mostly by
observations, and then the fundamental theoretical work, also known in solar modulation
as the {\it ab initio} approach. For the latter, the focus is on developing a sound theoretical
basis for turbulence, diffusion and particle drift theories. In the end, observations have to be
reproduced by using these two approaches in numerical models based on solving the heliospheric
transport equation (TPE) for cosmic rays (CRs) as proposed by \cite{Parker-1965}. This equation
basically describes four major processes, outward convection, inward diffusion both parallel and
perpendicular to the magnetic field lines, particle drifts (consisting of gradient, curvature and
current sheet drifts), and adiabatic energy changes. Utilizing only these four processes has done
amazingly well in explaining and understanding what causes the global modulation of CRs,
from $\sim 1$~MeV up to 50~GeV, over 11-year and 22-year cycles, see the review by \cite{Potgieter-2013}.
However, when shorter scale changes in the lower energy ranges are studied, for example, the acceleration
effect of travelling shocks in the heliosphere, focusing and momentum diffusion also come into play
(for a theoretical overview of these processes see \cite{Schlickeiser-2002}). Diffusive shock
acceleration of CRs, for instance at the TS, is also contained in
the TPE, although more subtle to utilize in numerical models than the other mentioned processes.
These reasonably well-known aspects, together with lesser known aspects such as the projected
effects of magnetic reconnection on CR modulation, have also been invoked to explain what is happening
inside the inner heliosheath. This has been driven by observations of the ACRs
which deviate significantly from what established models had predicted. Evidently, the time has
come to study in more detail also the diffusion, adiabatic energy changes, drift and other processes inside
the heliosheath. This is a major theoretical and modelling challenge. Observationally, it is well established
that CRs inside the inner heliosheath  are modulated even to the point of being extraordinary, for instance,
the spectacular increase in low-energy galactic electrons \cite{Webber-McDonald-2013} from the TS to the
HP. Solving Parker's TPE for studies of solar modulation requires some crucial knowledge of the following:
\begin{enumerate}
\item {The {\it heliospheric structure} and geometric extent such as where is the HP located in all directions,
and, is the thickness of the inner heliosheath symmetrically orientated with respect to the Sun and does this
change with solar activity because the TS changes position, e.g., \cite{Manuel-etal-2015}.
MHD models, as described above, have contributed immensely so that we have a reasonable understanding
of the heliospheric extent. \emph{V1} observations \cite{Stone-etal-2013} have, of course, put a real value on
the modulation ``desk'' of where the HP is located.}

\item The {\it unmodulated input spectra}, better known as HP spectra or local interstellar spectra (LIS).
In this context, we had to rely on numerical models of galactic propagation to give an indication of what to use
below 20~GeV, where modulation makes a progressively important difference (e.g., \cite{Strauss-Potgieter-2014}).
Mostly, modelers simply guessed the spectral shapes at energies below 1~GeV, until Voyager~1 gave a real clue of
what it is between 5--50~MeV since it had moved away from the HP. There still is some controversy whether solar
modulation would stop abruptly at the HP, as has been widely assumed, or may continue beyond the HP into the
outer heliosheath, see \cite{Scherer-etal-2011}, \cite{Strauss-etal-2013}, \cite{Guo-Florinski-2014}, and \cite{Luo-etal-2015}.
This could affect the observed value of the LIS, especially at the lowest energy range.

\item The {\it solar wind velocity} and its time and spatial profile. Our corresponding knowledge is comparatively detailed as a result
of many \emph{in situ} observations, e.g. from \emph{Ulysses} and \emph{V2}, as well as of comprehensive MHD
modeling. The next step is to fully understand how the dominant radial velocity component upstream of the TS
is transformed into three components downstream towards the HP.  Subsequently, of additional importance is the
divergence of this velocity profile, because this determines the energy changes in the heliosphere and
heliosheath. Towards Earth, energy losses dominate to the extent that all modulated CR spectra, except
for electrons and positrons, have a characteristic $E^{+1}$ spectral shape below $\sim$500~MeV
(e.g., \cite{Strauss-etal-2011}). Inside the inner heliosheath, this is expected to be completely different and needs to
be determined (for different scenarios see \cite{Langner-etal-2006}).

\item The {\it HMF geometry}. In this context, the widely used Parker HMF,
with its perfect spirals and cones in the polar regions of the heliosphere is idealistic, owing to the fact
that it has only a radial and an azimuthal component. More complicated HMF models also contain a latitudinal
component (\cite{Fisk-1996}), which makes them very difficult to handle in most finite-difference based
numerical models. Only recently the stochastic differential equation (SDE) approach to numerical
modeling of solar modulation has presented a way around these difficulties. Unfortunately, observational
evidence for Fisk-typed fields, and their consequences for CR modulation, is not conclusive, possibly
because measurements are not made where it is necessary for verifying this, see \cite{Sternal-etal-2011}.
For CR drifts, the geometry of the HMF is very important because what is used in models prescribes how much
gradients and curvatures the CRs experience. Additionally, the wavy HCs is a
major feature, which plays an important role all over the solar cycle, starting from solar minimum conditions, when the tilt angle is small,
to solar maximum, when the tilt angle becomes very large, and contributes to the theoretically predicted
charge-sign dependence in CR modulation which now is an observational fact (e.g., \cite{Potgieter-2013}).
Inside the inner heliosheath, the HMF is surely more complicated than upstream of the TS, and as such a hard
problem to handle in numerical models.

\item The {\it propagation tensor in the TPE}. This tensor is the sum of a symmetrical diffusion tensor and
an asymmetrical drift tensor, containing the drift coefficient. In terms of HMF aligned elements, the
diffusion tensor contains one parallel and two perpendicular coefficients (in the radial and in the latitudinal
directions). If the TPE is solved in heliocentric spherical coordinate system, the geometry of the HMF comes
into play so that the nine elements of the tensor are then given as
\begin{eqnarray}
\kappa_{rr}          &=&  \kappa_{\|} \cos^2\psi + \kappa_{\perp r} \sin^2\psi,\nonumber\\
\kappa_{\perp\theta} &=&  \kappa_{\theta\theta},\nonumber\\
\kappa_{\phi\phi}    &=&  \kappa_{\perp r} \cos^2\psi + \kappa_{\|} \sin^2\psi,\\
\kappa_{\phi r}      &=&  \kappa_{r \phi} =(\kappa_{\perp r} - \kappa_{\|}) \cos\psi \sin\psi,\nonumber\\
\kappa_{\theta r}    &=&  \kappa_d \sin\psi\nonumber = -\kappa_{r\theta},\\
\kappa_{\theta\phi}  &=&  \kappa_d \cos\psi\nonumber = -\kappa_{\phi\theta},
\end{eqnarray}
where $\kappa_{rr}$ is the effective radial diffusion coefficient, thus a combination of the parallel diffusion
coefficient and the radial perpendicular diffusion coefficient $\kappa_{\perp r}$, with $\psi$ the spiral angle
of the average HMF; $\kappa_{\theta\theta} = \kappa_{\perp\theta}$ is the effective diffusion coefficient perpendicular
to the HMF in the polar direction; $\kappa_{\phi\phi}$ describes the effective diffusion in the azimuthal direction,
and so on. The four drift coefficients are given in the last two rows. Inspection shows that the five diffusion
coefficients are determined by what is assumed for parallel and perpendicular diffusion, and all of them depend
on the geometry of the assumed HMF. For instance, beyond $\sim$20~AU in the equatorial plane $\psi\rightarrow 90^0$,
so that $\kappa_{rr}$ is dominated by $\kappa_{\perp r}$ but by $\kappa_{\|}$ in the polar regions of the heliosphere,
whereas $\kappa_{\phi\phi}$ is dominated by $\kappa_{\|}$. This is true only if the HMF is Parkerian in its geometry.
These nine tensor elements become significantly more complicated if the HMF geometry is containing a latitudinal component,
see \cite{Effenberger-etal-2012}.

\item {The {\it heliospheric turbulence} is determining the elements of the diffusion tensor. Modelling the
      evolution of the turbulence forms the basis of the {\it ab initio} approach to solar modulation, see, e.g.,
      \cite{Engelbrecht-Burger-2013} and \cite{Wiengarten-etal-2016}. For the fundamental, theoretical principles
      involved see the comprehensive description in \cite{Shalchi-2009}. It suffices to say that it is quite
      complicated, perhaps mostly because there are still far too many unknowns in the various still developing
      theories so that the impression is given that the more complicated the theory gets, the more the confusion
      becomes of what exactly to use in modulation models.}
\end{enumerate}
The lack of global observations to support or to oppose new developments in
the fundamental theory is of course a fact of the matter. On the other hand, the empirical approach in determining
the diffusion coefficients, not paying attention to the fundamental reasons of what exactly in terms of turbulence
determines the rigidity and spatial dependence of the diffusion coefficients, has been quite robust. A main obstacle
has been that there is a limit to what standard numerical approaches allow modellers to do. This is slowly but
surely overcome by new approaches such as using SDEs (e.g., \cite{Strauss-etal-2011}) and is greatly supported
by the availability of powerful computer clusters.

This brings up the question of what is the mentioned tensor throughout the heliosheath? The short answer is
that we are still very unsure, because of the vastly more complex
(i) heliospheric structure with predicted large asymmetries with respect to the Sun, and in the nose-tail and north-south
directions,
(ii) solar wind profile,
(iii) corresponding HMF profile and wavy HCS, and (iv) turbulence which is clearly far more intricate than closer
to the Sun.
The turbulence should be expected to be different in the distant tail
of the heliosphere, in the nose direction, and at higher latitudes.

The drift scale $\kappa_{d}$ is commonly assumed to vanish in the heliosheath, most likely because it is the most
convenient option in numerical modeling (e.g., \cite{Florinski-Pogorelov-2009}). In contrast, it is assumed
that drifts still occur inside the heliosheath, similar to the inner heliosphere, with $\kappa_d$ scaling proportional
to radial distance, which was found as unlikely by \cite{Ngobeni-Potgieter-2014}. From these extreme differences
it is clear than much work is needed to sort out how particle drifts would change from the TS to the HP. Similar
to closer to the Sun, it is a matter of what the scattering parameter $\omega\tau$ globally is, with $\omega$ the
gyro-frequency of a CR particle and $\tau$ a time scale defined by its scattering, of whatever nature. When
$10 \leq \omega\tau \leq \infty$ particle drift assumes its maximal weak scattering value; whereas with
$\omega\tau\rightarrow0$ no particle drifts are present, and for $\omega\tau\rightarrow 1$ particle drifts are
reduced by half compared to the weak scattering value. This 50\% reduction was reported by several modeling
studies where reproducing and explaining the observations was of essence (e.g., \cite{Langner-Potgieter-2004},
\cite{Ngobeni-Potgieter-2014}). The latest publication that reported on observational evidence of drift effects in the
outer heliosphere was by \cite{Webber-etal-2005}. In the context of drifts, of major importance is the fate of the wavy HCS as it becomes compressed beyond the TS towards the surface of the HP (see, e.g., \cite{Florinski-2011}).



Concerning the above diffusion coefficients, the foremost conclusion about how they should behave globally at and
beyond the TS, is that they decrease considerably across the TS and stay at these low levels inside the inner heliosheath.
This is easily accomplished by assuming that these coefficients scale proportional to $1/B$, with $B$ the magnitude
of the HMF across the TS. In this context, the simulations by \cite{Nkosi-etal-2011} emphasized exactly this
behaviour for low-energy galactic electrons which had increased by almost a factor of 400 from the TS to the HP
at 10~MeV (\cite{Webber-McDonald-2013}). The inner heliosheath acts as an ever present modulation ``barrier,''
reducing CR fluxes significantly, depending on their rigidity, of course (\cite{Potgieter-Nndanganeni-2013}).
The observed occurrence inside the inner heliosheath of TS particles (TSP) and accelerated ACRs make the
estimations of the diffusion coefficients in this region far more difficult than closer to the Sun. From a
modelling point of view, what happens in the heliosheath is usually side-stepped, conveniently ignored or
treated explicitly as if similar to the inner heliosphere, clearly because of a lack of a proper theory
(e.g., \cite{Luo-etal-2013}). It has also become clear that close to the HP, even more complicated processes
could occur, adding to the difficulty of establishing the spatial dependence of the elements of the diffusion
tensor (e.g., \cite{Florinski-etal-2012}). Beyond the HP, what is assumed for the diffusion coefficients
depends on whether one accepts that Voyager~1 is already in the interstellar medium or not, or perhaps it is
in what can be called the very local interstellar medium or perhaps it simply is in the outer heliosheath which, in principle,
could be different from the pristine interstellar medium. For estimates, nothing more, of the value of these
diffusion coefficients see, e.g., \cite{Strauss-etal-2013} and \cite{Luo-etal-2015}.

In conclusion, knowledge about all the diffusion coefficients and the drift scale inside the inner heliosheath
is still in a rudimentary phase, but progress is made, inspired by Voyager~1 \&~2 observations. It is
already clear that establishing the rigidity and spatial dependence of the diffusion coefficients applicable to
the inner heliosheath is much more complicated than for the inner parts of the heliosphere, and that finding one set
of such parameters throughout the entire heliosheath may be wishful thinking. The significant differences in CR
observations between \emph{V1} and \emph{V2} (\cite{Webber-Intriligator-2015}) emphasize the latter statement.
\subsection{How can the ACR and GCR anisotropies be explained?}
Interestingly, the CR measurements in the heliosheath not only allow for a study of the spatial and rigidity
dependence of diffusion but also of its dependence on pitch-angle. This opportunity arises because,
after the crossing of the HP, Voyager~1 observed the ACR and GCR pitch-angle distributions to be anisotropic,
see Fig.~2 in \cite{Krimigis-etal-2013}. This anisotropy is different for both CR species: while the ACR distribution exhibits an
enhancements near 90$^\circ$, the GCR distribution shows the opposite, namely a depletion around that pitch-angle.

Given the anisotropic nature of the pitch-angle distributions, the often employed diffusion approximation and, thus,
the Parker transport equation cannot be used as a modelling basis. One must rather formulate the latter on a pitch-angle
resolving level, i.e.\ employ a variant of the so-called Skilling equation \cite{Skilling-1971}.
\cite{Strauss-Fichtner-2014} opted for a simplified
description by considering a two-dimensional, Cartesian box locally aligned with and enclosing a small section of the HP
surface. By additionally neglecting all processes other than spatial diffusion the Skilling equation reduces to:
\begin{equation}
\frac{\partial f}{\partial t} = -v\mu \frac{\partial f}{\partial y}
           + \frac{\partial }{\partial \mu} \left( D_{\mu \mu} \frac{\partial f}{\partial \mu} \right)
           + \frac{\partial}{\partial x} \left( \kappa_{\perp} \frac{\partial f}{\partial x} \right)
\label{eq-skilling}
\end{equation}
with $f$ denoting the pitch-angle dependent distribution function, $v$ the particle speed, $\mu$ the cosine of its
pitch-angle, and $x$ and $y$ the two spatial coordinates normal and tangential to the HP, respectively.

The central ingredients in this CR transport equation are the pitch-angle diffusion coefficient $D_{\mu\mu}$ and the
spatial diffusion coefficient $\kappa_{\perp}$. While the diffusion along the magnetic field is critically determined by
the $\mu$-dependence of the former, which can be computed from standard quasilinear theory (e.g., \cite{Schlickeiser-2002}),
the diffusion across the magnetic field and, thus, across the HP, is depending on the $\mu$-dependence of the latter, which
must be derived for the specific `magnetic' environment close to the HP. The corresponding derivation is subject of the
following section and results in a form that is principally similar to that suggested in an ad-hoc manner by
\cite{Droege-etal-2010}, namely $\kappa_\perp \sim (1-\mu^2)^{1/2}$.

\begin{figure}
\includegraphics[width=\textwidth]{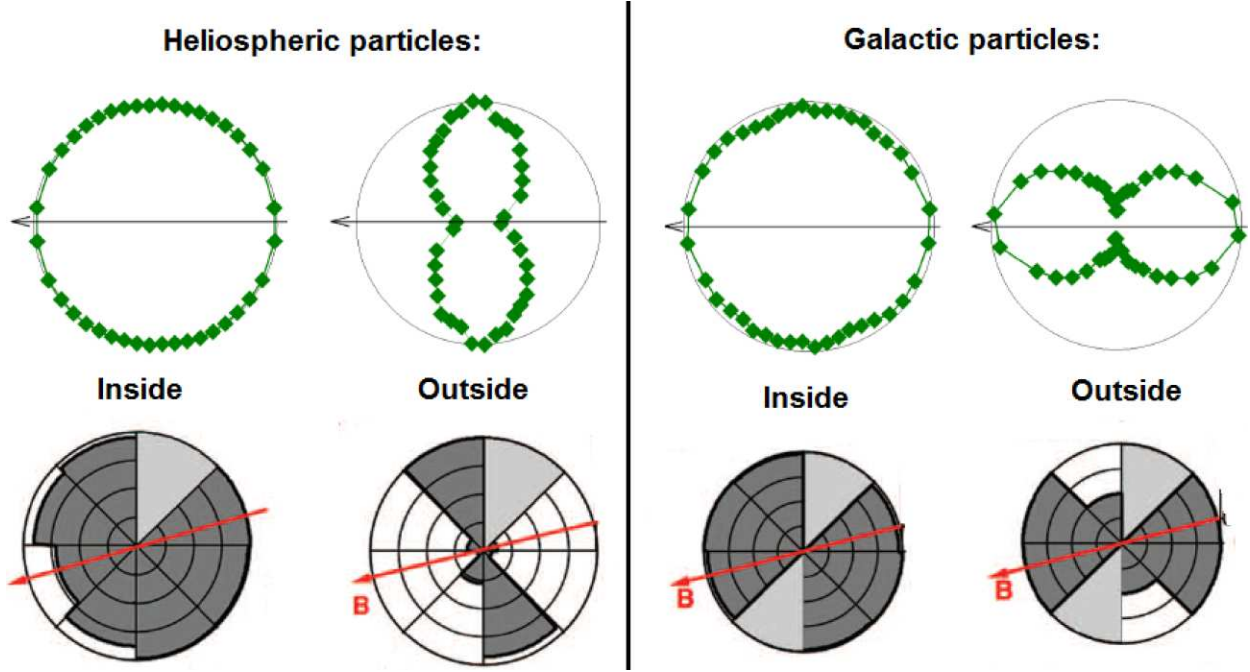}
\caption{
Comparison of the computed and observed ACR (the two upper and lower left panels) and GCR pitch-angle anisotropy
(the two upper and lower right panels) in the outer heliosheath close to the HP.
Taken from \cite{Strauss-etal-2015}.
}
\label{anis-comp}
\end{figure}

As is demonstrated in \cite{Strauss-Fichtner-2014} and \cite{Strauss-etal-2015}, using these diffusion coefficients in
the above transport equation suffices, at least qualitatively, to {\it simultaneously} explain the above-mentioned
anisotropies of the ACR and the GCR pitch-angle distributions, see the comparison of the computed with the observed
anisotropies in Fig.~\ref{anis-comp}.
\subsection{What is the pitch-angle dependence of perpendicular diffusion?}
As discussed in the previous section, the different pitch-angle anisotropies in ACRs and GCRs
beyond the HP can be explained if their perpendicular diffusion across the HP can be described with a
coefficient $\kappa_{\perp}$ that varies as $(1-\mu^{2})^{1/2}$ where $\mu$ is the cosine of a
particle's pitch angle. Then the question arises whether there is any theoretical support for such a dependence.
Typically, discussions in the literature involve pitch-angle averaged expressions for $\kappa_{\perp}$ of
energetic particles so that currently little is known about its pitch-angle dependence. Note
$\kappa_{\perp} \propto  (1-\mu^{2})^{0.5}$ was used by \cite{Droege-etal-2010} in a focused transport model to
study anisotropic solar energy particle transport, but the model for $\kappa_{\perp}$ was not derived from
first principles.

We discuss the possible pitch-angle dependence of $\kappa_{\perp}$ assuming that $\kappa_{\perp}$ originates
either from (i) cross-field scattering due to particles interacting gyroresonantly with Alfv\'{e}n waves
($r_{g} \approx l_{c}$, where $r_{g}$ is the particle gyroradius ignoring its $\mu$-dependence and $l_{c}$ is
the correlation length of the magnetic field waves and turbulence), or (ii) from particle guiding center (GC)
motion along and across relatively large-scale random walking magnetic field lines ($r_{g} \ll l_{c}$). For
the latter scenario, we investigate two further possibilities: Either purely transversal or purely longitudinal
fluctuations.

We consider first the $\mu$-dependence of perpendicular diffusion due to gyroresonant scattering across magnetic
field lines. A simple way to estimate the $\mu$-dependence is to specify a $\mu$-dependent version of the
expression for $\kappa_{\perp}$ found by modeling particle scattering using the BGK scattering term. Accordingly,
\begin{equation}
\kappa_{\perp}(\mu) =
\frac{\kappa_{\|}(\mu)}{1+\omega^{2}\tau^{2}_{sc}(\mu)},
\end{equation}
where $\kappa_{\|}(\mu)$ is the $\mu$-dependent parallel diffusion coefficient, $\omega$ is related to the
particle gyrofrequency and $\tau_{sc}(\mu)$ is the $\mu$-dependent particle scattering time. Observations show
that near the HP magnetic field turbulence is weak, so that $\omega\tau_{sc} \gg 1$. Thus,
\begin{equation}
\kappa_{\perp}(\mu) =
\frac{\kappa_{\|}(\mu)}{\omega^{2}\tau^{2}_{sc}(\mu)}.
\end{equation}
According to  standard quasi-linear theory for the gyroresonant interaction of energetic particles with the
inertial range of parallel propagating Alfv\'{e}n waves,
\begin{equation}
\displaystyle
\kappa_{\|}(\mu) \propto v^{2}(1-\mu^{2})\tau_{sc}(\mu),	
\end{equation}
where
\begin{equation}
\tau_{sc}(\mu)\propto  \frac{1}{\omega}\frac{B_{0}^{2}}{\langle\delta B_{\perp}^{2}\rangle_{A}^{res}(\mu)}.
\end{equation}
In this expression $\langle\delta B_{\perp}^{2}\rangle_{A}^{res}(\mu)$ is the wave magnetic field energy density
associated with the resonant wave number that can be expressed as
\begin{equation}
\langle \delta B^{2}_{\perp}\rangle_{A}^{res}(\mu) = \langle \delta B^{2}_{\perp}\rangle_{A}
                                                     \left(\frac{r_{g}|\mu-jV_{A}/v|}{l_{c\|}}\right)^{s-1},
\end{equation}
where $l_{c\|}$ is the correlation length for parallel-propagating Alfv\'{e}n wave turbulence, $V_A$ is the
Alfv\'en speed,  $j=+1(-1)$ for
forward (backward) propagating Alfv\'{e}n waves along ${\bf B}_{0}$, and $-s$ is the power-law index of the
Alfv\'{e}n wave turbulence spectral energy density in the inertial range. Upon inserting the expression for
$\tau_{sc}$ in $\kappa_{\perp}(\mu)$, we find that
\begin{equation}
\kappa_{\perp}(\mu) \propto (1-\mu^{2})vr_{g}\frac{\langle \delta B^{2}_{\perp}\rangle_{A}^{res}(\mu)}{B_{0}^{2}} \propto (1-\mu^{2})|\mu-jV_{A}/v|^{s-1}vr_{g}\frac{\langle \delta B^{2}_{\perp}\rangle_{A}}{B_{0}^{2}}\left(\frac{r_{g}}{l_{c\|}}\right)^{s-1}.
\end{equation}
Therefore, $\kappa_{\perp}(\mu) \propto (1-\mu^{2})|\mu|^{s-1}$ if $\mu \gg V_{A}/v$ and the required
$(1-\mu^{2})$-dependence is not achieved. To have $\kappa_{\perp}(\mu) \propto (1-\mu^{2})$ would either require
that $s=1$, or that $\tau_{sc}$  is independent of $\mu$. The value of $s=1$ is typically the power-law index associated with
the energy containing range of the wave turbulence spectral energy density, thus implying that CRs are
interacting resonantly with the energy-containing range of Alfv\'{e}n wave turbulence, whereas for $\tau_{sc}$ to
be approximately independent of $\mu$ would require sufficiently strong resonant broadening effects \cite{Lee-2005}.
It is not clear whether CRs have sufficiently large gyroradii to resonate with the energy-containing range,
but it cannot be ruled out.

Let us now estimate the $\mu$-dependence of $\kappa_{\perp}$ due to random-walking magnetic field lines. We define
the mean-square displacement of energetic particles across the mean magnetic field ${\bf B}_{0}$ due to this
interaction as
\begin{equation}
\langle \Delta x_{\perp}^{2} \rangle =  \int_{0}^{\Delta t} dt'
\int_{0}^{\Delta t} dt''   \langle  v_{\perp}(t')v_{\perp}(t'')\rangle,
\end{equation}
where  $\langle v_{\perp}(t')v_{\perp}(t'')\rangle$ is the perpendicular component of the two-time velocity
correlation function for energetic particles interacting with a turbulent magnetic field. It is also assumed that
the particles interact with stationary and homogeneous magnetic field turbulence so that the velocity correlation
function is stationary. It then follows that at late times we have a Kubo formula given by
\begin{equation}
\kappa_{\perp}(\mu) = \lim_{\Delta t \rightarrow \infty} \frac{\langle \Delta x_{\perp}^{2} \rangle}{2(\Delta t)} = \int_{0}^{\infty}dt'\langle v_{\perp}(0) v_{\perp}(t')\rangle.
\end{equation}
Upon assuming that the two-time perpendicular velocity correlation function decays exponentially when CRs
interact with magnetic field turbulence so that
\begin{equation}
\langle v_{\perp}(0)v_{\perp}(t)\rangle = \langle v_{\perp}^{2}(0)\rangle e^{-\frac{t}{\tau_{dec}}},
\end{equation}
where $\tau_{dec}$ represents the characteristic time scale of decay, the perpendicular coefficient can be expressed as
\begin{equation}
\kappa_{\perp}(\mu)_ = \langle v_{\perp}^{2}\rangle\tau_{dec}.
\end{equation}
If one interprets $v_{\perp}$ as the component of the GC velocity across the mean magnetic field, we can
make use of standard GC theory according to which
\begin{equation}
{\bf V}_{g} = v_{\|}{\bf b} + {\bf V}_{E} + \frac{p_{\perp}v_{\perp}}{2qB}\frac{{\bf B} \times \nabla B}{B^{2}}
            + \frac{p_{\|}v_{\|}}{qB}{\bf b}\times \nabla_{\|}{\bf b},
\end{equation}
where the first term is GC motion along the local magnetic field, the second term ${\bf V}_{E}$ is electric
field drift, the third term represents grad-B drift, and the last term is curvature drift, all across the local magnetic
field. The expression has been simplified by dropping the parallel drift term and by applying the fast particle limit
$v \gg V_{E}$.

As usual, it is assumed that the  total magnetic field can be decomposed into a mean field component ${\bf B}_{0}$
and a perpendicular random walking component ${\bf \delta B}$ so that ${\bf B} = {\bf B}_{0} + \delta {\bf B}$ and
that the fluctuations are weak, $\delta B/B_{0} \ll 1$.

We now consider two limits in this model. Firstly, we apply the standard assumption of dominating transversal
fluctuations, $\delta {\bf B_{\perp}} \gg \delta {\bf B}_{||}$, so that $\delta {\bf B} \approx \delta {\bf B}_{\perp}$.
The resulting GC velocity component projected in the direction perpendicular to ${\bf B}_{0}$, the mean
magnetic field in the plasma flow frame, is approximately
\begin{equation}
v_{\perp} \approx v\mu\frac{\delta B_{\perp}}{B_{0}} + V_{A}\frac{\delta B_{\perp}}{B}
                   + \frac{1}{2}v(1-\mu^{2})\frac{r_{g}}{l_{c\perp}}\frac{\delta B_{\perp}^{2}}{B_{0}^{2} }
                   + v\mu^{2}\frac{r_{g}}{l_{c\perp}}\frac{\delta B_{\perp}^{2}}{B_{0}^{2} },
\end{equation}
using dimensional analysis to approximate spatial derivatives. In this expression $l_{c\perp}$ is the perpendicular
turbulence correlation length.

After inserting the expression for $v_{\perp}$ into $\kappa_{\perp}(\mu)$ we find that
\begin{equation}
\kappa_{\perp}(\mu) \approx \left\langle v|\mu|\frac{\delta B_{\perp}}{B_{0}} + V_{A}\frac{\delta B_{\perp}}{B}
       + \frac{1}{2}v(1-\mu^{2})\frac{r_{g}}{l_{c\perp}}\frac{\delta B_{\perp}^{2}}{B_{0}^{2} }
       + v\mu^{2}\frac{r_{g}}{l_{c\perp}}\frac{\delta B_{\perp}^{2}}{B_{0}^{2} }\right\rangle^{2}\tau_{dec}.
\end{equation}
Keeping only first-order terms in $\delta B/B_0$, assuming $v \gg V_A$ and specifying the particle decorrelation time as
\begin{equation}
\tau_{dec} = \frac{l_{c\perp}}{v|\mu|\langle\delta B_{\perp}^{2}\rangle^{0.5}/B_{0}},
\end{equation}
we end up with
\begin{equation}
\kappa_{\perp}(\mu) \approx v|\mu|l_{c\perp}\frac{\langle\delta B_{\perp}^{2}\rangle^{0.5}}{B_{0}}.
\end{equation}
Thus, we recover the classical expression for field line random-walk (FLRW) perpendicular diffusion \cite{Jokipii-1966},
but with the $\mu$-dependence explicitly shown to be $\kappa_{\perp} \propto |\mu|$. The dependence required by
\cite{Strauss-Fichtner-2014} is therefore not obtained in this limit.

In the second limit, we assume dominating longitudinal fluctuations, $\delta {\bf B_{||}} \gg \delta {\bf B}_{\perp}$,
so that $\delta {\bf B} \approx \delta {\bf B}_{||}$. This is motivated by Voyager~1 observations of mainly compressive
fluctuations near the HP \cite{Burlaga-Ness-2014}, see also \cite{Florinski-etal-2013}. In this limit, the
GC drift velocity becomes
\begin{equation}
v_{\perp} \approx  \frac{vr_L}{2l_{c\perp}} \left( 1 - \mu^2 \right) \frac{\delta B_{||}^2}{B_0^2}.
\end{equation}
Assuming the decorrelation time is given by the time it takes a particle to drift across a perpendicular correlation scale
\begin{equation}
\tau_{\mathrm{dec}} = \frac{l_{c\perp}}{ v_{\perp} },
\end{equation}
we obtain for $\kappa_{\perp}(\mu)$
\begin{equation}
\kappa_{\perp} (\mu)= \frac{vr_L}{2}  \left( 1 - \mu^2 \right) \left\langle  \frac{\delta B_z^2}{B_0^2}    \right\rangle,
\end{equation}
{which turns out to validate the assumptions made by \cite{Strauss-Fichtner-2014}, or any equivalent form of it that
has maximum at $\mu = 0$, as, e.g., used in \cite{Droege-etal-2010}. This functional form, which has been
derived in \cite{Strauss-etal-2016}, not only explains the observed ACR and GCR anisotropies in a unified
treatment, as discussed in the previous section, but also predicts, via the ocurrence of the Larmor radius,
a linear dependence on rigidity.}
\begin{figure}
\includegraphics[width=0.95\textwidth]{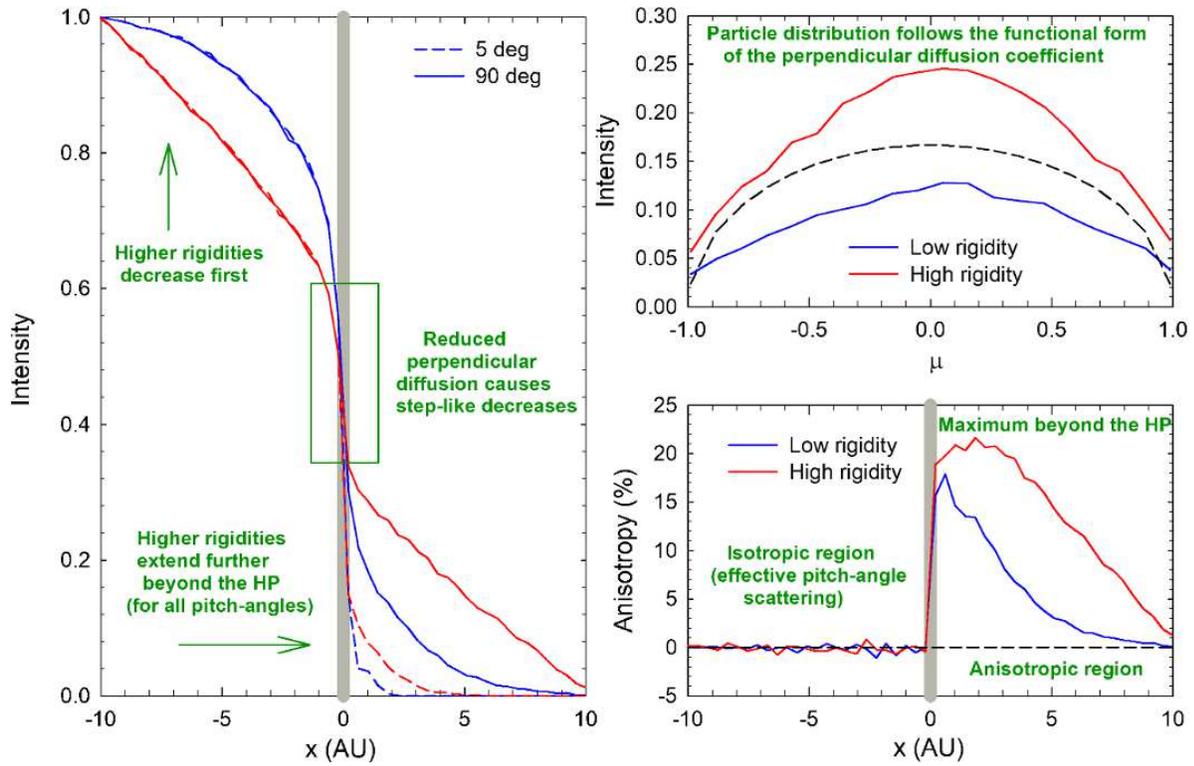}
\caption{
The model result for the ACR pitch-angle anisotropy for two different rigidities. The red curves are for the higher
rigidity.
Taken from \cite{Strauss-etal-2015}.
}
\label{anis-comp-rig}
\end{figure}
Solving the transport equation (\ref{eq-skilling}) for ACRs at two different rigidities results in the solutions
shown in Fig.~\ref{anis-comp-rig}. Evidently, on the inner heliosheath side of the HP the intensity of the ACRs of higher
rigidity decreases first, while it remains the higher one on the outer heliosheath side. These findings are in qualitative
agreement with V1 measurements as discussed in \cite{Florinski-etal-2015}.
%
%
%
\section{Facing the turbulent nature of of the media}
\label{sec:intro}

\subsection{Importance of turbulence}
It is well known that the interstellar medium (ISM) is magnetized and turbulent
\cite{armstrong81,armstrong95,verdini07,lazarian09,Chepurnov2010}.  A Kolmogorov-type
power law is measured with \emph{in situ} measurements in SW \cite{coleman68,Leamon1998}\footnote{More discussion of the SW turbulence can be found in
\cite{bruno05}}. Turbulent state of plasmas is expected in astrophysics. Indeed, magnetized astrophysical
plasmas generally have very
large Reynolds numbers due to the large length scales involved, as well as the
fact that the motions of charged particles in the direction perpendicular to
magnetic fields are constrained. Plasma flows at these high Reynolds
numbers $R=VL_f/\nu$, where $V$ and $L_f$ are the velocity and the scale of the
flow, $\nu$ is fluid viscosity,  are prey to numerous linear and
finite-amplitude instabilities, from which turbulent motions readily develop.

The LISM is expected to reflect the properties of the cascade of
turbulence in the larger volumes of the ISM.
For interstellar medium the drivers of turbulence include supernova explosions
that shape the interstellar medium \cite{mckee77,nakamura06}, accretion flows
\cite{klessen10}, magneto-rotational instability in the galactic disk
\cite{sellwood99}, thermal instability \cite{kritsuk02,koyama02},
collimated outflows \cite{nakamura07}, etc.  Similarly, the fast plasma flow and
plasma instabilities provide the natural environment for turbulence to develop
in the SW. In addition, turbulence is also expected to be produced by the heliosphere interaction with
the LISM.

Turbulence is known to affect most of properties of fluids, e.g., propagation of waves,
energetic particle behavior, magnetic field
generation, etc. \cite{moffatt78,dmitruk01,schlickeiser03,vishniac03,cranmer05,longair10}.
Similarly, \cite{lazarian99} shows that the constrains on the classical Sweet--Parker
reconnection are being lifted in the presence of turbulence and the reconnection
rate becomes fast, i.e., independent on resistivity. Plasma thermal conductivity is also being
radically changed \cite{Narayan2003,Lazarian2006}. Therefore, models that do not account for turbulent properties
may result in a significantly distorted picture of reality.

We note that the presence of a magnetic field makes MHD turbulence anisotropic \cite{oughton03,montgomery81,matthaeus83,shebalin83,higdon84,goldreich95}.  The relative
importance of hydrodynamic and magnetic forces changes with scale, so the
anisotropy of MHD turbulence does too. This scale-dependent change of anisotropy is
important for many astrophysical processes, e.g.  scattering and acceleration of energetic particles, and thermal
conduction. For a number of processes, e.g. magnetic reconnection (see \cite{lazarian99}), the Alfv\'enic component of the cascade is the most important,
for others, e.g. scattering, fast modes may be dominant  \cite{yan02}. The justification of the separate discussion of slow, fast and Alfv\'en modes follows, e.g., from
numerical studies \cite{cho02,cho03} that showed that the Alfv\'enic
turbulence develops an independent cascade which is marginally affected by the
fluid compressibility \cite{lithwick01}.

Below we discuss how turbulence affects the major processes under consideration, i.e., modeling energetic particles,
cosmic rays, and magnetic fields.

\subsection{Magnetic reconnection in turbulent media and particle acceleration}
Magnetic field embedded in a perfectly conducting fluid is generally believed
to preserve its topology for all time \cite{parker79}.  This definitely contradicts the existing evidence that
in almost perfectly
conducting plasmas, e.g., in stars and disks of galaxies, magnetic fields demonstrate the changes in topology, ``magnetic
reconnection'', on dynamical time scales \cite{parker70,lovelace76,priest02}.
Reconnection can be observed directly in the solar corona
\cite{innes97,yokoyama95,masuda94}. While a lot of work in the field has concentrated on showing how
reconnection can be rapid in plasmas with very small collisional rates
\cite{shay98,drake01,drake06,daughton06}, or can develop due to tearing instability, e.g. \cite{Bhattacharjee2016} and ref. therein,
the shortcoming of those studies is that they disregard pre-existing turbulence.

\begin{figure}[!t]
 \begin{center}
\includegraphics[width=0.5\textwidth]{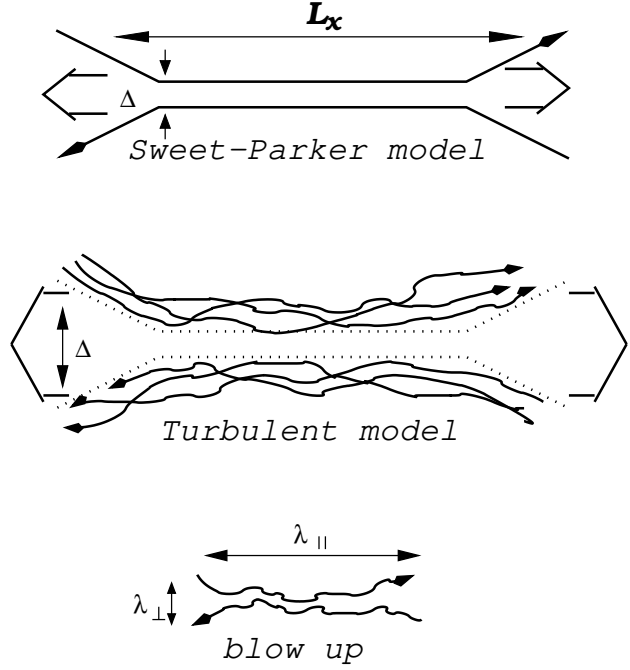}
\caption{
{\it Top panel}: Sweet--Parker model of reconnection. The outflow is limited by
a thin slot $\Delta$, which is determined by ohmic diffusivity. The other scale
is an astrophysical scale $L\gg \Delta$.
{\it Middle panel}: Reconnection of weakly stochastic magnetic field according to
\cite{lazarian99}. The model that accounts for the stochasticity of magnetic field lines. The
outflow is limited by the diffusion of magnetic field lines, which depends on
field line stochasticity.
{\it Bottom panel}: An individual small scale reconnection region. The reconnection
over small patches of magnetic field determines the local reconnection rate. The
global reconnection rate is substantially larger as many independent patches
come together (from \cite{lazarian04a}).
} \label{fig_rec}
 \end{center}
\end{figure}

 A model of turbulent reconnection that was suggested in \cite{lazarian99} is illustrated by Figure \ref{fig_rec}.  In this model, the outflow scale $\Delta$ is determined not by ohmic resistivity, as is the case of the Sweet--Parker model, but by wandering of magnetic field lines. Thus, the level of turbulence controls the reconnection speed $V_\mathrm{rec}\approx V_\mathrm{A} \times \Delta/L$ changes with the turbulence level: the stronger the turbulence, the larger the
reconnection speed. The model has been successfully tested numerically in \cite{kowal09,Kowal2012,Takamoto2015}. Such consequence of the model  the violation of flux freezing in turbulent media was tested in \cite{Eyink2013}. The comparison of the SW measurements and numerics can be found in \cite{Lalescu2015}, while other comparisons of theoretical predictions and observations can be found in \cite{lazarianetal16}. A notable example discussed in \cite{Lalescu2015} is the application of the model from~\cite{lazarian99} to the Parker spiral and heliospheric current sheet \cite{Eyink2015}.

In view of the simulations that have been performed or planned within our study of  the heliosheath processes and  structure of the heliopause, the presence of turbulent reconnection allows us not to worry about the exact reproduction of small-scale (microphysical) plasma processes. Indeed, the model of \cite{lazarian99} predicts reconnection rates that are independent of the microphysics, but only determined by the turbulence level.

We should note here that while the idea of the turbulent enhancement of  reconnection
rates was discussed earlier in  \cite{matthaeus85,matthaeus86} using assumptions clearly different
different from those in~\cite{lazarian99}.  For instance, such key process
as field wandering intrinsic to the model of \cite{lazarian99} has not been considered.
On the contrary, the components of the approach chosen in \cite{matthaeus85,matthaeus86}, e.g., the X-point and possible effects of heating and compressibility, are not used in \cite{lazarian99}.

\begin{figure}[!t]
\centering
\includegraphics[width=0.5\textwidth]{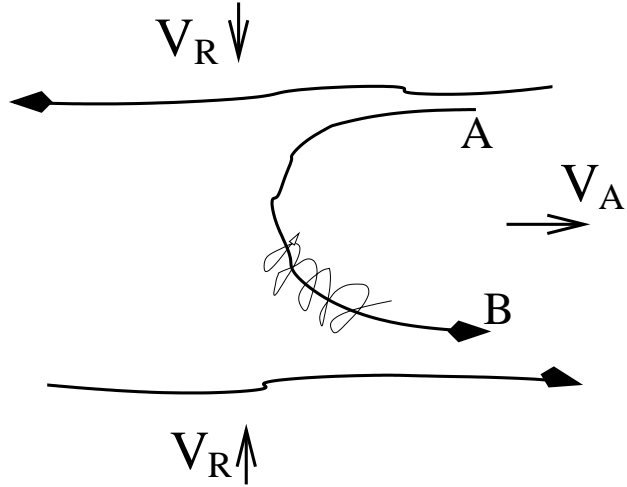}
\caption{ Cosmic rays spiral about a reconnected magnetic field line and bounce
back at points A and B. The reconnected regions move towards each other with the
reconnection velocity $V_{rec}$. The advection of cosmic rays entrained on magnetic
field lines happens at the outflow velocity, which is in most cases of the order
of $V_A$. Bouncing at points A and B happens because either of streaming
instability induced by energetic particles or magnetic turbulence in the
reconnection region. In reality, the outflow region gets filled in by the
oppositely moving tubes of reconnected flux which collide only to repeat on a
smaller scale the pattern of the larger scale reconnection. (From \cite{lazarian05}.})
\label{fig_recon}
\end{figure}

Acceleration of particles is natural within the reconnection model \cite{lazarian99}. Figure~\ref{fig_recon}
exemplifies the simplest scenario of acceleration within the reconnection region expected within model \cite{lazarian99}. As a particle bounces
back and forth between converging magnetic fluxes, it gains energy through the
first order Fermi acceleration as described in \cite{degouveia03,degouveia05,lazarian05}. Later on,  a similar process
was suggested in \cite{drake06} in the framework of tearing mode reconnection. The main
difference between the two processes that the one in Figure~\ref{fig_recon} takes place in 3D,
whereas the one in \cite{drake06} is two dimensional. The latter resulted in artificial constraints on the
acceleration. For instance, the acceleration would stop if magnetic islands produced by reconnection
get circular. In 3D, such reconnection of the line itself is highly improbable and the acceleration
proceeds more efficiently.

\begin{figure}[!t]
\centering
\includegraphics[width=0.8\textwidth]{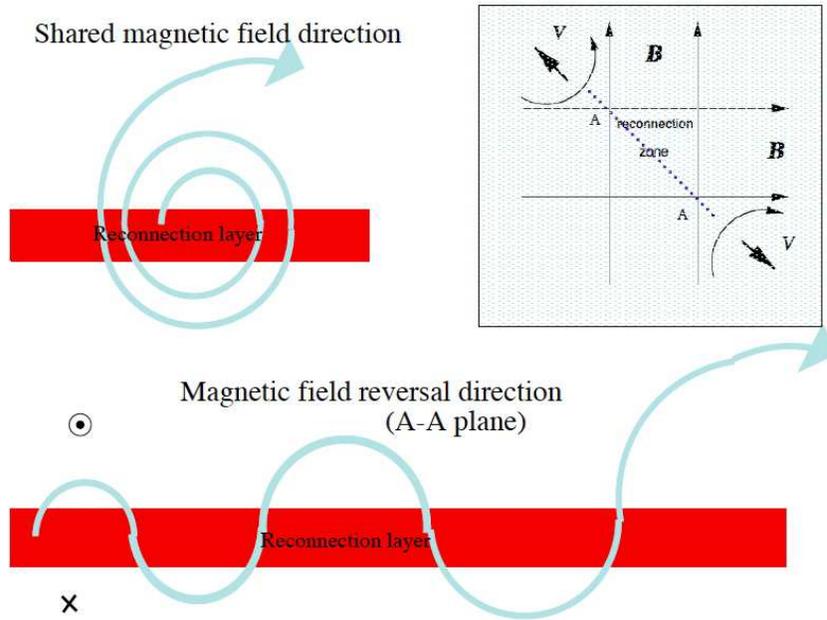}
\caption{Particles with a large Larmor radius gyrate about the magnetic field
shared by two reconnecting fluxes (the latter is frequently referred to as
``guide field''. As the particle interacts with converging magnetized flow
corresponding to the reconnecting components of magnetic field, the particle
gets energy gain during every gyration. (From \cite{lazarian10}.)
\label{fig_accel2}}
\end{figure}

Similarly, the first order Fermi acceleration can happen
in terms of the perpendicular momentum.  This is illustrated in
Figure~\ref{fig_accel2}.  A particle with the large Larmour radius is
bouncing back and forth between converging mirrors of reconnecting magnetic
field is systematically increasing the perpendicular component of its
momentum.  Both processes take place in reconnection layers.

Numerical studies of cosmic ray acceleration in reconnection regions
were performed in  \cite{kowal11,kowal12a}.
\begin{figure}[ht]
 \centering
 \includegraphics[width=0.48\textwidth]{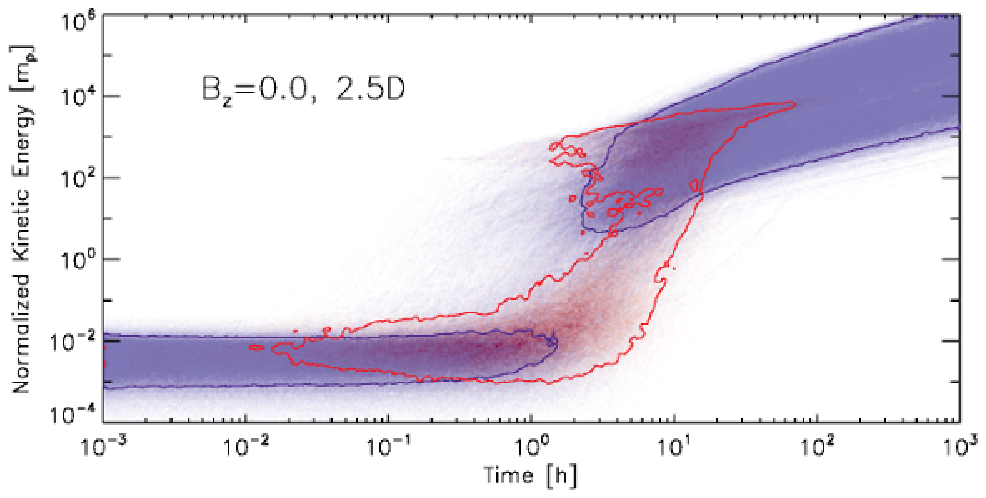}
 \includegraphics[width=0.48\textwidth]{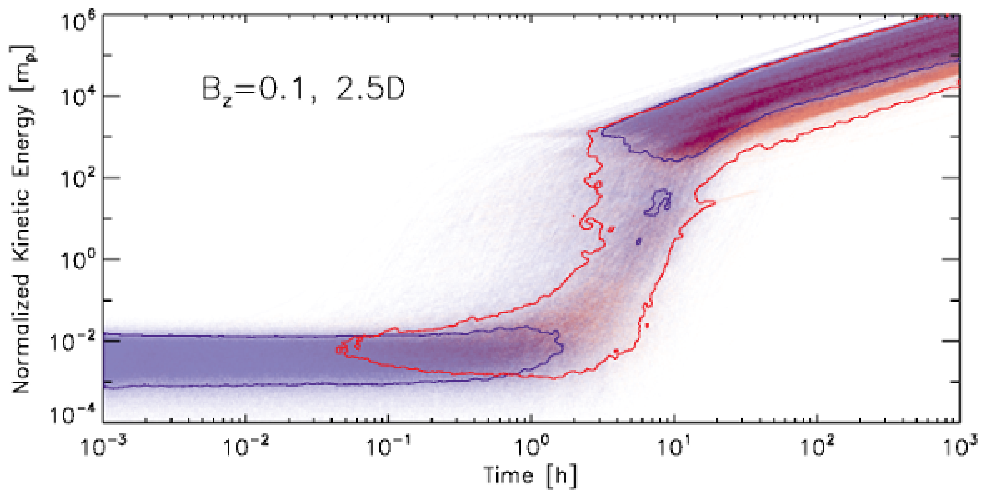}
 \includegraphics[width=0.48\textwidth]{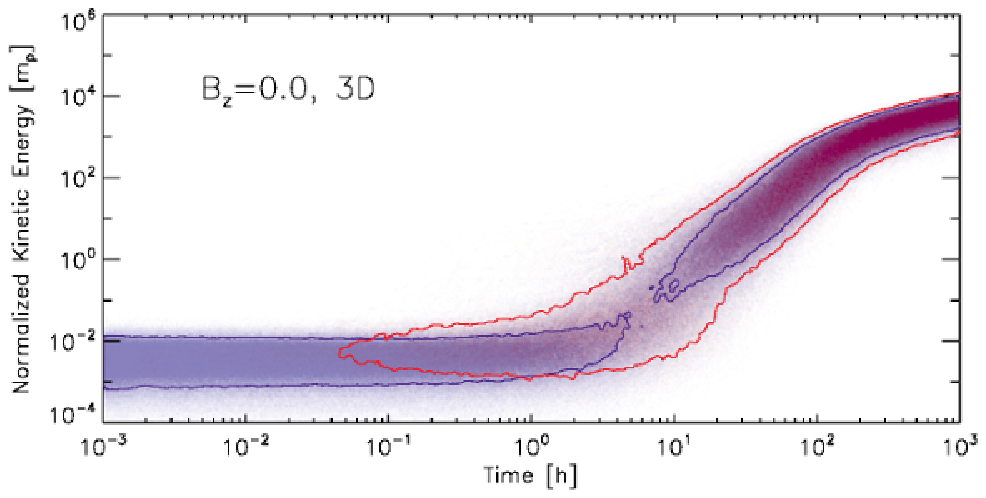}
 \caption{Kinetic energy evolution of a group of 10$^4$ protons in 2D models of
reconnection with a guide field $B_z$=0.0 and 0.1 (top panels, respectively). In
the bottom panel a fully 3D model with initial $B_z$=0.0 is presented.  The
colors show how the parallel (red) and perpendicular (blue) components of the
particle velocities increase with time. The contours correspond to values 0.1
and 0.6 of the maximum number of particles for the parallel and perpendicular
accelerations, respectively.  The energy is normalized by the rest proton mass
energy.  The background magnetized flow with multiple current sheet layers is at
time 4.0 in Alfv\'en time units in all models. From \cite{kowal11}.
\label{fig:energy_2d_3d}}
\end{figure}

Figure~\ref{fig:energy_2d_3d} illustrates the time evolution of the kinetic
energy of particles which have their parallel and perpendicular (red and
blue points, respectively) velocity components increased in three different models of
reconnection.  The upper left panel shows the energy evolution in a 2D model
without any guide field.
Initially, particles get accelerated by increasing their perpendicular
velocity component only.  Later on, an exponential growth of energy is observed
mostly due to the acceleration of the parallel component, which stops after the
energy reaches values of 10$^3$--10$^4$~$m_p$ (where $m_p$ is the proton rest
mass energy).  Finally, particles again increase their perpendicular velocity component,
only a a smaller linear rate.  In a 2.5D case, there is an additional, initially slow increase in
the perpendicular component followed by the
exponential acceleration of the parallel velocity component.  Due to the
effects of a weak guide field, the parallel component increases further to
higher energies at a  rate similar to the perpendicular rate.  This implies that
the presence of the guide field removes the restriction  typical of the 2D model
without guide field and allows particles to increase their parallel
velocity components as they travel along the guide field. This illustrates the advantage of  open loops compared to
2D islands.  This result is reconfirmed by the 3D model (see the
bottom panel of Figure~\ref{fig:energy_2d_3d}), where no guide field is necessary
as the MHD domain in our simulations is three-dimensional.  In this case, we observe a
continuous increase of both components, which suggests that, as expected, the particle
acceleration behavior changes significantly in 3D compared to 2D reconnection.

As far as the heliosphere and the heliotail are concerned, the process of acceleration via turbulent reconnection may
be responsible for the origin of anomalous cosmic rays \cite{lazarian09b} and
giving boost to acceleration of cosmic rays passing through the heliotail \cite{lazarian10}.

\subsection{Scattering and Second order Fermi acceleration by turbulence}
\label{sec:2nd_fermi}
The  process of scattering depends on the statistical
properties of magnetic turbulence that interacts with the particles.  Adopting
the decomposition of compressible MHD turbulence into Alfv\'enic, slow, and fast
\cite{yan02,yan04} identified the fast mode as the
principal mode responsible for scattering and turbulent acceleration of CRs in the
galactic environment.  Later, similar conclusions were made for the CR
acceleration in clusters of galaxies \cite{brunetti07}. We believe that the fast modes
are also very important for heliospheric scattering and acceleration.

The inefficiency of the resonant interaction  of slow and Alfv\'en modes
with cosmic rays \cite{chandran00,yan02} is due to the mode anisotropy, which increases with the scale decrease.  Indeed, the resonant interaction
of the CRs and Alfv\'enic perturbations occurs when the CR Larmor radius is of
the order of the parallel scale of the eddy.  As eddies of scales much less
than the injection scale are very elongated, the CR samples many uncorrelated
eddies, which significantly reduces the interaction efficiency.

\subsection{Perpendicular superdiffusion of cosmic rays}

On scales larger  than the injection scale, cosmic rays follow magnetic field lines that undergo the process of
accelerated divergence, i.e. Richardson diffusion \cite{Lazarian2014}. The characteristic scale of turbulence in the galaxy
is about 150 pc (see \cite{Draine2011,Chepurnov2010}). Therefore energetic particles in the LISM
definitely exhibit superdiffusion perpendicular to the local direction of magnetic field. In fact, as the particles move along
magnetic field lines the distance $s$, the Richardson diffusion causes their deviation in respect to the magnetic field direction
that grows as $\delta_{\bot}\sim s^3$. This is an essential process to take into account in modeling energetic particle behavior
in the LISM.

In addition, the Richardson superdiffusion can be very important for shock acceleration \cite{Lazarian2014}.
Papers \cite{Perri-Zimbardo-2012,Zimbardo-etal-2015}, on the other hand,  propose a hypothetical existence of the Levi flights for the dynamics of particles to make them superdiffusive. Superdiffusion mitigates
the difference between the parallel and perpendicular shock acceleration if magnetic turbulence is subAlfv\'enic. Indeed, the
possibility of returning of the energetic particles streaming along the magnetic field to the shock is significantly reduced for the
perpendicular shock due to the rapid growth of the perpendicular displacement $\delta_{\bot}$.

\section{R\'esum\'e}
With the scientific results presented and discussed in this paper, we have demonstrated the progress that has been made during recent years
in our understanding of the outer heliosphere, the heliopause, and the local interstellar medium. At the same time we have emphasized the
need of the constructive interplay between measurements and model simulations in order to continue to make progress.

Furthermore, we have identified key questions that should be answered by future investigations, namely: (1) What is the proper definition of
the heliopause, i.e.\ what is the true boundary between the heliosphere and the local interstellar medium?, (2) What is the influence of pickup
ions on the structure of the outer heliosphere?, (3) What is the nature of the turbulence in the inner and the outer heliosheath and how does it
influence the transport of energetic particles?, (4) What is the signficance of the multi-species structure of the local interstellar medium for
its interaction with the heliosphere?

Finally, we have pointed out various growing connections between heliospheric physics and astrophysics. On the one hand, they are of
conceptual nature, like the relation of the heliosphere to astrospheres or of the heliotail to astrotails. On the other hand, they
represent actual physical links, like the understanding the local interstellar medium as a representative for the general interstellar medium
or the signature of the heliotail in the flux of TeV cosmic rays. These connections demonstrate the significance of heliophysics research
for astrophysics.

\section*{Acknowledgements}
We are grateful to the International Space Science Institute (ISSI) in Bern, Switzerland, that supported
two meetings for an international team on the topic `Heliosheath Processes and Structure of the Heliopause:
Modeling Energetic Particles, Cosmic Rays, and Magnetic Fields' supported by the International Space Science
Institute (ISSI) in Bern, Switzerland. The work of the USA team was supported, in part, by NASA grants NNX14AJ53G, NNX14AF41G, NNX14AF43G, NNX15AN72G, and NNX16AG83G, and DOE Grant DE-SC0008334. It was also partially supported by the IBEX mission as a part of NASA's Explorer program. We acknowledge NSF PRAC award ACI-1144120 and related computer resources from the Blue Waters sustained-petascale computing project. Supercomputer time allocations were also provided on SGI Pleiades by NASA High-End Computing Program award SMD-15-5860 and on Stampede by NSF XSEDE project MCA07S033. The work of HF, MSP, KS and RDS was partly carried out within the
framework of the bilateral BMBF-NRF-project ``Astrohel'' (01DG15009) funded by the Bundesministerium f\"ur
Bildung und Forschung. The responsibility of the contents of this work is with the authors.

\bibliographystyle{spphys}        

\end{document}